\documentclass[11pt]{article}
\usepackage{jheppub} 

\usepackage{latexsym,amsmath,amsfonts,amssymb, tensor}
\usepackage{graphicx}
\usepackage{epsf}
\usepackage{makeidx}
\usepackage{multirow}
\usepackage[all]{xy}
\usepackage{hyperref}
\usepackage{verbatim} %for comments
\usepackage{rotating}
\usepackage{xcolor}
\usepackage{multirow}
\usepackage{bm}
\usepackage{cleveref}
\usepackage{slashed}
\usepackage[normalem]{ulem}

\usepackage{dirtytalk}

\usepackage{graphicx}
\usepackage{latexsym,amsmath,amsfonts,amssymb}
\usepackage{mathrsfs}
\usepackage{bbm}
\usepackage{bm}
\usepackage{subcaption}
\usepackage{paralist}
\usepackage{youngtab}
\usepackage{bclogo}
\usepackage{mathrsfs}

\usepackage{esvect} %for arrows \vv

\usepackage{framed}
 
\usepackage{mathtools} % for square underbraces

\usepackage{tikz}
\usepackage{tikz-cd}
\usepackage{diagbox}
\usetikzlibrary{positioning}
\usetikzlibrary{calc}
\usetikzlibrary{decorations.pathreplacing,calligraphy}

\usepackage{xstring}
\usetikzlibrary{decorations.pathmorphing} 
\usetikzlibrary{decorations.markings} 
\usetikzlibrary{snakes}
\usetikzlibrary{arrows} 
\usetikzlibrary{shapes} 
\usetikzlibrary{matrix} 
\usetikzlibrary{positioning} 
\usepackage[english]{babel} 
\usepackage[autostyle]{csquotes}
\usetikzlibrary{positioning,fit}

\tikzstyle{brane}=[draw]
\tikzset{D7/.style={circle, draw=black, inner sep=0pt, fill=white, minimum size=3mm}}
\tikzset{hasse/.style={circle, fill,inner sep=2pt}}
\tikzset{flavour/.style={regular polygon,fill=white,regular polygon sides=4,inner sep=2.5pt, draw}}
\tikzset{gauge/.style={circle, draw,inner sep=2.5pt}}
\tikzset{gaugeb/.style={circle, draw,fill=black,inner sep=2.5pt}}
\tikzset{gauger/.style={circle, draw,fill=cyan,inner sep=2.5pt}}
\tikzset{gaugeg/.style={circle, draw,fill=red,inner sep=2.5pt}}
\tikzset{bd/.style={circle, draw=black, inner sep=0pt, fill=black, minimum size=2mm}}
\tikzset{wd/.style={circle, draw=black, inner sep=0pt, fill=white, minimum size=2mm}}
\tikzset{SUd/.style={circle, draw=black, inner sep=0pt, fill=yellow, minimum size=2mm}}
\tikzset{Dynkin/.style={circle, draw=black, inner sep=0pt, fill=white, minimum size=2mm}}
\tikzstyle{ligne}=[draw, thick] 
\tikzset{doublearrow/.style={ draw=black!75, color=black!75, thick, double distance=3pt, }}

\tikzset{->-/.style={decoration={
  markings,
  mark=at position #1 with {\arrow{>}}},postaction={decorate}}}
\tikzset{->>-/.style={decoration={
  markings,
  mark= between positions #1-0.05 and #1+0.05 step 0.1 with {\arrow{>}}
  },postaction={decorate}}}
\tikzset{->>>-/.style={decoration={
  markings,
  mark= between positions #1-0.1 and #1+0.1 step 0.1 with {\arrow{>}}
  },postaction={decorate}}}
  \tikzset{->>>>-/.style={decoration={
  markings,
  mark= between positions #1-0.2 and #1+0.2 step 0.1 with {\arrow{>}}
  },postaction={decorate}}}
    \tikzset{->>>>>-/.style={decoration={
  markings,
  mark= between positions #1-0.3 and #1+0.2 step 0.1 with {\arrow{>}}
  },postaction={decorate}}}
  \tikzset{->>>>>>-/.style={decoration={
  markings,
  mark= between positions #1-0.1 and #1+0.5 step 0.1 with {\arrow{>}}
  },postaction={decorate}}}
\tikzset{node3d/.style={circle,draw,minimum size=1.3cm, very thick}}
\tikzset{node1d/.style={circle,draw,minimum size=1.2cm, thick}}
\tikzset{square/.style={regular polygon,regular polygon sides=4}}
\tikzset{node3ds/.style={circle,draw, minimum size=0.8cm, very thick}}
\tikzset{node1ds/.style={circle,draw, minimum size=0.8cm, thick}}
\tikzset{squareMini/.style={regular polygon,regular polygon sides=4,draw, minimum size=1.1cm, thick}}

\tikzset{cross/.style={cross out, draw=black, minimum size=2*(#1-\pgflinewidth), inner sep=0pt, outer sep=0pt},
cross/.default={1pt}}

\numberwithin{equation}{section}

\newcommand{\nn}{\nonumber}
\newcommand{\mat}[1]{\begin{pmatrix} #1 \end{pmatrix}}
\newcommand{\bmat}[1]{\begin{bmatrix} #1 \end{bmatrix}}
\newcommand{\be}{\begin{equation}} 
\newcommand{\ee}{\end{equation}}
\newcommand{\bea}{\begin{equation} \begin{aligned}} \newcommand{\eea}{\end{aligned} \end{equation}}

\newcommand{\bit}{\begin{itemize}} 
\newcommand{\eit}{\end{itemize}} 

\usepackage{ytableau}
\Ylinethick{0.3pt}
\Yboxdim{5pt}

\newcommand{\cO}{\mathcal{O}}

\newcommand{\Z}{\mathbb{Z}}
\newcommand{\C}{\mathbb{C}}
\newcommand{\K}{\mathbb{K}}

\def\bbC{{\mathbb{C}}}

\def\bbP{{\mathbb{P}}}
\def\bbQ{{\mathbb{Q}}}

\def\bbZ{{\mathbb{Z}}}

\renewcommand{\t}{\widetilde }
\renewcommand{\d}{\partial }

\newcommand{\half}{{1\over 2}}

\newcommand{\CC}{\mathcal{C}}

\newcommand{\CE}{\mathcal{E}}

\newcommand{\CH}{\mathcal{H}}
\newcommand{\CI}{\mathcal{I}}

\newcommand{\CM}{\mathcal{M}}
\newcommand{\CN}{\mathcal{N}}
\newcommand{\CO}{\mathcal{O}}

\newcommand{\CR}{\mathcal{R}}
\newcommand{\CS}{\mathcal{S}}

\newcommand{\CV}{\mathcal{V}}
\newcommand{\CW}{\mathcal{W}}

\newcommand{\FR}{\mathfrak{R}}

\newcommand{\m}{\mathfrak{m}}

\newcommand{\h}{\widehat}

\DeclareMathOperator{\Tr}{Tr}

\newcommand{\SL}{{\mathscr L}}

\newcommand{\ov}{\over}

\newcommand{\dilog}{{\text{Li}_2}}

\DeclareMathAlphabet{\pazocal}{OMS}{zplm}{m}{n}

\usepackage{diagbox}
\usepackage{array}
\makeatletter
\newcommand{\thickhline}{%
    \noalign {\ifnum 0=`}\fi \hrule height 1pt
    \futurelet \reserved@a \@xhline
}
\newcolumntype{"}{@{\hskip\tabcolsep\vrule width 1pt\hskip\tabcolsep}}
\makeatother

\makeatletter
\g@addto@macro{\endtabular}{\rowfont{}}% Clear row font
\makeatother
\newcommand{\rowfonttype}{}% Current row font
\newcommand{\rowfont}[1]{% Set current row font
   \gdef\rowfonttype{#1}#1%
}
\newcolumntype{L}{>{\rowfonttype}l}

\usepackage{multirow}
\usepackage{arydshln}
\begin{document}

% format
\baselineskip=18pt  % a la harvmac
\numberwithin{equation}{section}  % make eq labels (sec.num)
\allowdisplaybreaks  % allow page breaks in displayed eqs

%%%%%%%%%%%%%%%%%%%%%%%%%%%%%%%%%%%%%%%%%%%
%%%        TITLE BEGINS HERE
%%%%%%%%%%%%%%%%%%%%%%%%%%%%%%%%%%%%%%%%%%%

%% ========== title (note version) begins here ==========
%
%\vspace*{-1cm}
%\begin{center}
% {\Large\bf Title of the Document}
%\end{center}
%\vspace*{-.5cm}
%
%% ========== title (note version) ends here ==========

%% ========== title (paper version, a la harvmac) begins here ==========

\thispagestyle{empty}

\vspace*{0.8cm} 
\begin{center}
{{\Huge  
Grothendieck lines in 3d $\CN=2$ SQCD  % 3d $A$-model, \\
%\medskip 
and\\
\medskip 
the quantum K-theory of the Grassmannian %Gr$(N_c, n_f)$
}}

 \vspace*{1.5cm}
Cyril Closset,  Osama Khlaif

 \vspace*{0.7cm} 

 { School of Mathematics, University of Birmingham,\\ 
Watson Building, Edgbaston, Birmingham B15 2TT, United Kingdom}\\

\vspace*{0.8cm}
\end{center}
\vspace*{.5cm}

\noindent
We revisit the 3d GLSM computation of the equivariant quantum K-theory ring  of the complex Grassmannian from the perspective of line defects. 
 The 3d GLSM onto $X={\rm Gr}(N_c, n_f)$ is a circle compactification of the 3d  $\CN=2$ supersymmetric gauge theory with gauge group $U(N_c)_{k, k+l N_c}$ and $n_f$ fundamental chiral multiplets, for any  choice of the Chern-Simons levels $(k,l)$ in the `geometric window'. For $k=N_c-{n_f\ov2}$ and $l=-1$, the twisted chiral ring generated by the half-BPS lines wrapping the circle has been previously identified with the quantum K-theory ring QK$_T(X)$. We identify new half-BPS line defects in the UV gauge theory, dubbed  {\it Grothendieck lines}, which flow to the structure sheaves of the (equivariant) Schubert varieties of $X$. They are defined by coupling $\CN=2$ supersymmetric gauged quantum mechanics of quiver type to the 3d GLSM. We explicitly show that the 1d Witten index of the  defect worldline reproduces the Chern characters for the Schubert classes, which are written in terms of double Grothendieck polynomials. This gives us a physical realisation of the Schubert-class basis  for QK$_T(X)$.  We then use  3d $A$-model techniques to explicitly compute  QK$_T(X)$ as well as other K-theoretic enumerative invariants such as the topological metric.  We also consider the 2d/0d limit of our 3d/1d construction, which gives us local defects in the 2d GLSM, the {\it Schubert defects}, that realise equivariant quantum cohomology classes.

\newpage
%%%%%%%%%%%%%%%%%%%%%%%%%%%%%%%%%%%%%%%%%%%
%%%           TITLE ENDS HERE
%%%%%%%%%%%%%%%%%%%%%%%%%%%%%%%%%%%%%%%%%%%

\tableofcontents
%\printindex

%%%%%%%%%%%%%%%%%%%%%%%%%%%%%%%%%%%%%%%%%%%
%%%        MAIN TEXT BEGINS HERE
%%%%%%%%%%%%%%%%%%%%%%%%%%%%%%%%%%%%%%%%%%%

\section{Introduction}

Two-dimensional  $\CN=(2,2)$  supersymmetric gauge theories often give us useful UV completions of non-linear $\sigma$-models (NLSMs) onto a target space $X$, where $X$ is a K\"ahler manifold. Such 2d gauge theories are known as gauged linear $\sigma$-models (GLSM)~\cite{Witten:1993yc}. The target space $X$ is realised as a classical Higgs branch -- that is, as a Geometric Invariant Theory (GIT) quotient, $X= V /\!\!/ G$, 
where the vector space $V\cong \C^n$ is spanned by VEVs of chiral multiplets, and $G$ is the gauge group. When $G$ is abelian, $X$ is famously a toric manifold.%
\footnote{More generally, it could be a toric variety or even a toric stack -- see~{\it e.g.}~\protect\cite{Pantev:2005zs, Pantev:2005wj, Hellerman:2006zs}.}
In this paper, we consider one of the simplest non-abelian GLSMs, namely the one with a gauge group $G=U(N_c)$ coupled to $n_f$ fundamental chiral multiplets. The target space is then the complex Grassmannian of $N_c$-planes in $\C^{n_f}$:
\be\label{X def intro}
X = \C^{N_c n_f }/\!\!/ U(N_c) \cong {\rm Gr}(N_c, n_f)~.
\ee
Famously, the twisted chiral ring of this 2d $\CN=(2,2)$ gauge theory, $\CR^{\rm 2d}$, is isomorphic to the NLSM twisted chiral ring, which is itself identified with the small (equivariant) quantum cohomology ring of $X$ \cite{Witten:1993xi}:
\be
\CR^{\rm 2d} \cong {\rm Q}{\rm H}_T^{\bullet}(X)~.
\ee
The GLSM naturally encodes the ${ SL}(n_f, \C)$-equivariant deformation of $X$, which corresponds to turning on the twisted mass parameters $m_\alpha$ ($\alpha=1, \cdots, n_f$) for the $SU(n_f)$ flavour symmetry.
 The twisted chiral ring can be deduced most directly from the knowledge of the effective twisted superpotential $\CW(\sigma, m)$ on the 2d Coulomb branch. Moreover, we have a Frobenius algebra structure on $\CR^{\rm 2d}$. The Frobenius metric, also called the topological metric, is given by the two-point function of twisted chiral ring operators in the topological $A$-model on the sphere:
 \be
  \eta(\omega_\mu , \omega_\nu) \equiv  \langle  \omega_\mu \,\omega_\nu \rangle_{\mathbb{P}^1}~,
 \ee
 where $\omega_\mu$ form a $\K$-basis of $\CR^{\rm 2d}$, and we have the topological $A$-twist on  $\mathbb{P}^1\cong S^2$. Note that, in general, we are working over the field of equivariant parameters, which is $\K \equiv \bbQ[m_1, \cdots, m_{n_f}, q_{\rm 2d}]$ for this 2d field theory (here $q_{\rm 2d}$ denotes the 2d `quantum parameter').

In this work, we are interested in the 3d $\CN=2$ uplift of this 2d GLSM, building on a number of previous works in recent years~\cite{ 10.1215/00127094-2010-218,Jockers:2019lwe,Ueda:2019qhg,Gu:2020zpg, Jockers:2021omw, Gu:2022yvj}. To define the 3d theory, we must also specify the Chern-Simons (CS) levels $k$ and $l$ for the 3d gauge group:
\be
U(N_c)_{k, k+ lN_c}\equiv {SU(N_c)_k \times U(1)_{N_c (k+l N_c)}\ov \Z_{N_c}}~.
\ee
Following standard semi-classical methods~\cite{Intriligator:2013lca}, we recently analysed the vacuum structure of this theory as a function of $N_c, n_f$ and  $k, l$, which uncovered a rich structure of infrared vacua \cite{Closset:2023vos}. When the 3d Fayet-Iliopoulos (FI) parameter $\xi$ is positive, there always exists a maximal Higgs branch~\eqref{X def intro}, but there are also many more topological and hybrid vacua in general. In this paper, we will focus on the {\it geometric window} in the parameters $(k, l)$, at fixed $N_c, n_f$, such that the only vacua at $\xi >0$ are on the Higgs branch $X$. In this case, it makes sense to look for a purely geometric interpretation of our 3d GLSM. We consider the 3d $\CN=2$ theory on $\Sigma\times S^1$ and study the theory as an effective 2d $\CN=(2,2)$ theory on $\Sigma$ with a topological $A$-twist~\cite{Nekrasov:2014xaa} -- this is also called the 3d $A$-model~\cite{Closset:2017zgf, Closset:2019hyt}. The twisted chiral operators of this effective theory are half-BPS line operators, $\SL$, wrapped on the $S^1$ factor, and they again form a ring, $\CR^{\rm 3d}$, with the topological metric:
 \be\label{3d top met intro}
  \eta(\SL_\mu , \SL_\nu) \equiv  \langle  \SL_\mu \,\SL_\nu \rangle_{\mathbb{P}^1\times S^1}~.
 \ee
 It is expected that every such line $\SL$ flows to a element of the (equivariant) quantum K-theory of $X$. For instance, it has been proposed that the standard Wilson lines flow to locally free sheaves (that is, vector bundles) on $X$ \cite{Gu:2020zpg, Jockers:2021omw}. In this work, our main focus will be on constructing  new half-BPS operators in the UV gauge theory that flow to more general coherent sheaves with support on Schubert varieties inside $X$. We call these lines the {\it Grothendieck lines}, for reasons that will become clear momentarily.

 It is worth mentioning that there is another approach to quantum K-theory through the study of the partition function of the 3d $\CN=2$ theory on a $D^2\times S^1$~\cite{Jockers:2018sfl,Dedushenko:2023qjq}, where $D^2$ is topologically a disk, which more closely parallels the approach by Givental~\cite{givental2015permutationequivariant}.

\medskip
\noindent
{\bf Quantum K-theory ring from GLSM.} 
 Interestingly, the requirement of being inside the geometric window still allows for many different values of the CS levels $k, l$. For the specific choice:
 \be
 k= N_c- {n_f\ov 2}~, \qquad l=-1~,
 \ee
 it is known that the twisted chiral ring is to be identified with the `standard' (equivariant) quantum K-theory~\cite{givental2000wdvvequation, givental2001quantum, givental2021quantum}  of the Grassmannian~\cite{Ueda:2019qhg,Jockers:2019lwe}:
 \be
 \CR^{\rm 3d} \cong {\rm QK}_T(X)~.
 \ee
This quantum K-theory ring of $X$ was first computed explicitly in~\cite{10.1215/00127094-2010-218}. From the physics perspective, the twisted chiral ring~$ \CR^{\rm 3d}$ is naturally realised as the algebra of Wilson lines~\cite{Kapustin:2013hpk}.  In the present paper, we realise $ \CR^{\rm 3d}$ in terms of defect lines instead, and we show that these Grothendieck lines correspond to the structure sheaves of Schubert varieties of $X$, $\CO_\lambda$ (where the partition $\lambda$ indexes the Schubert varieties), which allows us to directly compare our physics computations to the mathematical results of Buch and Mihalcea~\cite{10.1215/00127094-2010-218}. Of course, we find perfect agreement. We also compute the twisted chiral ring for other values of $k, l$ in the geometric window, but we  leave a full mathematical interpretation of these results for future work --- it is expected that these 3d GLSMs are related to quantum K-theory with `level structure'~\cite{Ruan2018TheLS,RUAN2022108770}, but a more precise understanding remains lacking. 

The 3d $A$-model is best formulated in terms of an effective theory on the 2d Coulomb branch, in which case the ring structure is encoded in the effective twisted superpotential $\CW(x, y)$~\cite{Nekrasov:2009uh,Nekrasov:2014xaa}. In that context, we can write everything in terms of the single-valued gauge parameters $x_a$, $a=1, \cdots, N_c$, and of the flavour parameters $y_\alpha$. These are interpreted as the exponentiated Chern roots of the tautological bundle $S$ over $X$ and the $SU(n_f)$-equivariant parameters (with $y_\alpha= 1$ in the non-equivariant limit), respectively. Any (equivariant) coherent sheaf $\CE$ on $X$ enters the 3d $A$-model through its (equivariant) Chern character ${\rm ch}(\CE)$, which is a polynomial in the $x_a$'s (and rational in the $y_\alpha$'s, in our conventions). In particular, the Chern character of the (equivariant) sheaves $\CO_\lambda$ is given by the double Grothendieck polynomials~\cite{lascoux1982structure, 10.1215/S0012-7094-94-07627-8}:
\be\label{chTO intro}
{\rm ch}_T(\CO_\lambda) =  \mathfrak{G}_\lambda(x,y)~.
\ee
 As an application of the Gr\"obner-bases methods reviewed in~\cite{Closset:2023vos}, we will write down the ring QK$_T(X)$   directly in terms of formal variables $\CO_\lambda$ that represent the Schubert classes.  All obvervables are naturally valued in $\K\equiv \Z(y_1, \cdots, y_{n_f}, q)$, with $q$ the 3d `quantum parameter'.

\medskip
\noindent
{\bf Grothendieck lines as 1d SQM.}  We will define the Grothendieck lines in the UV 3d $\CN=2$ gauge theory in terms of an $\CN=2$ supersymmetric quantum mechanics (SQM) written as a 1d quiver gauge theory coupled to the 3d gauge fields. In particular, the 1d Witten index of the SQM gives us precisely the Chern character of the coherent sheaf to which the defect flows in the infrared. Using the localisation formula for the Witten index of a gauged SQM~\cite{Hori:2014tda}, we demonstrate that the Witten indices of the Grothendieck lines give us precisely the Grothendieck polynomials~\eqref{chTO intro}. 
 This provides a new physical realisation of the Grothendieck polynomials as supersymmetric path integrals of 1d $\CN=2$ quivers.

It is interesting to note that the (double) Grothendieck polynomials can also be realised as wavefunctions in certain integrable systems \cite{motegi2013vertex, Gorbounov:2014bra, wheeler2019littlewood}.  The two perspectives are conjecturally related by the Bethe/gauge correspondence~\cite{Nekrasov:2009uh, Nekrasov:2014xaa}. See~{\it e.g.}~\cite{Bullimore:2017lwu} for a similar construction of point defects in the 2d $\CN=(2,2)$ GLSM for $T^\ast \mathbb{P}^{n_f-1}$ which correspond to the Bethe wavefunctions in the XXX spin chain.

\medskip
\noindent
{\bf JK residues and enumerative invariants.} We will also revisit the computation of K-theoretic enumerative invariants using the 3d $A$-model on $\mathbb{P}^1\times S^1$ as well as the localisation formula for the twisted index~\cite{Benini:2015noa} (see also~\cite{Closset:2015rna, Benini:2016hjo,Closset:2016arn}) with insertion of Grothendieck lines,
\be
\Big\langle \CO_\lambda \, \CO_\mu \, \cdots \Big\rangle_{\mathbb{P}^1 \times S^1}~.
\ee
 All such observables can be computed either from the 3d $A$-model perspective or from a sum over Jeffrey-Kirwan (JK) residues, and we explore both perspectives. We show that, for $(k,l)$ in the geometric window, the JK residue formula only obtains contribution from poles associated with Higgs branch vacua. In particular, for $q=0$, this gives a simple integral representation of the K-theoretic Littlewood-Richardson coefficients. We also compute the topological metric~\eqref{3d top met intro} explicitly. These results are straightforward application of well-established supersymmetric localisation results, which also have a more rigorous mathematical counterpart~\cite{Kim:2016jye, givental2021quantum}.%
 \footnote{More precisely, the results of~\cite{givental2021quantum} should correspond to supersymmetric localisation of the 3d gauge theory on $\C \times S^1$ with the $\Omega$-background on $\C$. It would be interesting to rigorously establish the residue formulas we discuss section~\protect\ref{sec: QKGr}  from a mathematical perspective.}
 
 Finally, we explore the 2d limit of the 3d GLSM, which corresponds to the ordinary GLSM for the Grassmannian. The latter computes the equivariant quantum cohomology QH$^\bullet_T(X)$. In that limit, the Grothendieck lines become defect point operators that represent equivariant cohomology classes. They are realised in the $A$-twisted GLSM as double Schubert polynomials, which we reconstruct from the $\CN=2$ supersymmetric matrix models describing the point defects.

\medskip
\noindent
This paper is organised as follows. In section \ref{sec:3dAmod}, we review aspects of the 3d A-model for the 3d $\CN=2$ $U(N_c)_{k, k+lN_c}$ gauge theory coupled to $n_f$ fundamental chiral multiplets. We recall the JK residue formula for the $\mathbb{P}^1\times S^1$ correlation functions and show how these expressions simplify in the geometric window. In section \ref{sec:lines}, we explicitly construct the Grothendieck lines as 1d supersymmetric quivers, and we demonstrate that these lines flow to the structure sheaves of the Schubert subvarieties of $X={\rm Gr}(N_c, n_f)$. In section~\ref{sec: QKGr}, we revisit the computation of QK$_T(X)$ from the GLSM perspective. In section~\ref{sec:2dlimit}, we briefly discuss the 2d limit of our results. Various supplementary materials are provided in appendix.

 %%%%%%%%%%%%%%%%%%%%%%%%%%%%%%%
  \section{The 3d $A$-model onto the Grassmannian}\label{sec:3dAmod}
 
Let us consider the 3d $\CN=2$ SQCD$[N_c, k, l, n_f, 0]$ theory. That is, we have the gauge group $U(N_c)_{k, k+l N_c}$ coupled to $n_f$ fundamental chiral multiplets, with $k+ {n_f\ov 2}\in \Z$ due to the parity anomaly. Let us study this theory on $\Sigma\times S^1$, with the topological $A$-twist along the Riemann surface  $\Sigma$ . We are interested in the half-BPS line defects that preserve the $A$-twist supercharges, and which must then wrap the $S^1$. Such lines appear as twisted chiral local operators from the perspective of the effective 2d $\CN=(2,2)$ supersymmetric theory on $\Sigma$.
 This effective 2d theory is sometimes called the 3d $A$-model~\cite{Closset:2017zgf}. In the case at hand, it consists of an effective field theory on the classical 2d Coulomb branch (CB). We thus have $N_c$ abelian 2d vector multiplets $\CV_a$ ($a=1,\cdots, N_c$) for a maximal torus $\prod_{a=1}^{N_c} U(1)_a$ of the $U(N_c)$ gauge group, which interact through a one-loop-exact twisted superpotential $\CW(u)$.  
  
Owing to their 3d origin, the 2d CB scalars, denoted by $u_a$, are dimensionless and only defined up to the gauge equivalences $u_a\sim u_a+1$. Similar comments hold for background vector multiplets for 3d flavour symmetries, which give us twisted masses $\nu_\alpha$. It is convenient to use the single-valued variables:
 \be
 x_a\equiv e^{2\pi i u_a}~, \qquad\qquad y_\alpha \equiv e^{2\pi i \nu_\alpha}~, \qquad\qquad q\equiv e^{2\pi i \tau}~.
 \ee
The $SU(n_f)$ flavour fugacities $y_i$  satisfy the constraint $\prod_{\alpha=1}^{n_f} y_\alpha=1$, and  $q$ denotes the exponentiated FI parameter -- here $\tau\equiv \frac{\vartheta}{2\pi} +i\beta \xi$, with $\beta$ the radius of $S^1$ and $\xi$ the real FI parameter for the $U(N_c)$ gauge group. 
 Then, the effective twisted superpotential for SQCD$[N_c, k, l, n_f, 0]$ reads:
 \bea\label{CW full}
& \CW &=&\;  {1\ov (2\pi i)^2} \sum_{\alpha=1}^{n_f} \sum_{a=1}^{N_c}  \dilog(x_a y_\alpha^{-1})
+ \tau \sum_{a=1}^{N_c} u_a \\
 &&&\;   +{k+{n_f\ov 2}\ov 2}\sum_{a=1}^{N_c} u_a(u_a+1) + {l\ov 2} \left(\left(\sum_{a=1}^{N_c} u_a\right)^2 +\sum_{a=1}^{N_c} u_a \right)~.
 \eea
The so-called Bethe equations,  
\be\label{BE Pia}
\Pi_a(x, y, q) = 1~,
\ee
 determine the 2d vacua of the 3d $A$-model, also known as the Bethe vacua. The quantity $\Pi_a$ is  the gauge flux operator:
 \be
  \Pi_a(x, y, q) \equiv e^{2\pi i \d_{u_a} \CW} = q  (-\det{x})^l (-x_a)^{k+{n_f\ov 2}} \prod_{\alpha=1}^{n_f} {1\ov 1-x_a y_\alpha^{-1}}~.
 \ee
Here, we also introduced the notation $\det{x}\equiv \prod_{a=1}^{N_c} x_a$.  A Bethe vacuum, which we denote by $\h x\equiv (\h x_a)$, is defined as a solution to the Bethe equations that is acted on freely by the $U(N_c)$ Weyl group:
\be\label{SBE def}
\CS_{\rm BE} = \Big\{\; \h x \;  \; \Big| \;\Pi_{a}(\h x,  y, q) =1~, \, \forall a\; \;\; {\rm and} \;\; \h x_{a}\neq \h x_{b}~, \forall a\neq b\, \Big\}\Big/S_{N_c}~.
\ee
 Such solutions come in Weyl orbits of $N_c!$ solutions, obtained by permutation of the CB variables $x_a$, and we count the Bethe vacua up to such permutations. The number of Bethe vacua is equal to the Witten index of the 3d $\CN=2$ gauge theory. The latter was computed explicitly in~\cite{Closset:2023jiq} for all values of $N_c, k, l$ and $n_f$.

\subsection{Correlation functions of half-BPS lines and Frobenius algebra}\label{subsec: line observables}
We wish to compute the correlation functions of half-BPS lines, $\SL_\mu$. We will sometimes denote by $\SL$  any given product of such lines located at distinct points on $\Sigma$:
\be
\SL = \prod_s \SL_{\mu_s} (p_s)~.
\ee
Due to the topological $A$-twist, correlation functions do not depend on the positions $p_s\in \Sigma$, hence we will often omit these labels in the following. Parallel lines form a ring,  $\CR^{\rm 3d}$, which is defined physically as the twisted chiral ring obtained by fusion along $\Sigma$, with the product:
\be
\SL_\mu \SL_\nu = {\CN_{\mu\nu}}^\lambda \SL_\lambda~, \qquad \qquad {\CN_{\mu\nu}}^\lambda \, \in\,   \K\equiv \Z(y, q)~.
\ee 
Here, $\{\SL_\mu\}$ forms a $\K$-basis of $\CR^{\rm 3d}$, and the sum over repeated indices is understood. The structure coefficients are encoded by certain $3$-point functions. Moreover, the ring $\CR^{\rm 3d}$ is endowed with a Frobenius algebra structure because the $A$-model is a 2d TQFT on $\Sigma$ -- see~\cite{Ueda:2019qhg} for an explicit discussion in the present case. The non-degenerate Frobenius metric $\eta$, also known as the topological metric, is simply the two point function on the Riemann sphere:
\be
\eta_{\mu\nu} \equiv \eta(\SL_\mu, \SL_\nu) = \Big\langle \SL_\mu(p_1) \SL_\nu(p_2)\Big\rangle_{\mathbb{P}^1\times S^1}~.
\ee 
Let us denote by $\eta^{\mu\nu}$ the inverse metric. Then, the structure constants are obtained from the genus-0 three-point functions according to:
\be
 {\CN_{\mu\nu}}^\lambda =  \eta^{\lambda\delta} \CN_{\mu\nu\delta}~, \qquad 
 \qquad
 \CN_{\mu\nu\delta} = \Big\langle \SL_\mu(p_1) \SL_\nu(p_2) \SL_\delta(p_3)\Big\rangle_{\mathbb{P}^1\times S^1}~.
\ee
Here, the `expectation value' $\langle \cdots \rangle_{\Sigma\times S^1}$ denotes the unnormalised path integral of the 3d $\CN=2$ supersymmetric field theory on $\Sigma\times S^1$ with the $A$-twist along $\Sigma$ and with the periodic spin structure on $S^1$. 
Note that these observables are valued in the field $\K=\Z(y,q)$ of rational functions  with integer coefficients of the 3d flavour parameters $y_\alpha$ and $q$. The integrality property is related to the fact that we are considering a 3d $\CN=2$ theory on $\Sigma \times S^1$, hence the observables could also be computed, in principle, as traces over Hilbert spaces of certain effective supersymmetric quantum mechanics on $S^1$ -- see {\it e.g.}~\cite{Bullimore:2018yyb, Bullimore:2019qnt, Bullimore:2020nhv}. 

\medskip
\noindent
Let us now review two distinct ways by which we will explicitly compute these observables:

\medskip
\noindent
{\bf The Bethe-vacua formula.} From the perspective of the infrared Coulomb-branch theory   on $\Sigma$, one can compute any  correlator on the sphere as a sum over the Bethe vacua~\cite{Nekrasov:2014xaa}:
\be\label{sum bethe vacua formula}
       \Big\langle\SL\Big\rangle_{\bbP^1\times S^1}  =  \sum_{\h x \in \CS_{\rm BE} } \CH(\h x, y)^{-1}\,  \SL(\h x, y, q)~.
\ee
Here, $\CH$ is known as the handle-gluing operator. As an explicit rational function in the gauge parameters, it reads:
\be
\CH(x, y, q) = \det\left({\bf H}\right)\, e^{2\pi i \Omega}~,
\ee
with ${\bf H}$ the Hessian matrix:
\be
{\bf H}_{ab}(x, y) = {\d \CW\ov \d u_a \d u_b} =\delta_{ab} \left( k+{n_f\ov 2}+  \sum_{\alpha=1}^{n_f}{x_a y_\alpha^{-1}\ov 1- x_a y_\alpha^{-1}} \right)  + l~,
\ee
and $\Omega$ the effective dilaton potential, which is given by:
\be\label{Omega full}
e^{2\pi i \Omega} = \prod_{\alpha=1}^{n_f}\prod_{a=1}^{N_c} (1- x_a y^{-1}_\alpha)^{-r+1} \prod_{\substack{a,b\\ a\neq b}} (1- x_a x_b^{-1})^{-1} \, (\det x)^{K_{RG}}~.
\ee
Note the dependence on the $R$-charge $r\in \Z$ for the chiral multiplets and on the gauge-R bare CS level $K_{GR}$ -- we will shortly set $r=0$ and $K_{GR}=0$.  In writing down~\eqref{CW full} and \eqref{Omega full}, we set all bare flavour and R-symmetry CS levels to zero, $K_{FF}=K_{FR}=K_{RR}=0$, following the conventions of~\cite{Closset:2023vos}. In~\eqref{sum bethe vacua formula}, the product of lines, $\SL$, is represented by a polynomial $\SL(x, y, q)$. These line operators will be the studied in more detail in section~\ref{sec:lines}.

In principle, the sum over Bethe vacua~\eqref{sum bethe vacua formula} can be performed explicitly using computational algebraic geometry methods, as we explained in~\cite{Closset:2023vos}.  In practice,  this method relies on Gr\"obner basis algorithms that quickly become computationally prohibitive as we increase $n_f$ and $N_c$. Such methods can nonetheless be used to efficiently compute the quantum K-theory ring of the Grassmannian for small-enough values of $N_c$ and $n_f$, as we will discuss in section~\ref{sec: QKGr}.

\medskip
\noindent
{\bf The JK residue formula.} The other method to compute correlation functions on the sphere is through supersymmetric localisation in the UV theory, which leads to a JK residue formula~\cite{Benini:2015noa} (see also~\cite{Benini:2016hjo, Closset:2016arn}):  
    \begin{equation}\label{JK-residue-formula}
    \Big\langle\SL\Big\rangle_{\bbP^1\times S^1}   = \frac{1}{N_c!} \sum_{\mathfrak{m}\in \bbZ^{N_c}} \sum_{x_* \in \t{\mathfrak{M}}_{\text{sing}}^{\mathfrak{m}}} \underset{x = x_*}{\text{JK-Res}} \left[\textbf{Q}(x_*), \eta_\xi \right]\mathfrak{I}_{\mathfrak{m}}[\SL](x, y, q)~.
    \end{equation}
    This formula involves a  sum over all $U(N_c)$ magnetic fluxes $\m=(\m_a)\in \Z^{N_c}$ through $\mathbb{P}^1$, and a sum over JK residues in each flux sector.
 The JK residues are taken with respect to the $N_c$-form:
    \begin{equation}\label{JK form}
        \mathfrak{I}_\mathfrak{m}[\SL](x, y, q)  = (-2\pi i)^{N_c} Z_{\mathfrak{m}}(x, y, q) \SL(x, y, q)\; \frac{dx_1}{x_1} \wedge \cdots \wedge \frac{dx_{N_c}}{x_{N_c}}~,
\end{equation}
which has codimension-$N_c$ singularities (including at infinity) denoted by $\t{\mathfrak{M}}_{\text{sing}}$. For each magnetic  flux $\m\in \Z^{N_c}$, the factor $Z_{\mathfrak{m}}$ is given in terms  of the effective dilaton potential and gauge flux operators as:
\be
Z_{\mathfrak{m}}(x, y, q) =e^{-2\pi i \Omega} \prod_{a=1}^{N_c} \Pi_a(x, y, q)^{\m_a}~.
\ee
To each singularity $x_\ast\in \t{\mathfrak{M}}_{\text{sing}}$, one assigns a charge vector ${\bf Q}(x_\ast)$ which determines whether or not the singularity contributes non-trivially to the JK residue, given a choice of the auxiliary parameter $\eta_\xi$.  In appendix~\ref{app: JK residue}, we further study this JK residue formula for the SQCD$[N_c, k, l, n_f, 0]$ theory. In particular, we show that, for a certain choice of $\eta_\xi$,  the sum over singularities $x_*$  is closely related to the sum over 3d vacua that we recently studied in~\cite{Closset:2023jiq}.

   %%%%%%%%
 \subsection{The geometric window and the Grassmannian 3d GLSMs}\label{subsec:geom window}
For positive FI parameters and vanishing masses,  $\xi>0$ and $\nu_\alpha=0$, the SQCD$[N_c, k, l, n_f, 0]$ theory has a pure Higgs branch given by the Grassmannian manifold:
\be\label{CMHiggs}
\CM_{\rm Higgs} = {\rm Gr}(N_c, n_f)~,
\ee
 For generic values of $k$ and $l$, the theory also has a number of additional topological and hybrid vacua~\cite{Closset:2023jiq}, which precludes a purely geometric interpretation of the infrared physics. At fixed $N_c$ and $n_f$ with $n_f \geq N_c$, we will say that the $U(N_c)$ gauge theory is in the {\it geometric window} if and only if its Witten index is equal to the Euler characteristic of the Grassmannian:
 \be
 {\bf I}_W[N_c, k, l, n_f, 0]= \chi({\rm Gr}(N_c, n_f))= \mat{n_f \\ N_c}~,
 \ee
 with ${\bf I}_W[N_c, k, l, n_f, 0]$ the index computed in~\cite{Closset:2023jiq}.
The geometric window intersects the subset of theories with $l=0$ at $|k|\leq {n_f\ov 2}$. For $l\neq 0$, we find a larger but finite number of theories in the geometric window.%
 \footnote{We exclude the case $N_c=1$ from this discussion, since $l$ is a redundant parameter in that case. A completely explicit (though unwieldy) formula for ${\bf I}_W[N_c, k, l, n_f, 0]$ is given in~\protect\cite{Closset:2023jiq}. We do not have any more elegant description of the geometric window at the moment. 
 %\CyC{give examples in appendix?}
} Let $n_{\rm gw}(N_c, n_f)$ denote the number of such theories with $l<0$. Then, the number of distinct theories in the geometric window is:
 \be
 N_{\rm gw}^{\rm tot}(N_c, n_f) = 2 n_{\rm gw}(N_c, n_f)+ n_f+1~.
 \ee 
 Using the known formula for the Witten index~\cite{Closset:2023jiq}, we can compute $n_{\rm gw}$ by brute force, as shown in table~\ref{tab: ngw}. For $N_c=2$ and $N_c=3$ (and $n_f>N_c$), we find the patterns:
 \bea
 &n_{\rm gw}(2, n_f)= {n_f(n_f+1)\ov 2} +1 + \delta_{n_f, 3}~, \\
&  n_{\rm gw}(3, n_f)={ \begin{cases}{n_f^2\ov 4} +2 + 3 \delta_{n_f, 4} \quad & \text{if}\; n_f \; \text{is even,}\\ 
  {n_f^2\ov 4} +{7\ov 4} +  \delta_{n_f, 5} \quad & \text{if}\; n_f \; \text{is odd.}\end{cases}}
 \eea
We did not find any clear pattern in general, however. 
    %%%%%
\begin{table}[t]
\renewcommand{\arraystretch}{1.1}
\centering{\footnotesize
\be\nn
 \begin{array}{|c||c|c|c|c|c|c|c|c|c|c|c|c|c|c|c|c|c|c|c|c|c|c|c|c|c|}
 \hline
 N_c \backslash n_f &3 & 4& 5 & 6 & 7 & 8 & 9 & 10 & 11 & 12 & 13 & 14 & 15 & 16 & 17 & 18
   & 19 & 20 & 21& 22& 23& 24& 25& 26& 27 \\ \hline \hline
2 & 8 & 11 & 16 & 22 & 29 & 37 & 46 & 56 & 67 & 79 & 92 & 106 & 121 & 137 & 154 & 172 & 191 & 211 & 232 &
   254 & 277 & 301 & 326 & 352 & 379 \\\hline
 3 & - & 9 & 9 & 11 & 14 & 18 & 22 & 27 & 32 & 38 & 44 & 51 & 58 & 66 & 74 & 83 & 92 & 102 & 112 & 123 &
   134 & 146 & 158 & 171 & 184 \\\hline
 4 & - & - & 12 & 10 & 11 & 12 & 15 & 18 & 21 & 25 & 29 & 33 & 38 & 43 & 48 & 54 & 60 & 66 & 73 & 80 & 87
   & 95 & 103 & 111 & 120 \\\hline
 5 & - & - & - & 17 & 12 & 12 & 13 & 14 & 16 & 19 & 22 & 25 & 28 & 32 & 36 & 40 & 44 & 49 & 54 & 59 & 64 &
   70 & 76 & 82 & 88 \\\hline
 6 & - & - & - & - & 23 & 15 & 13 & 14 & 15 & 16 & 18 & 20 & 23 & 26 & 29 & 32 & 35 & 39 & 43 & 47 & 51 &
   55 & 60 & 65 & 70 \\\hline
 7 & - & - & - & - & - & 30 & 19 & 16 & 15 & 16 & 17 & 18 & 20 & 22 & 24 & 27 & 30 & 33 & 36 & 39 & 42 &
   46 & 50 & 54 & 58 \\\hline
 8 & - & - & - & - & - & - & 38 & 23 & 19 & 17 & 17 & 18 & 19 & 20 & 22 & 24 & 26 & 28 & 31 & 34 & 37 & 40
   & 43 & 46 & 49 \\\hline
 9 & - & - & - & - & - & - & - & 47 & 28 & 22 & 20 & 19 & 19 & 20 & 21 & 22 & 24 & 26 & 28 & 30 & 32 & 35
   & 38 & 41 & 44 \\\hline
 10 & - & - & - & - & - & - & - & - & 57 & 33 & 26 & 23 & 21 & 21 & 21 & 22 & 23 & 24 & 26 & 28 & 30 & 32 &
   34 & 36 & 39 \\\hline
 11 & - & - & - & - & - & - & - & - & - & 68 & 39 & 30 & 26 & 24 & 23 & 23 & 23 & 24 & 25 & 26 & 28 & 30 &
   32 & 34 & 36 \\\hline
 12 & - & - & - & - & - & - & - & - & - & - & 80 & 45 & 34 & 29 & 27 & 25 & 25 & 25 & 25 & 26 & 27 & 28 &
   30 & 32 & 34 \\ \hline
   \end{array} 
\ee}
\caption{Some values of $n_{gw}(N_c, n_f)$, the number of theories in the geometric window with $l<0$.}
\label{tab: ngw}
\end{table}
%%%%%%%%
 It may be useful to further distinguish between theories in the geometric window which are maximally chiral ($|k|<{n_f\ov 2}$), marginally chiral ($|k|={n_f\ov 2}$), or minimally chiral ($|k|>{n_f\ov 2}$). We denote the number of such theories with $l<0$ by $n_{\rm gw}^+$,  $n_{\rm gw}^0$ and  $n_{\rm gw}^-$,  respectively, with $n_{\rm gw}= n_{\rm gw}^++n_{\rm gw}^0+n_{\rm gw}^-$. The three types of theories are distinguished by their infrared dual description~\cite{Nii:2020ikd, Amariti:2021snj, Closset:2023vos}.  
 In the maximally chiral case, we have the duality~\cite{Closset:2023vos}:
\be \label{max chiral dual}
U(N_c)_{k,\,k+l N_c}~, \;   n_f \,  {\tiny\yng(1)}   \qquad
\longleftrightarrow \qquad
U(n_f-N_c)_{-k, \, -k + l (n_f-N_c)}~, \;    n_f\, \overline{\tiny\yng(1)}~,
\ee
 in which case the Higgs branch vacua are easily matched as:
 \be
{\rm Gr}(N_c, n_f) \cong {\rm Gr}(n_f-N_c, n_f)~.
\ee
One can also check, in examples, %\CyC{table?}
 that we have $n_{\rm gw}^+(N_c, n_f)= n_{\rm gw}^+(n_f- N_c, n_f)$,
 as one would expect from the duality~\eqref{max chiral dual}. The dualities for the marginally and maximally chiral theories are more complicated, with the dual gauge group being $U(|k|+{n_f\ov 2} -N_c)\times U(1)$; in that case, in the classification of~\cite{Closset:2023jiq}, the Higgs branch~\eqref{CMHiggs} appears in the dual gauge theory as a hybrid Higgs-topological vacua such that the would-be TQFT factor has a single state. In this work, we shall focus on the `electric' SQCD description; it would certainly be interesting to understand better the duality map on defect lines (beyond the well-understood $l=0$ case~\cite{Closset:2016arn}).

 \medskip
 \noindent
 {\bf Grassmannian 3d $\CN=2$ GLSMs.} For the massless SQCD$[N_c, k, l, n_f, 0]$ theory with $(k,l)$ in the geometric window, by definition, we have a unique Higgs branch of vacua~\eqref{CMHiggs} when $\xi>0$. Turning on the $SU(n_f)$ mass parameters, $y_\alpha\neq 1$, along the maximal torus $T\subset U(n_f)$, one achieves a $T$-equivariant deformation of the geometry, and the Higgs branch collapses to  $\chi({\rm Gr}(N_c, n_f))$ massive vacua.
 
 Picking any $(k,l)$ in the geometric window, we then obtain a 3d GLSM which corresponds to the 3d gauge theory on $\Sigma\times S^1$. Following the 3d renormalisation group flow, it is expected that the gauge theory with $\xi\gg 0$ flows to an infrared 3d NLSM onto $X\equiv {\rm Gr}(N_c, n_f)$:  
\be
\text{3d NLSM} \; : \; \Sigma \times S^1 \longrightarrow X~.
\ee
We will not attempt to precisely define the NLSM infrared phase of our theory, however.
Instead, more conservatively, we shall define and study the 3d GLSM as an ordinary 2d GLSM for the effective 2d field theory on $\Sigma$ at scales $\mu \ll {1\ov \beta}$, with $\beta$ the radius of $S^1$. The 3d origin of this construction remains apparent in the 2d gauge-theory description through the non-trivial periodicities of the CB scalars $u_a$ and through the explicit form of the effective twisted superpotential and effective dilaton potential~\eqref{CW full} and~\eqref{Omega full}. Most importantly, the twisted chiral operators are now line defects, as emphasised above. In the next section, we study these lines operators in detail. We will come back to the study of the 3d GLSM observables in section~\ref{sec: QKGr}.

 \subsection{{$\mathbb{P}^1\times S^1$} correlation functions in the geometric window}\label{subsec: genus-0 correlator}

In the geometric window, the JK residue formula \eqref{JK-residue-formula} for the genus-$0$ correlators simplifies significantly. Then, as we explain in appendix~\ref{app: JK residue}, the only contributing singularities are ``Higgs branch'' singularities (where chiral multiplets go massless on the 3d Coulomb branch). Let us write the correlators as a sum over topological sectors, as:
\begin{equation}\label{KGW}
\Big\langle\SL\Big\rangle_{\bbP^1\times S^1}  (q,y) = \sum_{d= 0}^\infty  q^{d} \;\textbf{I}_{d}[{\SL}](y)~, 
   \end{equation}
where $d= |\m|\equiv \sum_{a=1}^{N_c} \m$ is the magnetic flux for the overall $U(1)\subset U(N_c)$. It corresponds to the degree of the holomorphic map $\phi\, : \, \Sigma \rightarrow X$ in the infrared NLSM realisation. 
 At each degree, we have the residue formula:
\be\label{Id explicit}
\textbf{I}_{d} [{\SL}](y) \equiv \sum_{\substack{\m_a \geq 0 \\ |\m|=d}}\;\; \sum_{1\leq \alpha_1<\cdots <\alpha_{N_c} \leq n_f} \underset{\{x_a = y_{\alpha_a}\}}{\text{Res}}\;\frac{(-1)^{|\m|(K+l) +N_c}\Delta(x) {\SL}(x,y)}{\prod_{a=1}^{N_c}x_a^{\mathfrak{r}_a}\prod_{\alpha=1}^{n_f}\left(1-x_a y_\alpha^{-1}\right)^{1+\mathfrak{m}_a}}~.
\ee
Note  that we have set the $R$-charge $r=0$, which is the natural choice from the GLSM point of view. We also defined the integers:%
% \footnote{Not to be confused with the labels $M_l$ used to define line defects in the previous section.}
\begin{equation}\label{ra}
    \mathfrak{r}_a \equiv N_c + K_{GR} - l |\mathfrak{m}| - K \mathfrak{m}_{a}~, \qquad a = 1, \cdots, N_c~,
\end{equation}
at fixed $\m$, and the Vandermonde determinant:
\begin{equation}\label{vandermonde}
    \Delta(x) \equiv \prod_{1\leq a\neq b\leq N_c} (x_a - x_b)~.
\end{equation}
The expression~\eqref{Id explicit} is given by a finite sum over all the magnetic fluxes $\m_a\geq 0$ at fixed degree $d=|\m|$. Then, in each flux sector $\m=\{\m_a\}$, we sum over all the ``Higgs-branch'' residues at:
\be\label{x to y HB sing}
(x_1, \cdots, x_{N_c})= (y_{\alpha_1}, \cdots, y_{\alpha_{N_c}})~, \qquad 
1\leq \alpha_1<\cdots <\alpha_{N_c} \leq n_f~.
\ee 
We use the $S_{N_c}$ gauge symmetry to order the singularities as indicated, thus cancelling out the $N_c!$ factor in~\eqref{JK-residue-formula}.%
\footnote{We also used the fact that singularities with $x_a=x_b=y_\alpha$ for som $a\neq b$ have vanishing residue due to the Vandermonde determinant in the numerator.}

\medskip
\noindent
{\bf The non-equivariant limit.} The residue formula~\eqref{Id explicit}  is  valid for generic values of the equivariant parameters $y_\alpha$. The JK residue formula~\eqref{JK-residue-formula} holds more generally, however. In particular, let us consider the non-equivariant limit, setting $y_i=1$. Then, we have:
\be\label{Id explicit nonEq}
\textbf{I}_{d} [{\SL}] \equiv \sum_{\substack{\m_a \geq 0 \\ |\m|=d}} \underset{\{x_a = 1\}}{\text{Res}}\;\frac{(-1)^{\mathfrak{m}(K+l) +N_c}\Delta(x) {\SL}(x)}{N_c! \prod_{a=1}^{N_c}x_a^{\mathfrak{r}_a}\left(1-x_a \right)^{n_f(1+\mathfrak{m}_a)}}~.
\ee
with a single residue at the codimension-$N_c$ singularly $x_a=1$, in each flux sector. (Note the factor of $1/N_c!$ compared to~\eqref{Id explicit}.)

\medskip
\noindent
{\bf $U(1)$ example.} As a simple example, consider the GLSM onto $\mathbb{P}^{n_f-1}$, which is the $U(1)_k$ theory with $n_f$ charge-$1$ chiral multiplets ($K\equiv k+{n_f\ov 2}$), with the constraint $0\leq K \leq n_f$ so that we are in the geometric window. We also choose $K_{RG}=0$. Focussing on the non-equivariant limit, the formula~\eqref{Id explicit} gives us:
\bea
&\textbf{I}_{\m} [{\SL}]  &=&\; {(-1)^{\m K+1}\ov 2\pi i} \oint {dx\ov x^{1-K \m}} \,{\SL(x)\ov (1-x)^{n_f(\m+1)}}\\
& &=&\;{ (-1)^{n_f(\m+1)+\m K+1}\ov (n_f(\m+1)-1)!} \left[{d^{n_f(\m+1)-1}\ov dx^{n_f(\m+1)-1}} \, {\SL(x)\ov x^{1-K\m}}\right] \Bigg|_{x=1}~,
\eea
for the non-negative integer $\m=d$. In particular, we find that the partition function on the sphere is given by:
\be
Z_{\mathbb{P}^1 \times S^1} = \langle 1 \rangle_{\mathbb{P}^1 \times S^1} =
\begin{cases}
    1\quad  \qquad &\text{if}\;\; 0 <K \leq n_f~,\\
    {1\ov 1-q}\quad  \qquad & \text{if}\; \; K=0~.
\end{cases}~,
\ee
Indeed, we easily see that $\textbf{I}_{\m} [1]=1$ if $\m=0$ or if $K=0$ (for any $m\geq 0$), while $\textbf{I}_{\m} [1]=0$ when $0<K\leq n_f$ and $\m>0$ because $K\m -1 < n_f(\m+1)$ and $K\m-1\geq 0$ in that case.  
 
 %%%%%%%%%%%%%%%%%%%%%%%%%%%%%%%%%%%%
 \section{Wilson lines and Grothendieck lines}\label{sec:lines}
 
The 3d uplift of the standard 2d GLSM modifies the target-space interpretation of the twisted chiral operators. While the 2d local operators $\omega\in \CR^{\rm 2d}$ represent cohomology classes on the target space $X$, the 3d line operators wrapping the $S^1$ should be interpreted as coherent sheaves on $X$:
 \be\label{RG map lines}
 \SL\in \CR^{\rm 3d}~, \qquad\qquad\qquad {\rm RG:}\;   \SL \longrightarrow  \CE_\SL \in {\rm coh}(X)~.
 \ee
It is expected that the RG flow maps any half-BPS defect line $\SL$ defined in the UV 3d $\CN=2$ gauge theory to a coherent sheaf $\CE_\SL$ on the target space $X$.  More precisely, as we will discuss further below, the physical observables only depend on the Grothendieck group of the abelian category ${\rm coh}(X)$ -- that is, on the K-theory of $X$:
 \be
 \SL\; \rightarrow\; [\CE_\SL] \in {\rm K}(X)~.
 \ee
 A proper, first-principle understanding of this map would require a better understanding of the 3d NLSM phase itself.  We will not try to tackle this problem here. 
  Instead, we will take a pedestrian view of K-theory based on representing sheaves by their Chern character:
  \be
  {\rm ch}\, :\; {\rm K}(X)\rightarrow {\rm H}^\bullet(X)~:\; [\CE_\SL] \mapsto {\rm ch}(\CE_\SL)~. 
  \ee
This is particularly suited to the 3d $A$-model description, as we will review momentarily. From this perspective, it will become clear that the physical observables defined in section~\ref{subsec: line observables} can only depend on the K-theory class of the line $\SL$, because they only depend on the Chern characters.

A chief motivation to study this 3d GLSM into the Grassmannian is to give a physics derivation of the quantum K-theory ring of $X={\rm Gr}(N_c, n_f)$~\cite{10.1215/S0012-7094-04-12131-1, 10.1215/00127094-2010-218}. In the recent literature, this was achieved by studying Wilson lines in the UV gauge theory~\cite{Kapustin:2013hpk, Jockers:2018sfl, Jockers:2019lwe, Ueda:2019qhg}. Under the map~\eqref{RG map lines}, Wilson lines map to locally free sheaves  ({\it i.e.} to complex vector bundles) on $X$. In this section, we construct a new class of UV line defects, dubbed Grothendieck lines, which flow to sheaves with compact support on subvarieties of $X$. The Grothendieck lines are instances of vortex loops, as studied {\it e.g.} in~\cite{Kapustin:2012iw, Drukker:2012sr, Assel:2015oxa, Hosomichi:2021gxe}. They give us a much more natural basis to describe the quantum K-theory ring of $X$, in direct parallel with standard mathematical results~\cite{10.1215/00127094-2010-218}.

  \subsection{The Coulomb-branch perspective}
 In the 3d $A$-model description~\cite{Closset:2017zgf, Closset:2019hyt}, we deal with the effective field theory on the 2d $\CN=(2,2)$ Coulomb branch parameterised by the dimensionless quantities:
 \be
 u_a= i \beta \sigma_a -  a_{0,a}~, \qquad\qquad a_{0, a} \equiv {1\ov 2 \pi} \int_{S^1} A_a~,\qquad \qquad a=1, \cdots, N_c~,
 \ee
 where $\sigma_a$ and $A_a$ denote the real scalar and 1-form gauge field, respectively, in the 3d $\CN=2$ $U(1)_a$ vector multiplet. The single-valued parameters $x_a$ are best interpreted as the holonomies of the half-BPS Wilson loops of unit charge for these abelian vector multiplets:
 \be\label{exponential-of-sigma}
 x_a =e^{2\pi i u_a} = \exp\left( - i \int_{S^1} (A_a- i \sigma_a d\psi)\right)~.
 \ee
The abelianised theory retains the Weyl group $S_{N_c}$ as a residual gauge symmetry that acts by permutation on the $x_a$ variables.   
 
 Given the Grassmannian $X={\rm Gr}(N_c, n_f)$, we have the rank-$N_c$ tautological vector bundle (also called the universal subbundle), $S$, whose fiber at a point $p\in X$ is the linear subspace $\C^{N_c}\subset \C^{n_f}$ that this point represents:
 \be
 \C^{N_c} \rightarrow S \rightarrow X~.
 \ee
 In the 2d GLSM, the scalars $\sigma_a$ can be identified with the curvature of $S$~\cite{Witten:1993xi}.  By the same token, we can identify $2\pi i u_a$ with the Chern roots of $S$, so that:
 \be
 {\rm ch}(S)= \sum_{a=1}^{N_c} x_a~.
 \ee
 More generally, the Chern character of any vector bundle (or coherent sheaf) on $X$ can be written as a symmetric polynomial in the $x_a$'s.  Given any UV lines $\SL$, the $A$-model computation of their correlation functions only depends on the symmetric polynomials that give us the Chern classes of the coherent sheaves $\CE_\SL$:
 \be
 \SL(x)\equiv {\rm ch}(\CE_\SL) \, \in \,  {\rm K}(X)~,
 \ee
 schematically.  While we have set $y_i=1$ in this discussion, turning on the flavour fugacities simply corresponds to a $T$-equivariant deformation, as already mentioned, so that we have: 
  \be
 \SL(x, y) \, \in \,  {\rm K}_T(X)~,
 \ee
 with $ \SL(x, y)$ representing the equivariant Chern character of some $T$-equivariant coherent sheaf. 
  In the rest of this section, we identify the UV lines that flow to some important classes of coherent sheaves on the Grassmannian. 
 
 \subsection{Wilson lines and vector bundles}\label{sec: Grothend-lines}
 Given any representation $\FR$ of $U(N_c)$, we have the half-BPS Wilson line:
 \be
 W_\FR = \Tr_\FR \left( P e^{-i \int_{S^1} \left(A- i \sigma d\psi\right)}\right)~,
 \ee
 whose classical VEV gives us the Chern character:
 \be
  W_\FR(x)\equiv {\rm ch}(\CE_{W_\FR}) = \Tr_\FR(x)~.
 \ee
 In particular, the fundamental Wilson line, $W_ {\tiny\yng(1)}$, flows to the tautological line bundle:
 \be
\CE_{W_ {\tiny{\yng(1)}}} \cong S~, \qquad\qquad W_ {\tiny\yng(1)}(x)= \sum_{a=1}^{N_c} x_a~.
 \ee
 Recall that an irreducible representation of  $U(N_c)$ is specified by a partition $\rho=[\rho_1, \cdots, \rho_{N_c}]$ or, equivalently, by a Young tableau with $\rho_a$ boxes in the $a$-th row. The Wilson loop in the representation $\FR_\rho$ determined by $\rho$ is then represented in the 3d $A$-model by the Schur polynomial:
 \be
   W_\rho(x) = {\rm ch}(\CE_{W_\rho})=s_\rho(x)~. 
 \ee
In fact, one can directly argue that $W_\rho$ flows to the vector bundle $\CE_{W_\rho}$ of rank ${\rm dim}(\FR_\rho)$ obtained from tensor products of $S$ by using the Schur functor. The basic reason is that the local operators that can end on Wilson lines, which corresponds to sections of $\CE_{W_\rho}$, are local operators transforming in the gauge-representation $\FR_\rho$ that are built out of chiral-multiplet operators such as $\phi$ itself, which maps onto the full target space. To restrict the support of the would-be coherent sheaf $\CE_\SL$ in the infrared to a subvariety of $X$, we need to use a different construction of the line $\SL$, as we now explain.

 \subsection{Grothendieck lines and Schubert classes}\label{subsec: Groth-lines}
 We will now construct half-BPS lines that flow to coherent sheaves that are not necessarily locally free. In particular, we would like to construct lines that flow to the Schubert classes -- the structure sheaves of Schubert varieties --, denoted by $\CO_\lambda$, which form a more convenient basis for the K-theory ring of $X$. Any coherent sheaf has a free resolution: 
 \be\label{free res}
 \cdots \rightarrow W_{\FR_2}\rightarrow W_{\FR_{1}} \rightarrow \SL\rightarrow 0~,
 \ee
 written here, schematically, in terms of line operators in the UV gauge  theory, with the locally-free sheaves constructed from Wilson lines (the representations $\FR_i$ that appear in~\eqref{free res} are not necessarily reducible). 
 At the level of the Chern characters, we then have:
 \be
 \SL(x, y)= \sum_{i} (-1)^{i+1}\,  W_{\FR_i}(x, y)~.
\ee
In principle, to compute any set of  observables, we could work out the relevant polynomials $\SL(x, y)$ from the knowledge of the free resolutions~\eqref{free res} -- this was done {\it e.g.} in~\cite{Gu:2020zpg}, in some examples. For instance, we have:
\be\label{SES for O1}
0 \rightarrow \det{S} \rightarrow \CO \rightarrow \CO_{\tiny{\yng(1)}} \rightarrow 0~,
\ee
where $ \CO_{\tiny{\yng(1)}}$ is the codimension-one Schubert variety, to be introduced momentarily. 
 Instead of going down that route, we would prefer to directly define the defect line $\SL$ in the UV. This allows us to compute the polynomials $\SL(x, y)$ directly from the physics.

\subsubsection{Schubert cells and Schubert subvarieties}\label{subsec: schubert cells}
Before discussing the physical construction, let us briefly review some important geometric facts about  the Grassmannian -- we refer to \cite{brion2004lectures, griffiths2014principles} for further background. 

\begin{figure}[t]
    \centering
    \begin{subfigure}[b]{0.5\textwidth}
    \centering
\be\nn
\begin{array}{|c|c|c|c|}
\hline
w\in S_4& \lambda &   \phi  &\text{dim}(X_\lambda) \\
\hline\hline
 \{1,2,3,4\}    \;&\; [0,0]\; & \;   \mat{1 & 0 & \ast & \ast \\ 0 & 1 & \ast & \ast}\;  &4  \\
 \{1,3,2,4\}   &[1,0] &   \mat{1 & \ast & 0 & \ast \\ 0 & 0 & 1 & \ast}&3  \\
  \{2,3,1,4\}  &[1,1] &\mat{0 & 1 & 0 & \ast \\ 0 & 0 & 1 & \ast} &2\\
  \{1,4,2,3\}   &[2,0] &   \mat{1 & \ast & \ast & 0 \\ 0 & 0 & 0 & 1}  &2 \\
  \{2,4,1,3\}   &[2,1] &   \mat{0 &1 & \ast & 0 \\ 0 & 0 & 0 & 1}   &1\\
  \{3,4,1,2\}  &[2,2] &   \mat{0 & 0 &1 & 0 \\ 0 & 0 & 0 & 1}   &0\\  
  \hline
\end{array}
\ee
    \caption{Schubert cells of ${\rm Gr}(2,4)$.  \label{Gr24 Cells}}
    \end{subfigure}%
  %%%%%%%%
    \begin{subfigure}[b]{0.5\textwidth}
    \centering
%    \[
% \xymatrix{
%  &  X_{\yng(2,2)}   \ar@<0.5ex>[d]  & \\
% & X_{\yng(2,1)}   \ar@<0.5ex>[dr] \ar@<0.5ex>[dl]   &\\
% X_{\yng(1,1)}   \ar@<0.5ex>[dr] & &  X_{\yng(2)}  \ar@<0.5ex>[dl] \\
% &X_{\yng(1)}  \ar@<0.5ex>[d]  & \\
% &X_{[0,0]}\equiv X&
% }
% \]
 \[
\xymatrix{
 &    X_{\phantom{.}}    & \\
&  X_{\yng(1)}  \ar@<0.5ex>[u]   &\\
X_{\yng(1,1)}  \ar@<0.5ex>[ur]  & &  X_{\yng(2)} \ar@<0.5ex>[ul] \\
& X_{\yng(2,1)} \ar@<0.5ex>[ur] \ar@<0.5ex>[ul] & \\
&X_{\yng(2,2)} \ar@<0.5ex>[u] &
}
\]
    \caption{Hasse diagram of Schubert varieties. \label{Hasse Gr24}}
    \end{subfigure}
    \caption{\textsc{Left:}  Schubert cells for ${\rm Gr}(2,4)$, indexed by permutations or partitions, with the matrix $\phi_\lambda$ shown explicitly.  \textsc{Right:} The Hasse diagram of inclusion of Schubert varieties inside ${\rm Gr}(2,4)$, where the arrow $X \rightarrow Y$ denotes inclusion of $X$ inside $Y$.  \label{fig:Gr24}}
\end{figure}
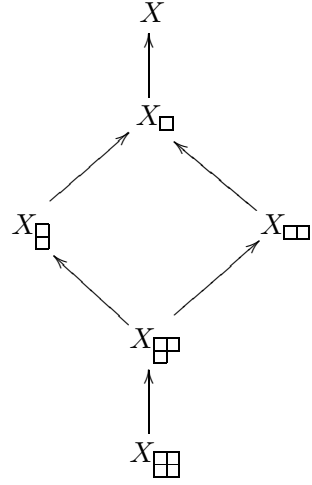%%

\medskip
\noindent
There is a natural action of ${ GL}(n_f, \C)$ on the Grassmannian $X\equiv {\rm Gr}(N_c, n_f)$. Indeed, any $N_c$-plane in $\C^{n_f}$ can be represented by the $N_c\times n_f$ matrix:
\be
\phi= \left({\phi^a}_\alpha\right)~, \qquad \qquad a=1, \cdots, N_c~, \quad \alpha=1, \cdots, n_f~,
\ee
modulo the action of the complexified gauge group  ${GL}(N_c, \C)$ from the left, and with the symmetry ${GL}(n_f, \C)$ acting from the right. We can use the gauge freedom to pick a representative:
\be\label{phi full Gr}
\phi=\mat{1 & 0 & \cdots & 0 & \star &\cdots& \star \\
                0 & 1 & \cdots & 0 & \star &\cdots & \star \\
               \vdots &  &  \ddots  &     &     &    \ddots&     \\
                0 & 0 & \cdots & 1 & \star &\cdots & \star 
} 
= \left(   \mathbb{I}_{N_c} \quad \bigstar_{N_c\times (n_f- N_c)}\right)~.
\ee
Here and in the following, $\mathbb{I}_{k}$ denotes a $k\times k$ identity matrix and $\bigstar_{n\times m}$ denotes any $n\times m$ matrix with undetermined entries (similarly, $0_{n\times m}$ will denote a matrix with vanishing entries).

The homology of $X$ is generated by the Schubert varieties $X_\lambda\subseteq X$, which are the closure of the Schubert cells, $\CC_\lambda$.  The Schubert cells correspond to fixed loci under the maximal torus $T\subset {\rm GL}(n_f, \C)$. Each Schubert cell can be indexed by a choice of $N_c$ integers:
\be
 I= \{\alpha_1, \cdots, \alpha_{N_c}\}~,\qquad \text{such that:} \quad     1\leq \alpha_1 < \alpha_2 < \cdots < \alpha_{N_c} \leq n_f~.
 \ee 
Such a set defines a Grassmannian permutation in $S_{n_f}$, 
\be
w = (I  \, J)= (\alpha_1  \cdots  \alpha_{N_c} \, \gamma_1 \cdots \gamma_{n_f- N_c})~, \qquad\quad \text{with}\quad  \gamma_1< \gamma_2 < \cdots < \gamma_{n_f- N_c}
\ee
where the ordered set $J$ is the complement of $I$ inside $\{ 1, \cdots, n_f\}$. Equivalently, and more conveniently for our purpose, we shall index $\CC_\lambda$ by the corresponding partition $\lambda= [\lambda_1, \cdots, \lambda_{N_c}]$ with $\lambda_a \leq n_f-N_c$, which is related to the Grassmannian permutation $w$ by:
\be\label{lambda from w}
\lambda = [\alpha_{N_c} - N_c~,\, \cdots~, \, \alpha_1 -1]~.
\ee
This also corresponds to the Young tableaux that can fit inside a $N_c\times (n_f-N_c)$ rectangle.
The Schubert cell $\CC_\lambda$ can be represented by a $N_c\times n_f$ matrix $\phi_\lambda$ in which the $\alpha_n-1$ first entries of the $n$-th row vanish, the $\alpha_n$-th entry is equal to one, and the entries above this latter entry also vanish. For $I=(1, 2, \cdots, N_c)$, we have the trivial partition, $\lambda= [0, \cdots, 0]$, and the Schubert cell is the full Grassmannian, with $\phi_\lambda$ given by~\eqref{phi full Gr}.  
The codimension of the corresponding Schubert variety is given by the length of the partition, $|\lambda|\equiv \sum_{a=1}^{N_c} \lambda_a$, hence the dimension is:
\be\label{dimXlambda}
{\rm dim}(X_\lambda)= N_c(n_f-N_c)- |\lambda|~.
\ee
Finally,  the Schubert variety $X_\lambda$ is also given by the disjoint union of all the `smaller' Schubert cells:
\begin{equation}
    X_{\lambda}  \equiv \overline{\CC}_{\lambda} = \bigsqcup_{\nu\supseteq\lambda} \CC_\nu~,
\end{equation}
where it is understood that we consider all partitions $\nu$ whose Young tableau contains the Young tableau of $\lambda$. As an example, the Schubert cells and Schubert varieties inside ${\rm Gr}(2,4)$ are shown in figure~\ref{fig:Gr24}.

The cohomology of the Grassmannian is freely generated (as a vector space) by the cocycles $[X_\lambda]$ Poincar\'e dual to the Schubert varieties:
\be
{\rm H}^\bullet(X, \Z)\cong \Z\big\langle [X_\lambda]  \big\rangle~.
\ee
In particular, the K\"ahler class of $X$ in ${\rm H}^{1,1}(X) \cong {\rm H}^2(X, \C)$ is proportional to $[X_{[1,0,\cdots, 0]}]$. The classical cohomology ring,
\be
[X_\lambda]  \cup [X_\mu] = \sum_{\nu} {c_{\lambda\mu}}^\nu\ [X_\nu] ~,
\ee
can be worked out using Schubert calculus (or, equivalently, from the representation theory of the symmetric group $S_{n_f}$). The structure coefficients ${c_{\lambda\mu}}^\nu\in \Z$ are known as the Littlewood–Richardson (LR) coefficients. Finally, we denote by $\CO_\lambda$ the structure sheave of $X_\lambda$. The classical $K$-theoretic product takes the form:
\be\label{classical Kring}
[\CO_\lambda]  \cdot [\CO_\mu] = \sum_{\nu} {C_{\lambda\mu}}^\nu\, [\CO_\nu] ~.
\ee
Here, the integers ${C_{\lambda\mu}}^\nu$ are the $K$-theoretic LR coefficients. These coefficients can only be non-vanishing if:
\be
  |\nu| \geq |\lambda| + |\mu|~,
\ee
with the equality being saturated in the cohomological case (in which case ${C_{\lambda\mu}}^\nu={c_{\lambda\mu}}^\nu$).%
\footnote{This is true in the non-equivariant case. In the equivariant case, the selection rules are weaker.}
By a slight abuse of notation, we will often denote the K-theory class  $[\CO_\lambda]$ by $\CO_\lambda$.

%%%%%%%%%
\subsubsection{Line defects and 1d $\CN=2$ supersymmetric gauge theories}\label{subsec: gen-line-op}

We would like to construct the line defects of the 3d $\CN=2$ gauge theory on $\Sigma\times S^1$ in the UV which flow to the Schubert classes $\CO_\lambda$ in the IR description. These UV defects, denoted by $\SL_\lambda$, wrap the $S^1$ and are localised at a point $p\in \Sigma$. In the GLSM description,  we have the maps:
 \be
 \Phi \, : \, \Sigma \rightarrow X~,
 \ee
given by the $n_f$ fundamental chiral-multiplet scalars of the UV gauge theory. In particular, we have $\Phi=\phi$ for the constant maps, with $\phi$ given by~\eqref{phi full Gr} up to gauge transformations. We then need to construct a defect $\SL_\lambda$ such that its insertion at the point $p$ restricts the constant maps onto $X_\lambda\subset X$. We schematically write this as:
\be
\SL_\lambda \Phi(p) : p \rightarrow X_\lambda~.
\ee
In practical terms, this means that, in the presence of the defect, $\phi$ should be restricted to take value in the Schubert cell $C_\lambda$, namely $\phi =\phi_\lambda$ as defined in subsection~\ref{subsec: schubert cells}.

This can be achieved very naturally using a standard construction for defects in supersymmetric gauge theories -- see {\it e.g.}~\cite{Gaiotto:2013sma, Gaiotto:2014ina, Bullimore:2014awa, Assel:2015oxa}. We simply couple the 3d $\CN=2$ gauge theory to a 1d $\CN=2$ gauge theory --- a gauged supersymmetric quantum mechanics (SQM) --- with $U(N_c)$ flavour symmetry that lives on the line and couples to the 3d gauge fields.  

\begin{figure}[t]
    \centering
    \begin{tikzpicture}[baseline=1mm]
\node[node3d] (1) []{$N_c$};
\node[square,draw,minimum size=1.7cm,very thick] (2) [below= 1.5 of 1]{$n_f$};
\node[node1d] (rn) [left= 1.5 of 1]{$r_n$};
\node[] (inter1) [left= 1.2 of rn]{$\;\cdots\;$};
\node[node1d] (rl) [left= 1.2 of inter1]{$r_l$};
\node[] (inter2) [left= 1.2 of rl]{$\;\cdots\;$};
\node[node1d] (r1) [left= 1.2 of inter2]{$r_1$};
\draw[->-=0.6,very thick] (1) --  (2) node[midway,right]{$\; \; \phi$};
\draw[->-=0.6, thick] (rn) --  (1) node[midway,above]{$\varphi_n^{n+1}$};
\draw[->-=0.6, thick] (inter1) --  (rn) node[midway,above]{$\varphi_{n-1}^{n}$};
\draw[->-=0.6, thick] (rl) -- (inter1) node[midway,above]{$\varphi_l^{l+1}$};
\draw[->-=0.6, thick] (inter2) -- (rl) node[midway,above]{$\varphi_{l-1}^{l}$};
\draw[->-=0.6, thick] (r1) -- (inter2) node[midway,above]{$\varphi_1^{2}$};
\draw[->-=0.6,red, dashed, thick] (r1) -- (2) node[midway,below]{$M_1\;$};
\draw[->-=0.6,red, dashed, thick] (rl) -- (2) node[midway,above]{$\; M_l$};
\draw[->-=0.6,red, dashed, thick] (rn) -- (2) node[midway,above]{$\;\; \; M_n$};
  \node[fit=(r1)(rn), dashed, blue, draw, inner sep=10pt, minimum width=12cm,minimum height=4.5cm, shift={(0.5cm,-1.1cm)}] (rn) {};
\end{tikzpicture}
    \caption{1d $\CN=2$ linear quiver coupled to 3d $\CN=2$ Gr$(N_c,n_f)$ GLSM. The 1d quiver is inscribed inside the dashed blue rectangle. The horizontal solid arrows denote bifundamental 1d $\CN=2$ chiral multiplets, and the dashed red arrows denote the $M_l$ 1d $\CN=2$ fermi multiplets coupled to each $U(r_l)$ gauge group.}
    \label{fig:line defect}
\end{figure}
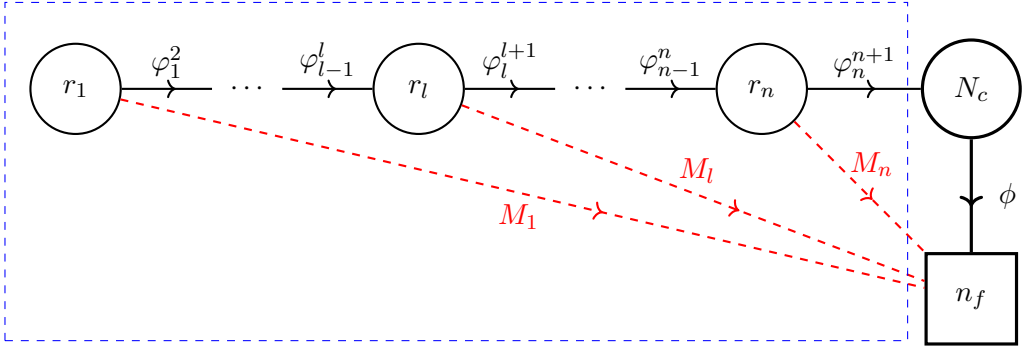

\medskip
\noindent
{\bf Defining the 1d defect.}
The gauged SQM we choose to consider is a 1d $\CN=2$ linear quiver with 1d gauge group:
\be
G_{\rm 1d} = \prod_{l=1}^n U(r_l)~, 
\ee
with $n \leq n_f$, as shown in figure~\ref{fig:line defect}.  Here, each node represents a 1d gauge group $U(r_l)$. Let us recall that the 1d vector multiplet consists of the 1d fields:
\begin{equation}\label{1d vector}
    \mathcal{V}^{(l)}_{\mathcal{N}=2} = (\sigma^{(l)}, v^{(l)}_{t},\lambda^{(l)}, \overline{\lambda}^{(l)}, D^{(l)})~, \quad\qquad l = 1, \cdots, n~,
\end{equation}
where $\sigma^{(l)}$ is a real scalar field, $v^{(l)}_t$ is the gauge connection, $\lambda^{(l)}, \overline{\lambda}^{(l)}$ are gauginos, and $D^{(l)}$ is an auxiliary field. The fields~\eqref{1d vector} transform in the adjoint representation of $U(r_l)$.

Matter fields of 1d $\CN=2$ gauge theories are organised into chiral multiplets and fermi multiplets -- see {\it e.g.}~\cite{Hori:2014tda} for a detailed account.%
\footnote{These are the straightforward dimensional reduction of the familiar 2d $\CN=(0,2)$ supermultiplets~\protect\cite{Witten:1993yc}.} As shown in figure~\ref{fig:line defect}, our 1d quiver has bifundamental chiral multiplets connecting the gauge nodes:
\begin{equation}
    \varphi_{l}^{l+1} \;:\;  U(r_{l}) \rightarrow U(r_{l+1}) ~, \qquad l = 1, \cdots, n ~,
\end{equation}
meaning that $\varphi_{l}^{l+1}$ transforms in the fundamental representation of $U(r_{l})$ and in the antifundamental representation of $U(r_{l+1})$. The corresponding scalar fields are naturally represented by $r_{l}\times r_{l+1}$ matrices. From now on, it is understood that $r_0\equiv 0$ and $r_{n+1} \equiv N_c$.  We will also assume that:
\be\label{condition-on-ranks}
    0 \leq r_1 \leq r_2 \leq \cdots \leq r_n \leq N_c~.
\ee 
% as well as:
% \be
% M_l \geq r_{l+1} - r_l~, \qquad l=1, \cdots, n~.
% \ee
% It turns out that these conditions are necessary in order to preserve supersymmetry in the infrared.%

We also couple $M_l$ fermi multiplets to the gauge group $U(r_l)$.  The fermi multiplet, denoted by $\Lambda^{(l)}_{\alpha^{(l)}}$, transforms in the fundamental of $U(r_l)$ and is charged under some subgroup $U(1)\subset U(1)^{n_f}\subset U(n_f)$ of the 3d flavour symmetry. For each $l$, 
 the coupling of the $M_l$ fermi multiplets to the $U(n_f)$ flavour group is determined by an indexing set:
\begin{equation}\label{Il def for M}
    I_{l} = \{\alpha^{(l)}_1~,\, \cdots~,\, \alpha^{(l)}_{M_l}\} \subset \{1~, \, 2~,  \cdots~,\, n_f-1\}~.
\end{equation}
We will further assume that $M_l \geq r_{l+1} - r_l~$ for all $l$,\footnote{It turns out that this condition and \eqref{condition-on-ranks} are necessary in order to preserve supersymmetry in the infrared. This can be understood through studying the Witten index of these theories, as we will briefly mention below. We leave a more thorough study of these 1d $\CN=2$ gauge theories for future work.}
 and $I_l \cap I_{l'} = \emptyset$ if $l\neq l'$. 
 The fermi multiplet also couple directly to the 3d chiral multiplet $\phi$ due to the presence of the following  1d $\CN=2$ superpotential (also known as $J$-terms):
\begin{equation}\label{J-superpotential}
  L_J=   \int d\theta J^{(l)}_{\alpha^{(l)}}(\varphi; \phi) \Lambda^{(l)}_{\alpha^{(l)}}~, \qquad \quad
    J^{(l)}_{\alpha^{(l)}}(\varphi; \phi) \equiv \varphi_{l}^{l+1}\cdot \varphi_{l+1}^{l+2} \cdots \varphi_{n-1}^{n}\cdot\varphi_{n}^{n+1} \cdot \phi^{(l)}_{\alpha^{(l)}}~,
\end{equation}
where $\cdot$ denotes matrix multiplication.
 This is the most natural superpotential consistent with 1d and 3d gauge invariance. Note that the 1d $\CN=2$ quiver gauge theory breaks the $U(n_f)$ flavour symmetry explicitly. This is necessary in order to construct defects that flow to the Schubert varieties. Any choice of Schubert cell similarly breaks ${\rm GL}(n_f, \C)$-covariance.

The line defect defined by this 1d-3d coupled system will be denoted by:
\begin{equation}\label{1d-defect}
    \SL \begin{bmatrix}
        \mathbf{r}\\
        \mathbf{M}
    \end{bmatrix}~, \qquad\quad \text{with}\quad
    \begin{bmatrix}
        \mathbf{r}\\
        \mathbf{M}
    \end{bmatrix} \equiv \begin{bmatrix}
        r_1\; & r_2\; & \cdots & r_n\\
        M_1\; & M_2\; & \cdots & M_n\\
    \end{bmatrix}~.
\end{equation}
Here, $\mathbf{r}$  denotes the ranks of the 1d gauge groups along the quiver, and  $\mathbf{M}$ gives us the distribution of fermi multiplets at each quiver node. Without loss of generality, we pick the indexing  sets~$I_l$ in~\eqref{Il def for M} as follows:
\be\label{Il def Gline}
	I_l = \left\{1 + \sum_{k=l+1}^{n}M_k,~ 2+\sum_{k=l+1}^{n}M_k, \cdots,~ M_l + \sum_{k=l+1}^{n}M_{k}\right\}~, \qquad l = 1, \cdots, n~.
\ee
That is, we start with $I_n= \{1, \cdots, M_n\}\subset \{\alpha\}_{\alpha=1}^{n_f}$ and we keep coupling the fermi multiplets to the next available $U(1)_\alpha\subset T$ factors as we go towards the left tail of the quiver in figure~\ref{fig:line defect}. Let us also mention that, to fully specify the 1d quiver~\eqref{1d-defect}, we also need to specify 1d bare Chern-Simons levels $
\kappa_l$ -- {\it i.e.} 1d Wilson lines -- for each $U(r_l)$ factor. Here, we set the 1d CS levels to $\kappa_l=0$ in the ``$U(1)_{-\half}$ quantisation convention'', as we will explain below. We will briefly discuss the effect of turning on $\kappa_l \neq 0$ at the end of this section.

\medskip
\noindent
{\bf The 1d vacuum equations.}  By assumption, the CS levels of the 3d gauge theory are chosen to be in the geometric window. Then, for positive 3d FI parameter, $\xi>0$, the theory flows to the Higgs branch $\CM_H\cong X$, and the matrix $\phi$ for the 3d chiral multiplets describe the Grassmannian as in~\eqref{phi full Gr}, up to gauge transformation. After coupling this 3d theory to the 1d defect  \eqref{1d-defect}, further constraints need to be imposed at $p\in \Sigma$, which corresponds to imposing the 1d vacuum equations. We expect that this further constrains the map $\phi$ at the point $p$ to describe a subvariety of the full 3d Higgs branch:
\begin{equation}\label{constr-space}
    \text{V} \begin{bmatrix}
        \textbf{r}\\
        \textbf{M}
    \end{bmatrix} \subseteq \text{Gr}(N_c, n_f)~.
\end{equation}
Indeed, due to the superpotential terms \eqref{J-superpotential}, we have to impose the $J$-term equations:
\begin{equation}\label{J-equaions}
    J^{(l)}_{\alpha^{(l)}}(\varphi) \equiv \varphi_{l}^{l+1}\cdot \varphi_{l+1}^{l+2}\, \cdots\, \varphi_{n}^{n+1} \cdot \phi^{(l)}_{\alpha^{(l)}} = 0~, \qquad\quad  l =1, \cdots, n~, \quad\quad \alpha^{(l)} \in I_l~.
\end{equation}
Here, for convenience of notation,  we decomposed the $N_c\times n_f$ matrix $\phi$ into the following blocks:
\begin{equation}\label{block-decomp}
\phi = \large \left(
    \renewcommand{\arraystretch}{1.2}
    \begin{array}{c|c|c|c|c}
    \phi^{(n)}\; &\; \phi^{(n-1)}\;&\; \cdots\; &\; \phi^{(1)}\; & \;\phi^{(0)} \\
    \end{array}
\right)~,
\end{equation}
where  $\phi^{(l)}$ is an $N_c\times M_l$ matrix consisting of the $M_l$ $N_c$-vectors $\phi^{(l)}_{\alpha^{(l)}}$, for ${\alpha^{(l)}\in I_l}$, with:
\begin{equation}\label{M0}
    M_0 \equiv n_f - \sum_{l=1}^{n} M_l~.
\end{equation}
We also need to impose the 1d $D$-term equations. Since we are interested in the geometric phase of the 1d theory, let us set to zero the scalars of the 1d gauge vector multiplets, $\sigma^{(l)}=0$. At each 1d gauge node $U(r_l)$, we then have the constraint:
\begin{equation}\label{D-equations}
    \varphi_{l}^{l+1}\cdot \varphi_{l}^{l+1~\dagger} - \varphi_{l-1}^{l~\dagger} \cdot\varphi^{l}_{l-1} = \zeta_l~ \mathbb{I}_{r_{l}}~,\quad \qquad l = 1, \cdots , {n}~,\\
\end{equation}
with the understanding that $r_0 \equiv 0$. Here, $\zeta_l$ is the 1d FI parameter associated with $U(1)\subset U(r_l)$, and $\mathbb{I}_{r_l}$ is the identity matrix on $\bbC^{r_l}$.

\medskip
\noindent
{\bf Solving for $\phi^{(n)}$.}   Let us first consider the $J$-equation~\eqref{J-equaions} for $l=n$. We have:
\begin{equation}\label{Jn-equations}
     \left(\varphi_{n}^{n+1}\right)^{i_n}\cdot \phi^{(n)}_{\alpha^{(n)}} = 0~, \qquad\quad \alpha^{(n)} = 1, \cdots, M_n~, \qquad i_n=1, \cdots, r_n~, 
\end{equation}
which  give us $r_n$ equations for each $\phi_{\alpha^{(n)}}^{(n)}\in \bbC^{N_c}$, $\alpha^{(n)}=1, \cdots, M_n$. We can then choose the $N_c\times M_n$ matrix $\phi^{(n)}$ to be:
\begin{equation}\label{block-n-final}
  \phi^{(n)} = \large\left(
    \renewcommand{\arraystretch}{1.2}
    \begin{array}{c|c}  
    \mathbb{I}_{N_c-r_n} & \bigstar_{N_c-r_n,\; M_n - (N_c - r_n)} \\
    &\\
    \hline
    &\\
    0_{r_n,\; N_{c} - r_n} & 0_{r_n,\; M_n - (N_c - r_n)} \\
    \end{array}\right)~,
  \end{equation}
with $\bigstar$ denoting the undetermined elements. Here we used the $U(N_c)$ gauge freedom to fix the first $N_c-r_n$ vectors, while the fact that the bottom part of~\eqref{block-n-final} is vanishing is due to~\eqref{Jn-equations} together with a convenient choice of $\varphi_{n}^{n+1}$ that is allowed thanks to the 1d gauge freedom.  The number of undetermined entries in~\eqref{block-n-final} is equal to:
\begin{equation}
    \text{F}_{n} \equiv \left(M_n - (N_c-r_n)\right) (N_c-r_n)~.
\end{equation}

 Plugging back the expression~\eqref{block-n-final} in~\eqref{Jn-equations}, we see that the first $(N_c-r_n)$ entries of each row of the matrix $\varphi^{N_c}_n$ must vanish. In addition, we can use the gauge actions on $\varphi_n^{N_c}$ to diagonalise the non-trivial block, so that we have: 
\begin{equation}\label{Cn-matrix}
    \varphi^{N_c}_n = \left(\begin{array}{c|c}
        0_{r_n ,\; N_c-r_n} & \text{C}^{(n)} \\
    \end{array}\right)~,
\end{equation}
 for  $\text{C}^{(n)}$ a diagonal matrix, $\text{C}^{(n)} = \text{diag}(\text{c}_1^{(n)}, \text{c}_2^{(n)},\cdots,\text{c}_{r_n}^{(n)})$ for $\text{c}^{(n)}_{i_n} \in \bbC^*$.
 Note that we assumed that we are on the 1d Higgs branch -- namely, that the vacuum expectation value~\eqref{Cn-matrix} Higgses $U(r_n)$ entirely. This is enforced by the $D$-term equations, given an appropriate choice of 1d FI parameters, as we will discuss momentarily. 

\medskip
\noindent
{\bf Solving for $\phi^{(n-1)}$.} Before looking at the next node of the 1d quiver, for $l=n-1$, let us first consider the $D$-term equations \eqref{D-equations} associated with $U(r_n)$, which reads:
\begin{equation}
    \varphi_{n}^{n+1}\cdot \varphi_{n}^{N_c~\dagger} - \varphi_{n-1}^{n~\dagger} \cdot\varphi^{n}_{n-1} = \zeta_n~ \mathbb{I}_{r_{n}}~.
\end{equation}
This, along with \eqref{Cn-matrix}, implies that the $r_n$ columns of the matrix $\varphi_{n-1}^{n}$ are orthogonal.  Since these columns are vectors in $\bbC^{r_{n-1}}$ (and given the assumption $r_n\geq r_{n-1}$), we have $r_{n} - r_{n-1}$ linearly dependent columns. This constrains the matrix $\varphi_{n-1}^{n}$ to be of the form:
\begin{equation}\label{Cn-1-matrix}
    \varphi_{n-1}^{n} = \left(\begin{array}{c|c}
        0_{r_{n-1},~ r_{n}-r_{n-1}} &  \text{C}^{(n-1)}\\ 
    \end{array}\right)~,
\end{equation}
where  $\text{C}^{(n-1)}$ can be gauge-fixed to be a diagonal matrix.
 
We then consider the $J$-equations \eqref{J-equaions} associated with the fermi multiplets in the fundamental of $U(r_{n-1})$. It reads:
\begin{equation}
    \varphi_{n-1}^{n}\cdot \varphi_n^{N_c} \cdot \phi^{(n-1)}_{\alpha^{(n-1)}} = 0~, \qquad \alpha^{(n-1)} = 1, \cdots, M_{n-1}~.
\end{equation}
The explicit expressions \eqref{Cn-matrix} and \eqref{Cn-1-matrix} imply that, for each $\alpha^{(n-1)}$, there are $r_{n-1}$ linear relations  amongst the $N_c$ components of the vector $\phi^{(n-1)}_{\alpha^{(n-1)}}$. One can further use the leftover 3d gauge symmetry to choose  $r_n-r_{n-1}$ vector consistent with the equations. We then find the explicit expression:
\begin{equation}
    \phi^{(n-1)} = \left(\begin{array}{c|c}
         0_{N_c - {r_n},\; r_n-r_{n-1}}& \bigstar_{N_c-r_n,\; M_{n-1}-(r_n-r_{n-1})}  \\
         &\\
         \hline
         &\\
         \mathbb{I}_{r_{n}-r_{n-1}}& \bigstar_{r_n-r_{n-1},\; M_{n-1}-(r_{n}-r_{n-1})} \\
         &\\
         \hline
         &\\
         0_{r_{n-1},\;r_n-r_{n-1}}& 0_{r_{n-1},\; M_{n-1}-(r_n-r_{n-1})}~.
    \end{array}\right)~.
\end{equation}
The number of free entries is given by:
\begin{equation}
    \text{F}_{n-1} \equiv \left(M_{n-1} - (r_n-r_{n-1})\right) (N_c - r_{n-1})~.
\end{equation}

\medskip
\noindent
{\bf Solving for $\phi^{(l)}$.} We can follow the same procedure repeatedly for $l=n-1, n-2, \cdots, 1$.  
 After fixing the matrix $\varphi_{l}^{l+1}$ using the associated 1d $U(r_l)$ and $U(r_{l+1})$ gauge symmetries, equations \eqref{J-equaions} become $r_{l}$ equations in $N_c$ variables fixing the lower $r_l\times M_l$ part of the block to be trivial. Furthermore, using the leftover 3d gauge symmetry, we can fix $r_{l+1}-r_l$ vectors of $\phi^{(l)}$. This then leads us to the explicit solution:
\begin{equation}\label{block-l-final}
    \phi^{(l)}  = \left(
    \begin{array}{c|c}
         0_{N_{c}-r_{l+1},\;r_{l+1}-r_l} & \bigstar_{N_{c}- r_{l+1},\; M_{l} - (r_{l+1}-r_l)}  \\
         \hline
         & \\
        \mathbb{I}_{r_{l+1} - r_l} & \bigstar_{r_{l+1}-r_l,\;M_{l} - (r_{l+1}-r_l)}\\
        &\\
        \hline
        &\\
        0_{r_{l},\;r_{l+1} - r_l} & 0_{r_{l},\; M_{l} - (r_{l+1}-r_l)}
    \end{array}
    \right)~,
\end{equation}
with the following numbers of undetermined entries:
\begin{equation}
    \text{F}_{l} \equiv \left(M_l - (r_{l+1} - r_l)\right)(N_c-r_l)~, \qquad l = 1, \cdots, n-2~.
\end{equation}
After gauge fixing, the matrix $\varphi_l^{l+1}$ takes the form:
\begin{equation}\label{Cl-matrix}
    \varphi_l^{l+1} = \left(\begin{array}{c|c}
        0_{r_{l}, ~r_{l+1}-r_l} & \text{C}^{(l)}  \\
    \end{array}\right)~, 
\end{equation}
with the diagonal matrix $\text{C}^{(l)} = \text{diag} (\text{c}^{(l)}_1, \text{c}^{(l)}_2, \cdots, \text{c}^{(l)}_{r_l})$, for $\text{c}_{i_l}^{(l)} \in \bbC^*$.

\medskip
\noindent
{\bf Solving for $\phi^{(0)}$.} For the last block in~\eqref{block-decomp}, denoted by $\phi^{(0)}$, we do not have any constraints coming from the 1d vacuum structure equations. In general, however, we can still fix $r_1$ vector using the 3d gauge freedom, so that we have:
\begin{equation}\label{block-0-final}
    \phi^{(0)} = \left(
    \begin{array}{c|c}
        0_{N_{c}-r_{1},\;r_{1}} & \bigstar_{N_c-r_1,\;M_0-r_1}  \\
         &\\
         \hline
         & \\
        \mathbb{I}_{r_{1}} & \bigstar_{r_1,\; M_0-r_1}\\
    \end{array}
    \right)~.
\end{equation}
Note that the number of undetermined component is given by $ \text{F}_{0} \equiv N_c \;(M_0 - r_1)$,  where $M_0$ is defined in \eqref{M0}.

\medskip
\noindent
{\bf Final result for $\phi$.}
Putting the above results together, we find that the line defect~\eqref{1d-defect} constrains the matrix $\phi$ to take the form:
\begin{equation}\label{phi-final}
    \phi ={ \large\left(
    \begin{array}{c|c|c|c|c|c|c|c|c|c|c|c}
        \mathbb{I}_{N_c-r_n} &\bigstar&0&\bigstar& \cdots & 0&\bigstar&\cdots & 0 & \bigstar&0&\bigstar  \\
        \hline
         0&0&\mathbb{I}_{r_n-r_{n-1}}&\bigstar&\cdots & 0&\bigstar&\cdots &0&\bigstar&0&\bigstar \\
         \hline
         0&0&0&0&\cdots & 0&\bigstar&\cdots &0&\bigstar&0&\bigstar \\
         \hline
         \vdots&\vdots&\vdots&\vdots&\ddots &\vdots&\vdots &\ddots & \vdots&\vdots&\vdots&\vdots\\
         \hline
         0&0&0&0&\cdots& \mathbb{I}_{r_{l+1}-r_l}& \bigstar&\cdots & 0&\bigstar&0&\bigstar\\
         \hline
    0&0&0&0&\cdots& 0& 0&\cdots & 0&\bigstar&0&\bigstar\\
    \hline
     \vdots&\vdots&\vdots&\vdots&\cdots &\vdots&\vdots &\cdots & \vdots&\vdots&\vdots&\vdots\\
         \hline
          0&0&0&0&\cdots& 0& 0&\cdots & \mathbb{I}_{r_2-r_1}&\bigstar&0&\bigstar\\
          \hline
          0&0&0&0&\cdots& 0& 0&\cdots & 0&0&\mathbb{I}_{r_1}&\bigstar\\
    \end{array}
    \right)}~.
\end{equation}
This parameterizes the subspace \eqref{constr-space}  of $X={\rm Gr}(N_c, n_f)$ of dimension:
\begin{equation}\label{full-dimension-V}
    \dim_{\bbC} \text{V}\bmat{
        \textbf{r}\\
        \textbf{M}
 } = \sum_{l=0}^{n} \text{F}_l = \sum_{l=0}^n \left(M_{l} - (r_{l+1} - r_l)\right) (N_c-r_l)~.
\end{equation}
In fact, we see that the matrix~\eqref{phi-final} describes a Schubert cell:
\be
C_\lambda =\text{V}\bmat{
        \textbf{r}\\
        \textbf{M}} ~,
\ee
with the partition~\eqref{lambda from w} given by:
\be\label{sol for lambda M r}
\lambda =\big[\underbrace{\t M_1 + r_1 -N_c}_{r_1}~,\, \cdots~,\,  \underbrace{\t M_l + r_l -N_c}_{r_l-r_{l-1}}~,\,  \cdots~,\, 
\underbrace{\t M_n+ r_n -N_c}_{r_n-r_{n-1}}~,\,  \underbrace{0}_{N_c-r_n}  \big]~,
\ee
where $\underbrace{a}_b$ means that the integer $a$ repeats $b$ times, and we defined the quantities:
\be
\t M_l \equiv \sum_{k=l}^n M_k~,
\ee
for ease of notation. One can check that~\eqref{full-dimension-V} indeed equals~\eqref{dimXlambda}.

%%%%%%%%%%
%%%%%%%%%%
\subsubsection{Grothendieck lines and Schubert classes}

Given a choice of Schubert variety $X_\lambda \subseteq X$, we can construct a line defect~\eqref{1d-defect} that flows to $X_\lambda$ in the infrared, simply by fixing the parameters $r_l$ and $M_l$ in terms of $\lambda$ as in~\eqref{sol for lambda M r}. Whenever several possible solutions for $(\mathbf{r}, \mathbf{M})$ exist, the corresponding 1d quivers are related by dualities, as we will briefly discuss below. 
 We then claim that the UV line defect flows to the structure sheaf of $X_\lambda$, $\CO_\lambda$, by giving us the so-called Schubert class $[\CO_\lambda]$ in K-theory. We write this as:
\be\label{SL to Olambda}
\SL_\lambda \equiv  \SL\begin{bmatrix}
        \mathbf{r}\\
        \mathbf{M}
    \end{bmatrix} \cong \CO_\lambda~.
\ee
These line defects thus give us the Grothendieck lines. We will verify this claim by explicit computation of the 1d Witten index.

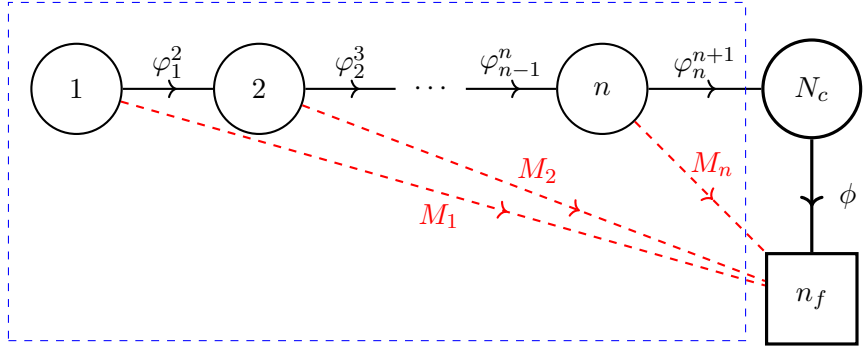
\begin{figure}
    \centering
    \begin{tikzpicture}[baseline=1mm]
\node[node3d] (1) []{$N_c$};
\node[square,draw,minimum size=1.7cm,very thick] (2) [below= 1.5 of 1]{$n_f$};
\node[node1d] (rn) [left= 1.5 of 1]{$n$};
\node[] (inter1) [left= 1.2 of rn]{$\;\cdots\;$};
\node[node1d] (r2) [left= 1.2 of inter1]{$2$};
\node[node1d] (r1) [left= 1.2 of r2]{$1$};
\draw[->-=0.6,very thick] (1) --  (2) node[midway,right]{$\; \; \phi$};
\draw[->-=0.6, thick] (rn) --  (1) node[midway,above]{$\varphi_n^{n+1}$};
\draw[->-=0.6, thick] (inter1) --  (rn) node[midway,above]{$\varphi_{n-1}^{n}$};
\draw[->-=0.6, thick] (r2) -- (inter1) node[midway,above]{$\varphi_2^{3}$};
\draw[->-=0.6, thick] (r1) -- (r2) node[midway,above]{$\varphi_1^{2}$};
\draw[->-=0.6,red, dashed, thick] (r1) -- (2) node[midway,below]{$M_1\;$};
\draw[->-=0.6,red, dashed, thick] (r2) -- (2) node[midway,above]{$\; M_2$};
\draw[->-=0.6,red, dashed, thick] (rn) -- (2) node[midway,above]{$\;\; \; M_n$};
  \node[fit=(r1)(rn), dashed, blue, draw, inner sep=10pt, minimum width=9.8cm,minimum height=4.5cm, shift={(0.5cm,-1.1cm)}] (rn) {};
\end{tikzpicture}
    \caption{Generic Grothendieck defect $\SL_\lambda$ with $n\leq N_c$. The numbers of fermi multiplets, $M_l$, are given in terms of the partition $\lambda$ as explained in the main text.}
    \label{fig: groth-defect}
\end{figure}

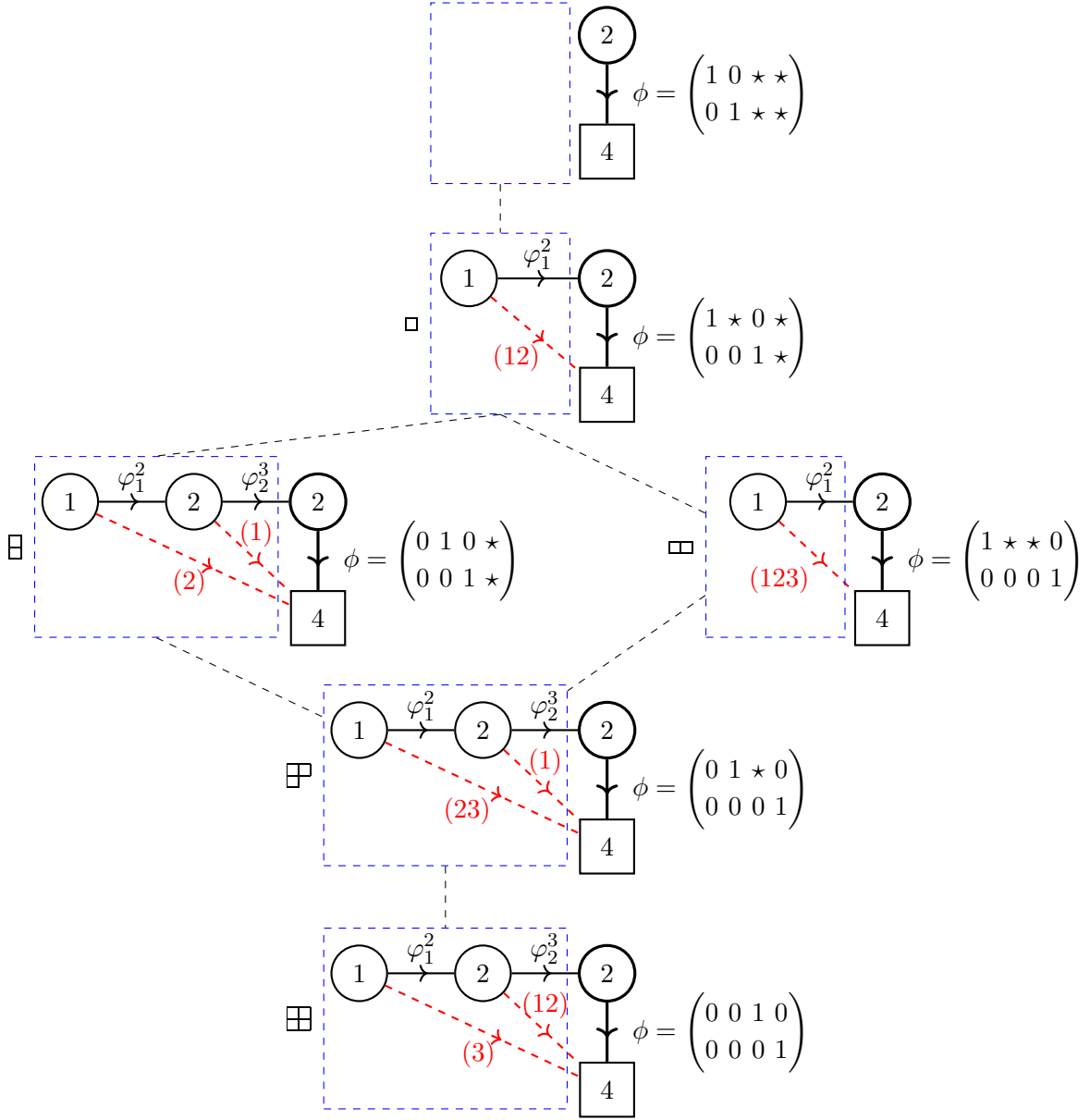
\begin{figure}
    \centering
 \begin{tikzpicture}[baseline=3mm]
 
%%%%%%% For partition 00
\node[node3ds] (001) []{$2$};
\node[squareMini] (002) [below= 0.85 of 001]{$4$};
\draw[->-=0.6,very thick] (001) --  (002) node[midway,right]{$\; \; \phi =  \begin{pmatrix} 1&0&\star&\star\\0&1&\star&\star\end{pmatrix}$};

%%%%%%%%%%%%%For partition 10
\node[node3ds] (101) [below = 1 of 002]{$2$};
\node[squareMini] (102) [below= 0.85 of 101]{$4$};
\draw[->-=0.6,very thick] (101) --  (102) node[midway,right]{$\; \; \phi =  \begin{pmatrix} 1&\star&0&\star\\0&0&1&\star\end{pmatrix}$};

\node[node1ds] (10r1) [left= 1.15 of 101]{$1$};
\draw[->-=0.6, thick] (10r1) --  (101) node[midway,above]{$\varphi_1^{2}$};
\draw[->-=0.6,red, dashed, thick] (10r1) -- (102) node[midway,below]{$(12)\;\;\;\;\;$};
 \node[fit=(10r1), dashed, blue, draw, inner sep=9pt, minimum width=2cm,minimum height=2.6cm, shift={(0.45cm,-0.65cm)}] (101d) {};

\node[] (yng10) [left = 0.05 of 101d]{$\yng(1)$};

\node[ dashed, blue, draw, inner sep=9pt, minimum width=2cm,minimum height=2.6cm, shift={(0cm,-0.65cm)}] (001d) [above = 1.35 of 101d] {};

%%%%%%%%%%%%%%%%%%%

\node[](1120mid) [below = 1 of 102, shift={(0.4cm,0cm)}] {};

%%%%%%%%%%%%%For partition 11
\node[node3ds] (111) [left = 4 of 1120mid]{$2$};
\node[squareMini] (112) [below= 0.85 of 111]{$4$};
\draw[->-=0.6,very thick] (111) --  (112) node[midway,right]{$\; \; \phi =  \begin{pmatrix} 0&1&0&\star\\0&0&1&\star\end{pmatrix}$};

\node[node1ds] (11r2) [left= 0.95 of 111]{$2$};
\node[node1ds] (11r1) [left= 0.95 of 11r2]{$1$};
\draw[->-=0.6, thick] (11r2) --  (111) node[midway,above]{$\varphi_2^{3}$};
\draw[->-=0.6, thick] (11r1) -- (11r2) node[midway,above]{$\varphi_1^{2}$};
\draw[->-=0.6,red, dashed, thick] (11r1) -- (112) node[midway,below]{$(2)\;$};
\draw[->-=0.6,red, dashed, thick] (11r2) -- (112) node[midway,above]{$\; (1)$};
 \node[fit=(11r1)(11r2), dashed, blue, draw, inner sep=9pt, minimum width=3.5cm,minimum height=2.6cm, shift={(0.35cm,-0.65cm)}] (111d) {};

\node[] (yng11) [left = 0.05 of 111d]{$\yng(1,1)$};

%%%%%%%%%%%%%For partition 20
\node[node3ds] (201) [right= 3 of 1120mid]{$2$};
\node[squareMini] (202) [below= 0.85 of 201]{$4$};
\draw[->-=0.6,very thick] (201) --  (202) node[midway,right]{$\; \; \phi =  \begin{pmatrix} 1&\star&\star&0\\0&0&0&1\end{pmatrix}$};

\node[node1ds] (20r1) [left= 0.95 of 201]{$1$};
\draw[->-=0.6, thick] (20r1) --  (201) node[midway,above]{$\varphi_1^{2}$};
\draw[->-=0.6,red, dashed, thick] (20r1) -- (202) node[midway,below]{$(123)\;\;\;\;\;\;\;\;\;\;$};
 \node[fit=(20r1), dashed, blue, draw, inner sep=9pt, minimum width=2cm,minimum height=2.6cm, shift={(0.25cm,-0.65cm)}] (201d) {};

\node[] (yng20) [left = 0.05 of 201d]{$\yng(2)$};

%%%%%%%%%%%%%For partition 21
\node[node3ds] (211) [below= 4 of 102]{$2$};
\node[squareMini] (212) [below= 0.85 of 211]{$4$};
\draw[->-=0.6,very thick] (211) --  (212) node[midway,right]{$\; \; \phi =  \begin{pmatrix} 0&1&\star&0\\0&0&0&1\end{pmatrix}$};

\node[node1ds] (21r2) [left= 0.95 of 211]{$2$};
\node[node1ds] (21r1) [left= 0.95 of 21r2]{$1$};
\draw[->-=0.6, thick] (21r2) --  (211) node[midway,above]{$\varphi_2^{3}$};
\draw[->-=0.6, thick] (21r1) -- (21r2) node[midway,above]{$\varphi_1^{2}$};
\draw[->-=0.6,red, dashed, thick] (21r1) -- (212) node[midway,below]{$(23)\;\;\;\;$};
\draw[->-=0.6,red, dashed, thick] (21r2) -- (212) node[midway,above]{$\;(1)$};
 \node[fit=(21r1)(21r2), dashed, blue, draw, inner sep=9pt, minimum width=3.5cm,minimum height=2.6cm, shift={(0.35cm,-0.65cm)}] (211d) {};

\node[] (yng21) [left = 0.05 of 211d]{$\yng(2,1)$};

%%%%%%%%%%%%%For partition 22
\node[node3ds] (221) [below= 1 of 212]{$2$};
\node[squareMini] (222) [below= 0.85 of 221]{$4$};
\draw[->-=0.6,very thick] (221) --  (222) node[midway,right]{$\; \; \phi =  \begin{pmatrix} 0&0&1&0\\0&0&0&1\end{pmatrix}$};

\node[node1ds] (22r2) [left= 0.95 of 221]{$2$};
\node[node1ds] (22r1) [left= 0.95 of 22r2]{$1$};
\draw[->-=0.6, thick] (22r2) --  (221) node[midway,above]{$\varphi_2^{3}$};
\draw[->-=0.6, thick] (22r1) -- (22r2) node[midway,above]{$\varphi_1^{2}$};
\draw[->-=0.6,red, dashed, thick] (22r1) -- (222) node[midway,below]{$(3)\;$};
\draw[->-=0.6,red, dashed, thick] (22r2) -- (222) node[midway,above]{$\; (12)$};
 \node[fit=(22r1)(22r2), dashed, blue, draw, inner sep=9pt, minimum width=3.5cm,minimum height=2.6cm, shift={(0.35cm,-0.65cm)}] (221d) {};

\node[] (yng22) [left = 0.05 of 221d]{$\yng(2,2)$};
%%%%%%%%%%%connecting them
\draw[-, black, dashed] (211d.south) -- (221d);
\draw[-, black, dashed] (201d) -- (211d);
\draw[-, black, dashed] (111d.south) -- (211d);
\draw[-, black, dashed] (101d.south) -- (201d);
\draw[-, black, dashed] (101d.south) -- (111d.north);
\draw[-, black, dashed] (001d) -- (101d);
 \end{tikzpicture}
    \caption{Generic Grothendieck defects $\SL_\lambda$ for ${\rm Gr}(2,4)$. The index set $I_l=(\alpha^{(l)})$ for the fermi multiplets coupling to $U(r_l)$ is displayed next to each red dashed arrow. Note that the 1d quivers for $\lambda=[1,1]$ and for $\lambda=[2,2]$ can be simplified by simply removing the $U(1)$ node, as explained in the main text.}
    \label{fig: Gr24quivers}
\end{figure}

\medskip
\noindent
{\bf The generic Grothendieck lines.} It is convenient to define a `generic' Grothendieck line that corresponds to the partition:
\be
    \lambda = [\lambda_1, \cdots, \lambda_{n}, \underbrace{0, \cdots, 0}_{N_c - n}]~, \qquad \lambda_1\leq n_f - N_c~,
\ee
with generic non-zero $\lambda_a$. Then, the relation~\eqref{sol for lambda M r} gives us:
\bea\label{Ml-groth}
 & r_l = l~, &&l = 1, \cdots, n~,\\
    &M_l = \lambda_l - \lambda_{l+1} +1 ~,\qquad &&l = 1, \cdots, n-1~,\\
    &M_n = \lambda_n - n + N_c~.
\eea
This Grothendieck line defect is a `complete flag' quiver, as displayed in figure~\ref{fig: groth-defect}.
As an example, all the 1d quivers giving us the Grothendieck lines $\SL_\lambda$ for ${\rm Gr}(2,4)$ are shown in figure~\ref{fig: Gr24quivers}. One can write down similar Hasse diagrams for any Grassmannian variety -- the example of ${\rm Gr}(3,5)$ is worked out in appendix~\ref{app: Gr(3,5)}, see figure~\ref{fig:Gr35 hass diag}.

\medskip
\noindent
{\bf Duality moves.} In general, the generic Grothendieck line is not the most efficient presentation of the defect line $\SL_\lambda$. Indeed, it is clear from~\eqref{sol for lambda M r} that the most `efficient' 1d quiver has $n$ nodes where $n$ is the number of {\it distinct} non-zero values for $\lambda_a$. The quiver simplification can be realised in terms of the following duality move. Whenever we have a node such that:
\be
r_l = r_{l+1}-M_l~,
\ee
we can remove the node, reconnect the $U(r_{l-1})$ and $U(r_{l+1})$ with a chiral multiplet arrow, and shift $M_{l-1}$ to $M_{l-1}+M_l$. Pictorially, we have:
\bea\label{duality move}
\begin{tikzpicture}[baseline=3mm]
%%%%%%%%%%%%%%%%%%%
\node[](mid)  {};
%%%%%%%%%%%%% quiver
\node[node1ds] (l3) [left = 4 of mid]{$r_{l+1}$};
\node[red] (M3) [below= 0.85 of l3]{$M_{l+1}$};
\node[node1ds] (l2) [left= 0.95 of l3]{$r_l$};
\node[red] (M2) [below= 0.85 of l2]{$M_{l}$};
\node[node1ds] (l1) [left= 0.95 of l2]{$r_{l-1}$};
\node[red] (M1) [below= 0.85 of l1]{$M_{l-1}$};
\draw[->-=0.6, thick] (l2) --  (l3) node[midway,above]{$\varphi_l^{l+1}$};
\draw[->-=0.6, thick] (l1) -- (l2) node[midway,above]{$\varphi_{l-1}^{l}$};
\draw[->-=0.6,red, dashed, thick] (l3) -- (M3);
\draw[->-=0.6,red, dashed, thick] (l2) -- (M2);
\draw[->-=0.6,red, dashed, thick] (l1) -- (M1);
\node[ ] (l3) [left = 2 of mid]{$\Rightarrow$};
 %
%%%%%%%%%%%%% quiver
\node[node1ds] (r3) [right = 2 of mid]{$r_{l+1}$};
\node[red] (rM3) [below= 0.85 of r3]{$M_{l+1}$};
\node[node1ds] (r1) [left= 1.5 of r3]{$r_{l-1}$};
\node[red] (rM1) [below= 0.85 of r1]{$M_{l-1}{+}M_l$};
\draw[->-=0.6, thick] (r1) --  (r3) node[midway,above]{$\varphi_{l-1}^{l+1}$};
\draw[->-=0.6,red, dashed, thick] (r3) -- (rM3);
\draw[->-=0.6,red, dashed, thick] (r1) -- (rM1);
 \end{tikzpicture}
\eea
This is a special example of a Seiberg-like duality for 1d $\CN=2$ gauge theories, applied at the $U(r_l)$ node%
\footnote{These dualities are generalisations of the Gadde-Gukov-Putrov 2d $\CN=(0,2)$ trialities~\protect\cite{Gadde:2013lxa}. They will be discussed more thoroughly in future work.} -- here, the dual gauge group is trivial (it ``condenses''), leaving us with the new chiral and fermi-multiplet  ``mesons''  that are shown in~\eqref{duality move}. Note that the duality move~\eqref{duality move} also holds for $l=1$ if $r_1=r_2-M_1$, with the net effect being simply to remove the leftmost node of the quiver.

\medskip
\noindent
{\bf Chern character and 1d Witten index.} 
  Given the above discussion, we expect that any Grothendieck line $\SL_\lambda$, defined as the above 1d $\CN=2$  quiver coupled to the 3d $U(N_c)$ gauge fields, flows to a coherent sheave with support on $X_\lambda \subset X$. We further claimed in~\eqref{SL to Olambda} that $\SL_\lambda$ exactly gives us the structure sheaves $\CO_\lambda$. To verify this claim at the level of the 3d $A$-model, we need to understand how the insertion of the line affects the low-energy 2d Coulomb branch description. This can be worked out simply by considering the path integral over the 1d fields along the $S^1$, which computes the Witten index of the 1d $\CN=2$ quiver. We denote this 1d index by $\SL_\lambda$ again, by abuse of notation:
  \begin{equation}\label{1d-partition-function}
    {\SL}_{\lambda}(x,y) \equiv \mathcal{I}_{W}^{\rm 1d}\begin{bmatrix}
       \mathbf{r} \\              \mathbf{M} 
    \end{bmatrix}(x,y)~.
\end{equation}
Note that it depends on the 3d gauge and flavour parameters, $x_a$ and $y_\alpha$.
  In the NLSM interpretation, this 1d path integral should give us the Chern character of the coherent sheaf that the line defect flows to. Indeed, we will now check that:
  \be
      {\SL}_{\lambda}(x,y) ={\rm ch}_T(\CO_\lambda)~,
  \ee
The equivariant Chern character of $\CO_\lambda$ is known to be given in terms of the {\it double Grothendieck polynomial} associated with the partition $\lambda$ \cite{lascoux1982structure, Lascoux2007, 10.1215/S0012-7094-94-07627-8}:
\be\label{chO as GrPol}
{\rm ch}_T(\CO_\lambda)= \mathfrak{G}_{\lambda} (x,y)~.
\ee
The latter can be written in the following explicit form~\cite{ikeda2013k}:%
\footnote{We use conventions compatible with our physical realisation. The variables $x$ and $b$ of~\protect\cite{ikeda2013k} correspond to   $1-x$ and $1-y^{-1}$, respectively,  in our variables (and after setting their parameter $\beta$ to $\beta=-1$). }
\begin{equation}\label{doub-groth-poly}
\mathfrak{G}_{\lambda} (x,y) = \frac{\det_{1\leq a, b \leq N_c} \left(x^{b-1}_{a} \prod_{\alpha=1}^{\lambda_b+N_c-b} \left(1-x_a y_{\alpha}^{-1}\right)\right)}{\prod_{1\leq a< b\leq N_c}(x_a - x_b)}~.
\end{equation}
%as reviewed in appendix~\ref{app:all polys}. 
In the non-equivariant limit, $y\rightarrow 1$, one obtains the ordinary Grothendieck polynomials:
\begin{equation}
    \mathfrak{G}_{\lambda} (x) = \frac{\det_{1\leq a, b \leq N_c} \left(x^{{N_c-b}}_{a} \left(1-x_a\right)^{\lambda_{N_c - b +1} - n +N_c}\right)}{\prod_{1\leq a< b\leq N_c}(x_a - x_b)}~,
\end{equation}
for the partition $\lambda  = [\lambda_1, \cdots, \lambda_n, 0, \cdots, 0]$. 
For $\lambda=[1,0,\cdots, 0]$,  for instance, we have:
\be
\mathfrak{G}_{\tiny{\yng(1)}}(x,y)=1-\det{x} = {\rm ch}(\CO_{\tiny{\yng(1)}})~,
\ee
in agreement with~\eqref{SES for O1}.
  
\medskip
\noindent
{\bf Example.} The double Grothendieck polynomials associated to  the Schubert subvarieties of Gr$(2,4)$ read:
\begin{align}\label{grothendieck-polys-Gr24}
   \begin{split}
    %    &\mathfrak{G}_{[0,0]}(x,y) = 1~,\\
        &\mathfrak{G}_{\yng(1)}(x,y) = 1-\frac{x_1 x_2}{y_1 y_2}~,\\
        &\mathfrak{G}_{\yng(1,1)}(x,y) =1-\frac{x_1+x_2}{y_1}+\frac{x_2 x_1}{y_1^2}~,\\
        &\mathfrak{G}_{\yng(2)}(x,y) = 1 -\frac{x_1 x_2}{y_1 y_2}-\frac{x_1 x_2}{y_1 y_3}-\frac{x_1
   x_2}{y_2 y_3}+\frac{x_1^2 x_2+x_1 x_2^2}{y_1 y_2 y_3}~,\\
        &\mathfrak{G}_{\yng(2,1)}(x,y) =1-\frac{x_1}{y_1}-\frac{x_2}{y_1}-\frac{x_1 x_2
   }{y_2 y_3}+\frac{ x_1 x_2}{y_1^2}+\frac{x_1 x_2^2 }{y_1 y_2 y_3}+\frac{x_1^2 x_2 }{y_1 y_2 y_3}-\frac{x_1^2 x_2^2 }{y_1^2 y_2 y_3}~,\\
        &\mathfrak{G}_{\yng(2,2)}(x,y) = 1-\frac{x_1}{y_1}-\frac{x_1}{y_2}-\frac{x_2}{y_1}-\frac{x_2}{y_2}+
        \frac{x_1^2}{y_1 y_2}+\frac{2x_1  x_2 }{y_1 y_2}+\frac{x_1x_2 }{y_1^2}+\frac{x_1x_2 }{y_2^2}+\frac{x_2^2}{y_1 y_2}\\
   &\qquad\qquad\quad-\frac{x_1x_2^2 }{y_1
   y_2^2}-\frac{x_1^2 x_2 }{y_1^2 y_2}-\frac{x_1^2 x_2 }{y_1 y_2^2}+\frac{x_1^2 x_2^2
   }{y_1^2 y_2^2}-\frac{x_1 x_2^2 }{y_1^2
   y_2}~.\\
    \end{split}
\end{align}
In the non-equivariant limit, we get the corresponding Grothendieck polynomials:
\begin{align}\label{noneq-grothendieck-polys-Gr24}
   \begin{split}
    %    &\mathfrak{G}_{[0,0]}(x,y) = 1~,\\
        &\mathfrak{G}_{\yng(1)}(x) = 1-{x_1 x_2}~,\\
        &\mathfrak{G}_{\yng(1,1)}(x) =(1-x_1)(1-x_2)~,\\
        &\mathfrak{G}_{\yng(2)}(x) =  1+ x_1 x_2 \left(x_1+x_2-3\right)~,\\
        &\mathfrak{G}_{\yng(2,1)}(x) =\left(1-x_1\right) \left(1-x_2\right) \left(1-x_1 x_2\right)~,\\
        &\mathfrak{G}_{\yng(2,2)}(x) = 
        (1-x_1)^2(1-x_2)^2~.\\
    \end{split}
\end{align}

%In the 2d limit \eqref{2dlimit-Groth}, we indeed get the corresponding Schubert polynomials which have the explicit form \eqref{schubet-poly-Gr24}.

\medskip
\noindent
{\bf Computing the 1d partition function.} The Witten index of any 1d $\CN=2$ gauge theory can be written in terms of the JK residue on the 1d complexified Coulomb branch\; \cite{Hori:2014tda}, which is parameterised by complex variables $z\in \C^\ast$. The 1d Witten index~\eqref{1d-partition-function} is then written as a nested contour integral over these 1d gauge variables:%
\footnote{Here, in quantising the theory,  we appropriately cancelled the 1d parity anomaly with a natural choice of 1d CS levels -- see section\;\ref{subsec:add 1d CS} below for more details.}
\begin{equation}\label{cont-int-Z}
    {\SL}_{\lambda}(x,y) =  \oint_{\text{JK}} \left[\prod_{l=1}^{n}\frac{1}{r_l ! } \prod_{i_l=1}^{r_l} \frac{-d z_{i_l}^{(l)}}{2\pi i z_{i_l}^{(l)}} \prod_{1\leq i_l \neq j_l \leq r_l} \left(1-\frac{z_{i_l}^{(l)}}{z_{j_l}^{(l)}}\right)\right] \text{Z}^{\rm 1d}_{\text{matter}}(z,x,y)~,
\end{equation}
where, 
\begin{equation}\label{Z-matter}
	\text{Z}^{\rm 1d}_{\text{matter}}(z,x,y) \equiv \prod_{l=1}^{n-1} \prod_{i_{l}=1}^{r_l} \frac{\prod_{\alpha^{(l)}\in I_l}\left(1-\frac{z_{i_l}^{(l)}}{y_{\alpha^{(l)}}}\right)}{\prod_{j_{l+1}=1}^{r_{l+1}}\left(1- \frac{z_{i_l}^{(l)}}{z_{j_{l+1}}^{(l+1)}}\right)} \prod_{i_n=1}^{r_n} \frac{\prod_{\alpha^{(n)}\in I_n} \left(1-\frac{z_{i_n}^{(n)}}{y_{\alpha^{(n)}}}\right)}{\prod_{a=1}^{N_c}\left(1-\frac{z_{i_n}^{(n)}}{x_a}\right)}~.
\end{equation}
The measure factor in \eqref{cont-int-Z} arises from 1d W-bosons for each gauge node along the quiver, and the matter multiplets contribute to \eqref{Z-matter}, with the numerator and denominator contributions arising from one-loop determinants of the 1d chiral and fermi multiplets, respectively. Note that the coupling to the 3d parameters $x_a$ only arises as the contribution from the $N_c$ fundamental chiral multiplets $\varphi_{n}^{n+1}$ of the $U(r_n)$ node.

The JK residue in~\eqref{cont-int-Z} is determined by the choice of 1d FI parameters. In order to solve the $D$-term equations on the 1d ``Higgs branch'' as above, we choose the $\zeta_l$ parameters along the 1d quiver to be all positive.
% with: %\CyC{check.}
%\be
%0 <\zeta_1 < \cdots < \zeta_l <\cdots \zeta_n~.
%\ee
 Then, the integration contour corresponds to iteratively picking the poles from the chiral multiplets from the denominator of \eqref{Z-matter}, and then  performing the $U(r_l)$ integrations in increasing order, from $l=1$ to $l=n$. 

\medskip
\noindent
{\bf Explicit computations.} For definiteness, let us compute the JK residue~\eqref{cont-int-Z} with the `generic' choice of linear quiver~\eqref{Ml-groth}. 
 To isolate the relevant poles, we can rewrite the index as:
\begin{equation}\label{cont-int-Z-simp}
		{\SL}_{\lambda}(x,y) = (-1)^{\mathfrak{n}(N_c,n)} (\det x)^n\oint\prod_{l=1}^{n}\left[\frac{d^l z^{(l)}  {{\Delta}^{(l)}(z)}}{l!(2\pi i)^l \det z^{(l)}}\right]\t {\text{Z}}^{\rm 1d}_{\text{matter}}(z,x,y)~,
\end{equation}
where the Vandermonde determinant $\Delta^{(l)}(z)$ is given by:
\begin{equation}\label{vand-det-fac}
    \Delta^{(l)} (z) \equiv \prod_{1\leq i_l\neq j_l \leq l} \left(z^{(l)}_{i_l} - z^{(l)}_{j_l}\right)~.
\end{equation}
Here we used the obvious shorthand notations 
  $ \det x \equiv \prod_{a=1}^{N_c} x_a$ and   $\det z^{(l)} \equiv \prod_{i_l=1}^l z_{i_l}^{(l)}$, and we introduced a sign factor $ \mathfrak{n}(N_c, n) \equiv n(N_c+1) + \sum_{l=1}^{n-1} l^2$. The matter contributions to~\eqref{cont-int-Z-simp} reads:
  \begin{equation}
    \t {\text{Z}}^{\rm 1d}_{\text{matter}}(z, x,y)\equiv \prod_{l=1}^{n-1}\prod_{i_l=1}^{l} \frac{\prod_{\alpha^{(l)}\in I_l}\left(1-\frac{z_{i_{l}}^{(l)}}{y_{\alpha^{(l)}}}\right)}{\prod_{j_{l+1}=1}^{l+1}\left(z_{i_l}^{(l)} - z^{(l+1)}_{j_{l+1}}\right)} \prod_{i_n=1}^{n} \frac{\prod_{\alpha^{(n)}\in I_n}\left(1-\frac{z_{i_n}^{(n)}}{y_{\alpha^{(n)}}}\right)}{\prod_{a=1}^{N_c}\left(z_{i_n}^{(n)}-x_a\right)}~.
\end{equation}

\medskip
\noindent
{\it Abelian 1d quiver.} As a warm-up exercice,  let us consider the case when the $N_c$-partition $\lambda$ consists of a single   row, $\lambda = [\lambda_1, 0, \cdots, 0]$. Then, the linear quiver consist of a single $U(1)$ node, and   \eqref{cont-int-Z-simp} gives us:
\begin{align}
\begin{split}
	{\SL}_{[\lambda_1, 0, \cdots, 0]} (x,y)&= (-1)^{\mathfrak{n}(N_c,1)}{\det x} \oint \frac{dz}{2\pi i  z}\frac{\prod_{\alpha=1}^{\lambda_1-1+N_c}\left(1-z y_\alpha^{-1}\right)}{\prod_{a=1}^{N_c}(z-x_a)} \\&= (-1)^{N_c+1}{\det x} \,\sum_{a=1}^{N_c} \frac{\prod_{\alpha=1}^{\lambda_1-1+N_c}\left(1-x_a y_\alpha^{-1}\right)}{x_a\prod_{b\neq a}(x_a-x_b)}~,
	\end{split}
\end{align}
where we picked the poles $z = x_a$ with $a=1, \cdots, N_c$. A little algebra shows that this expression reproduces the double Grothendieck polynomial \eqref{doub-groth-poly} associated with this partition.

\medskip
\noindent
{\it The general case.} Performing the contour integrals of \eqref{cont-int-Z-simp} recursively, one ends up with the following explicit formula: 
\begin{equation}\label{Z-after-cont-int}
	{\SL}_{\lambda} (x,y) = (-1)^{\mathfrak{n}(N_c,n)}{(\det x)^n}\sum_{\mathcal{J}}\prod_{l=1}^{n} {\Delta}^{(J_l)}(x) \prod_{i_l\in J_l} \frac{\prod_{\alpha^{(l)}\in I_l}\left(1-x_{i_l}y_{\alpha^{(l)}}^{-1}\right)}{x_{i_l}\prod_{\substack{j_{l+1}\in J_{l+1}\\j_{l+1}\neq i_{l}}}\left(x_{i_{l}} - x_{j_{l+1}}\right)}~,
\end{equation} 
with the index sets $\mathcal{J}$ defined as:
\begin{equation}\label{J-index-set}
	\mathcal{J} = \{J_1,\; J_2,\; \cdots,\; J_n\}~, 
\end{equation}
such that:
\begin{equation}
    J_1 \subset J_2 \subset \cdots \subset J_n \subseteq   \{1, \cdots, N_c\}~, \qquad |J_l| = l ~.
\end{equation}
Here, we introduced the Vandermonde-like   factors:
\begin{equation}\label{Jl-vandermonde}
	{\Delta}^{(J_l)}(x) \equiv \prod_{\substack{i_{l}, j_{l}\in J_l\\ i_{l}\neq j_{l}}} \left({x_{i_{l}}}- {x_{j_{l}}}\right)~, \qquad l =1, \cdots, n~.
\end{equation}
The expression \eqref{Z-after-cont-int} can be further massaged to:
\begin{equation}\label{Z-final-form}
	{\SL}_{\lambda}(x,y) = (-1)^{\mathfrak{n}(N_c,n)} \sum_{\mathcal{J}} \prod_{l=1}^{n} \left[\left(\prod_{j_{l}\in  \t {J}_{l}} x_{j_{l}}\right) \prod_{i_{l}\in J_l} \frac{\prod_{\alpha^{(l)}\in I_l}\left(1-x_{i_{l}} y_{\alpha^{(l)}}^{-1}\right)}{\prod_{j_{l+1}\in J_{l+1}\smallsetminus J_{l}}\left(x_{i_l} - x_{{j}_{l+1}}\right)}\right]~,
\end{equation}
where $\t {J}_l \equiv \{1, 2, \cdots, N_c\} \smallsetminus J_l$. This expression is actually the cofactor expansion of the following determinant:
\be\label{GrP from res}
	{\SL}_{\lambda}(x,y) = \frac{\det_{1\leq a, b \leq N_c} \left(x^{{N_c-b}}_{a} \prod_{l=N_c-b+1}^{N_c}\prod_{\alpha^{(l)}\in I_{l}} \left(1-x_a y_{\alpha^{(l)}}^{-1}\right)\right)}{\prod_{1\leq a< b\leq N_c}(x_a - x_b)}=  \mathfrak{G}_{\lambda}(x,y)~,
\ee
with the index sets $I_l$ defined exactly as in \eqref{Il def Gline} for $l = 1, \cdots, n$, with $I_l=\emptyset$ for $l>n$. It is then easy to see that this is exactly the same expression as in~\eqref{doub-groth-poly}, by redefining $b\rightarrow N_c-b+1$. Hence, we have shown that the Witten index of our Grothendieck defect lines precisely reproduces the double Grothendieck polynomial~\eqref{doub-groth-poly}.

%%%%%%%%%%
%%%%%%%%%%
\subsubsection{Effect of the 1d CS levels}\label{subsec:add 1d CS}

Similarly to how 3d $\CN=2$ GLSMs are only fully determined once we specify all the Chern-Simons levels, the 1d $\CN=2$ defects discussed above also allow for the presence of one-dimensional Chern-Simons levels, $\kappa_l$, associated to the $U(1)\subset U(r_l)$ factors. Equivalently, this corresponds to adding an abelian Wilson line with charge $\kappa_l$. Note that these CS levels must be integer-quantised:
\be
\kappa_l\in \bbZ~.
\ee
We call these the bare CS levels. There are also ``effective'' 1d CS levels that we chose so that all 1d fermions are in the ``$U(1)_{-\half}$ quantisation'', in order to cancel the 1d parity anomaly -- see {\it e.g.}~\cite{Closset:2019hyt, Closset:2023vos} for a detailed review in the 3d context; we pick the same conventions in 1d.

Thus, the general line defect of figure~\ref{fig: groth-defect} should be refined to include this additional data:
\begin{equation}\label{SL new with kappa}
	{\SL}^{({\kappa})}_\lambda\begin{bmatrix}
		1&\;2&\hdots&\;n-1&\;n\\
		M_1&\; M_2&\hdots&\;M_{n-1}&\; M_n\\
		\kappa_1&\;\kappa_2&\hdots&\;\kappa_{n-1}&\;\kappa_n
	\end{bmatrix}~.
\end{equation}
The addition of these 1d Wilson lines does not affect the structure of the Schubert variety that ${\SL}^{({\kappa} =0)}_\lambda$ maps into, hence we expect that there exists a distinct non-locally free sheaf with support on $X_\lambda$ that~\eqref{SL new with kappa} maps into.  On the other hand, at the level of the 1d partition function, the inclusion of a Wilson line in the representation $(\boldsymbol{\det})^{\kappa_l}$  of $U(r_l)$ amounts to adding the factor $(-\det z^{(l)})^{\kappa_l}$ to the integrand of the expression \eqref{cont-int-Z} for the Witten index.

Following the same procedure as we did for the case with $\kappa_l = 0$, we find that \eqref{Z-final-form} is generalised to:
\begin{equation}\label{Z-q-final-form}
	{\SL}^{(\kappa)}_{\lambda}(x,y) = (-1)^{\mathfrak{n}(N_c,n, \kappa)} \sum_{\mathcal{J}} \prod_{l=1}^{n} \left[\left(\prod_{j_l\in \overline{J}_{l}} x_{j_l}\right) \prod_{i_{l}\in J_l} \frac{x_{i_l}^{\kappa_l}\prod_{\alpha^{(l)}\in I_l}\left(1-x_{i_l} y_{\alpha^{(l)}}^{-1}\right)}{\prod_{j_{l+1}\in J_{l+1}\smallsetminus J_l}\left(x_{i_l} - x_{{j}_{l+1}}\right)}\right]~,
\end{equation}	
where $\mathfrak{n}(N_c,n, \kappa)= \mathfrak{n}(N_c,n)+ \sum_{l=1}^n \kappa_l$, and with the index sets $\mathcal{J}$ as defined in \eqref{J-index-set}.
 The expression \eqref{Z-q-final-form} can be written more compactly as:
\bea\label{q-doub-groth-poly}
	&{\SL}^{({\kappa})}_{\lambda} (x,y) =\\
&	 \frac{\quad(-1)^{\sum_{l=1}^n \kappa_l} \det_{1\leq a, b \leq N_c} \left(x^{N_c - b + \sum_{c=N_c-b+1}^{N_c} \kappa_c}_{a} \prod_{l=N_c-b+1}^{N_c}\prod_{\alpha^{(l)}\in I_{l}} \left(1-x_a y_{\alpha^{(l)}}^{-1}\right)\right)}{\prod_{1\leq a< b\leq N_c}(x_a - x_b)}~,
\eea
where it is understood  that $\kappa_l = 0$ if $l>n$. This obviously reduces to  \eqref{doub-groth-poly} if $\kappa_l=0$ for every $l$.

\medskip
\noindent
{\bf Example: Twisting the structure sheaf on $\bbP^{n_f-1}$.}  As a simple example, consider the $\bbP^{n_f-1}$ GLSM, corresponding to $N_c=1$. In this case we have $n_f$ Schubert varieties indexed by one-row partitions $\lambda=[\lambda]$ of length at most $n_f-1$. Therefore, the Grothendieck line defects are realised by 1d quivers with a single node at most.  Following the discussion above, we find that the characteristic polynomial \eqref{q-doub-groth-poly} associated with the Schubert variety  $X_\lambda\subset \bbP^{n_f-1}$ is given by:
\begin{equation}
    {\SL}_{\lambda}^{\kappa}(x, y) = (-x)^\kappa \prod_{\alpha=1}^{\lambda} (1-x y_\alpha^{-1})~, \qquad \lambda  = 0, \cdots, n_f-1~,
\end{equation}
where $\kappa\in \bbZ$ is the 1d CS level that we can attach to the 1d $U(1)$ gauge group in the line defect. The interpretation of this defect is that it flows to the structure sheaf on $X_\lambda$ twisted by the locally-free sheaf $\CO(-\kappa)$, namely:
\be
    {\SL}_{\lambda}^{\kappa} \cong \CO_\lambda(-\kappa)~,
\ee
up to a shift functor. It would be interesting to explicitly construct the coherent sheaves corresponding to~\eqref{SL new with kappa} for any $\kappa_l$ when $N_c>1$. We leave this as a challenge for the interested reader.

%%%%%%%%%%%%%%%%%%%%%%%%%%%%%%%%%%%%
\section{The quantum K-theory of ${\rm Gr} (N_c, n_f)$, revisited}\label{sec: QKGr}

In this section, we revisit the GLSM computation of the standard quantum K-theory  of the Grassmannian. As first discussed in~\cite{Jockers:2019lwe, Ueda:2019qhg}, ${\rm QK}({\rm Gr}(N_c, n_f))$ is naturally realised by the $U(N_c)$ gauge theory discussed in section~\ref{sec:3dAmod} for a specific choice of the CS levels, namely:%
\footnote{These levels are the ones we would obtain if we start by considering the 3d $\CN=4$ $U(N_c)$ gauge theory with $n_f$ fundamental hypermultiplets, which desribes the total space of the cotangent bundle over Gr$(N_c, n_f)$,  and if we then turn on a real mass term that triggers an RG flow which `integrates out' the non-compact fibers. Note also that choosing $K_{RG}\neq 0$ is simply equivalent to inserting the K-theory class corresponding to the line bundle $(\det S)^{-K_{RG}}$ -- this follows from the way this CS levels appears in~\eqref{Omega full}. The K-theoretic viewpoint on the 3d $\CN=4$ case has been studied in a varieties of works, see~{\it e.g.}~\protect\cite{Gaiotto:2013bwa, Pushkar:2016qvw, Koroteev:2017nab, Smirnov:2020lhm}.}
\be\label{QK levels}
   k = N_c - \frac{n_f}{2}~, \qquad\quad l = -1~, \qquad\quad K_{RG}=0~.
\ee
We also set $r=0$ for the $R$-charge of the chiral multiplets.
 It is indeed known that the algebra of Wilson loops of this specific 3d $\CN=2$ gauge theory is isomorphic to the quantum K-theory ring of ${\rm Gr}(N_c, n_f)$~\cite{Jockers:2019lwe, Ueda:2019qhg}. Here, we simply wish to insist on the simple fact  that we are free to choose any convenient basis for the chiral ring $\CR^{\rm 3d}$ of this 3d $\CN=2$ gauge theory compactified on a circle. In particular, we saw in the previous section that the Grothendieck lines $\SL_\lambda \cong \CO_\lambda$ can be naturally defined as line defects, giving us a physically and mathematically natural basis in terms of Schubert classes. At the level of the 3d $A$-model, we should simply represent the Schubert classes by the corresponding (double) Grothendieck polynomials, as in~\eqref{chO as GrPol}.

\subsection{QK ring from the Bethe ideal}
The simplest way to compute the 3d chiral ring, and hence the quantum K-theory ring of $X$, is to compute the so-called Bethe ideal~\cite{Closset:2023vos}. Schematically, we have:
\be\label{CR3d gen}
\CR^{\rm 3d}= {\K[x_1, \cdots, x_{N_c}]^{S_{N_c}}\ov (\d\CW)}~,
\ee
where $(\d\CW)$ denotes the algebraic ideal generated by the Bethe equations~\eqref{BE Pia}. This is slightly imprecise for $N_c>1$, because we need to properly account for the non-abelian sector. This can be done by a standard symmetrisation trick~\cite{Jiang:2017phk, Closset:2023vos}. Let us write the Bethe equations as polynomial equations in the gauge variables $x=\{x_a\}$:
\be
P_a(x)=0~, \qquad P_a\in \K[x]~, \qquad a=1, \cdots, N_c~.
\ee
We define the symmetrised polynomials:
\be
\h P_{ab} = {P_a- P_b\ov x_a-x_b} \in \K[x]~,\qquad a >b~,
\ee
and the Bethe ideal $\CI_{\rm BE}^{(x)}= (P, \h P)$ generated by the polynomials $P_a$ and $\h P_{ab}$. The next step is to change variables to take care of the residual gauge symmetry $S_{N_c}$ on the Coulomb branch. At this stage, let us introduce the formal variables $\CO_\lambda$, which are to be identified with the (double) Grothendieck polynomials $\mathfrak{G}_\lambda(x)$ (for all allowed partitions $\lambda$), as well as an additional variable $w$, and the corresponding polynomials:%
\footnote{The constraint $\h W=0$ ensures that non-physical solutions of the Bethe equations such that $x_a=0$ (for any gauge index $a$) are disallowed. }
\be\label{introduce Ol vars}
\h G_\lambda (x, \CO_\lambda) \equiv \mathfrak{G}_\lambda(x) - \CO_\lambda~, \qquad\qquad
\h W(x, w)= w \det{x} - 1~.
\ee
This gives us a new Bethe ideal of a larger polynomial ring:
\be
\CI_{\rm BE}^{(x, w, \CO)}= (P, \h P, \h G, \h W) \subset \K[x, w, \CO]~.
\ee
Since the Grothendieck polynomials are symmetric polynomials in $x_a$ and the Bethe ideal $\CI_{\rm BE}^{(x)}$ is $S_{N_c}$-invariant, we can reduce this larger Bethe ideal to an ideal in terms of the formal variables $\CO_\lambda$ only, using the relations $\h G_\lambda=0$:
\be
\CI_{\rm BE}^{(\CO)}= \CI_{\rm BE}^{(x,w,\CO)}\big|_\text{reduce} \subset \K[\CO]~.
\ee
This ideal, which we shall dub {\it the Grothendieck ideal}, can be computed very efficiently using Gr\"obner basis methods, as explained at length in~\cite{Closset:2023vos}. 
In this way, we arrive at a completely gauge-invariant description of the twisted chiral ring~\eqref{CR3d gen}, thus giving us an explicit presentation of the (equivariant) quantum K-theory ring directly in terms of the variables $\CO_\lambda$:
\be
\CR^{\rm 3d}\cong {\rm QK}_T(X) \cong {\K[\CO]\ov \CI_{\rm BE}^{(\CO)}}~.
\ee
The computation of the Grothendieck ideal is completely equivalent to deriving the quantum product structure in terms of the Schubert classes:
\be
\CO_\mu\; \CO_\nu = {\CN_{\mu\nu}}^\lambda\, \CO_\lambda~, \qquad \qquad  {\CN_{\mu\nu}}^\lambda\in \K~.
\ee
Obviously, we could also present the ring $\CR^{\rm 3d}$ in terms of any  complete set of symmetric polynomials we might be interested in, simply by changing the equations~\eqref{introduce Ol vars}.

\subsubsection{Example of $\mathbb{P}^{n_f-1}$} 

Let us start with the trivial example of the 3d $\CN=2$ $U(1)_{-{n_f\ov 2}}$ gauge theory with $n_f$ chiral multiplets of charge $1$. In this case, the 3d twisted chiral ring has the simple description:
\be\label{R3d U1 x}
\CR^{\rm 3d} \cong {\C[x] \ov  \left(\prod_{\alpha=1}^{n_f} (1-x y_\alpha^{-1}-q) \right)}~,
\ee
since we do not need to worry about the effect of the non-abelian gauge symmetry. This ring is generated by a single variable, $x$, which corresponds to the tautological line bundle $\CO(1)$ -- the is the Wilson line of unit charge in the gauge theory. The Schubert classes $\CO_\lambda$, on the other hand, are indexed by a one-dimensional partition, $\lambda= 1, \cdots, n_f-1$, and are represented by the double Grothendieck polynomials:
\be\label{Ol to x U1}
\CO_\lambda = \prod_{\alpha=1}^\lambda (1- x y_\alpha^{-1})~.
\ee
Treating $\CO_\lambda$ as a formal variable, the identification~\eqref{Ol to x U1} is imposed by the relation $\h G_\lambda =0$ in the quotient ring. Indeed, we are interested in writing the ring~\eqref{R3d U1 x} as:
\be\label{R3d U1 Ol}
\CR^{\rm 3d}  \cong {\K[\CO_1, \cdots, \CO_{n_f-1}]\ov \CI_{\rm BE}^{(\CO)}}~.
\ee
Of course, this is a redundant parameterisation, since one generator would suffice. For instance, we could write the QK ring entirely in terms of the variable:
\be
\CO_1 = 1- x y_1^{-1} \qquad \leftrightarrow \qquad x= y_1(1- \CO_1)~,
\ee
with the understanding that~\eqref{Ol to x U1} gives us an expression for $\CO_\lambda$ in terms products of $\CO_1$. The presentation~\eqref{R3d U1 Ol} is most useful if we are interested in working out the product structure, however. We give a few examples below. Let us also note that, in the non-equivariant limit, the QK ring is simply:
\be
\CO_\lambda\; \CO_{\lambda'} =q^{\lfloor \lambda+\lambda' \rfloor} \CO_{\lambda + \lambda' \, {\rm mod}\, n_f}~, 
\ee
for $0 \leq \lambda, \lambda' <n_f$. This follows from the fact that the Bethe equation is simply $(1-x)^{n_f}= q$ when $y_\alpha=1$. 

\medskip
\noindent
{\bf Equivariant QK ring of $\mathbb{P}^1$.} For $n_f=2$, the only non-trivial product is:
\be
\CO_1\; \CO_1 = \left(1-{y_2\ov y_1}\right) \CO_1 + {y_2 \ov y_1} q~,
\ee
which reduces to $\CO_1^2=q$ in the non-equivariant limit.

\medskip
\noindent
{\bf Equivariant QK ring of $\mathbb{P}^2$.} For $n_f=3$, we find:
\bea
& \CO_1\;\CO_1 &=&\;  \left(1-{y_2\ov y_1}\right) \CO_1 + {y_2 \ov y_1} \CO_2~,\\
& \CO_1\; \CO_2 &=&\;  \left(1-{y_3\ov y_1}\right) \CO_2 + {y_3 \ov y_1} q~,\\
& \CO_2\;\CO_2 &=&\;  \left(1-{y_3\ov y_1}\right)\left(1-{y_3\ov y_2}\right) \CO_2 + {y_3 \ov y_2} q \CO_1 + \left(1-\frac{y_3}{y_2}\right)\frac{y_3}{y_1} q~.\\
\eea
This is in perfect agreement with the results of~\cite{10.1215/00127094-2010-218} for the equivariant QK product.

\subsubsection{Example of ${\rm Gr}(2,4)$}

The quantum K-theory of the simplest non-trivial Grassmannian, ${\rm Gr}(2,4)$, is obtained from the $U(2)_{0, -2}$ gauge theory with $n_f=4$ fundamentals. The Bethe equations that gives us the QK ring are:
\be
P_1\equiv x_2 \prod_{\alpha=1}^4(1- x_1 y_\alpha^{-1}) + q x_1 =0~,\qquad 
P_2\equiv x_1 \prod_{\alpha=1}^4(1- x_2 y_\alpha^{-1}) + q x_2 =0~.
\ee
In this case, the minimal set of generators of the twisted chiral ring contains two non-trivial elements. For instance, we can write the ring in terms of the Wilson loops:
\be
W_{\yng(1)}= x_1+x_2~, \qquad W_{\yng(1,1)}=x_1 x_2~,
\ee
and work out the algebra of Wilson loops from there, as in~\cite{Gu:2020zpg}. Here, instead, we directly compute the Grothendieck ideal, which gives us the  products displayed in table~\ref{tab:Gr24equivQK}. Recall that $(y_1, \cdots, y_4)$ are parameters for the $SU(4)$ flavour symmetry; indeed, the  structure constants ${\CN_{\mu\nu}}^\lambda \in \K$ are invariant under the overall rescaling $(y_\alpha)\rightarrow (\lambda y_\alpha)$, which is a gauge transformation.
\begin{table}[t]
{\small
\begin{align} \nn
    \begin{split}
         &\mathcal{O}_{\yng(1)} \; \mathcal{O}_{\yng(1)} = \left(1-\frac{y_3}{y_2}\right) \mathcal{O}_{\yng(1)} + \frac{y_3}{y_2} \cO_{\yng(1,1)} + \frac{y_3}{y_2} \cO_{\yng(2)} - \frac{y_3}{y_2} \cO_{\yng(2,1)}~,\\
         &\cO_{\yng(1)} \; \cO_{\yng(1,1)} =  \left(1-\frac{ y_3}{y_1}\right)\cO_{\yng(1,1)} + \frac{y_3}{y_1}\cO_{\yng(2,1)}~,\\
         &\cO_{\yng(1)}\; \cO_{\yng(2)} = \left(1-\frac{y_4}{y_2}\right) \cO_{\yng(2)} + \frac{y_4}{y_2}\cO_{\yng(2,1)} ~,\\
           &\cO_{\yng(1)}\; \cO_{\yng(2,1)} =q\frac{y_4}{y_1}   -q\frac{y_4}{y_1} \cO_{\yng(1)}  + \left(1-\frac{y_4}{y_1}\right)\cO_{\yng(2,1)} +\frac{y_4}{y_1} \cO_{\yng(2,2)}~,\\
         &\cO_{\yng(1)}\; \cO_{\yng(2,2)} = q\frac{y_3 y_4}{y_1 y_2} \cO_{\yng(1)}  +\left(1-\frac{ y_3 y_4}{y_1 y_2}\right) \cO_{\yng(2,2)}~,\\
            &\cO_{\yng(1,1)}\; \cO_{\yng(1,1)} = \left(1-\frac{y_2}{y_1}\right)\left(1-\frac{y_3}{y_1}\right)\cO_{\yng(1,1)} +  \left(1-\frac{y_2}{y_1}\right) \frac{y_3}{y_1}\cO_{\yng(2,1)} +\frac{y_2}{y_1} \cO_{\yng(2,2)}~,\\
         &\cO_{\yng(1,1)}\; \cO_{\yng(2)} = q\frac{y_4}{y_1}  +\left(1-\frac{y_4}{y_1} \right)\cO_{\yng(2,1)}~,\\
           &\cO_{\yng(1,1)}\; \cO_{\yng(2,1)} = q\left(1-\frac{y_2}{y_1}\right)\frac{y_4}{y_1}  + q\frac{y_2y_4}{y_1^2}\cO_{\yng(1)}  +\left(1-\frac{y_2}{y_1} - \frac{y_4}{y_1} + \frac{y_2 y_4}{y_2^2}\right)\cO_{\yng(2,1)}
           %\\&\qquad \qquad
           + \left(\frac{y_2}{y_1}-\frac{y_4}{y_1}\right)\cO_{\yng(2,2)}~,\\
         &\cO_{\yng(1,1)}\; \cO_{\yng(2,2)} = q\left(1-\frac{y_3}{y_1}\right)\frac{y_4}{y_1} \cO_{\yng(1)}  +q\frac{y_3}{y_1} \cO_{\yng(2)}  +\left(1-\frac{y_3}{y_1}-\frac{y_4}{y_1}+\frac{y_3y_4}{y_1^2}\right)\cO_{\yng(2,2)}~,\\
        &\cO_{\yng(2)}\; \cO_{\yng(2)} = \left(1-\frac{y_4}{y_3}\right)\left(1-\frac{y_4}{y_2}\right) \cO_{\yng(2)} + \left(1-\frac{y_4}{y_3}\right)\frac{y_4}{y_2}\cO_{\yng(2,1)} + \frac{y_4}{y_3}\cO_{\yng(2,2)}~,\\
        &\cO_{\yng(2)}\; \cO_{\yng(2,1)} = q\left(1-\frac{y_4}{y_3}\right)\frac{y_4}{y_1}  + q\frac{y_4^2}{y_1 y_3}\cO_{\yng(1)}  +\left(1 - \frac{y_4}{y_3} -\frac{y_4}{y_1} + \frac{y_4^2}{y_1y_3}\right) \cO_{\yng(2,1)} 
        %\\ &\qquad \qquad 
        +\left(1-\frac{y_4}{y_1}\right)\frac{y_4}{y_3} \cO_{\yng(2,2)}~,\\
        &\cO_{\yng(2)}\; \cO_{\yng(2,2)} = q \left(1-\frac{y_4}{y_2}\right)\frac{y_4}{y_1}\cO_{\yng(1)} +q \frac{y_4}{y_2} \cO_{\yng(1,1)} + \left(1 - \frac{y_4}{y_2} - \frac{y_4}{y_1} + \frac{y_4^2}{y_1 y_2}\right)\cO_{\yng(2,2)}~,\\
            &\cO_{\yng(2,1)}\; \cO_{\yng(2,1)} =q\left(1-\frac{y_2}{y_1}\right)\left(1-\frac{y_4}{y_3}\right)\frac{y_4}{y_1}   +q\left(\frac{y_2}{y_1} - \frac{y_4}{y_1} + \frac{y_4}{y_3}-\frac{y_2 y_4}{y_1 y_3}\right) \cO_{\yng(1)} + q\frac{y_4}{y_1}\cO_{\yng(1,1)}+ q\frac{y_4}{y_1}\cO_{\yng(2)} \\           
            &\qquad\qquad+ \left[\left(1-\frac{y_2}{y_1}\right)\left(1-\frac{y_4}{y_1}\right)\left(1-\frac{y_4}{y_3}\right)-q\frac{y_4}{y_1}\right]\cO_{\yng(2,1)}
            %\\&\qquad \qquad 
            + \left(1-\frac{y_4}{y_1}\right)\left(\frac{y_2}{y_1} -\frac{y_4}{y_1} + \frac{y_4}{y_3}-\frac{y_2y_4}{y_1y_3}\right)\cO_{\yng(2,2)}~,\\
            &\cO_{\yng(2,1)}\; \cO_{\yng(2,2)} =q\left(1-\frac{y_3}{y_1}\right)\left(1-\frac{y_4}{y_2}\right)\frac{y_4}{y_1}\cO_{\yng(1)} +q\left(1-\frac{y_3}{y_1}\right)\frac{y_4}{y_2} \cO_{\yng(1,1)} +q \left(\frac{y_3}{y_1} - \frac{y_3 y_4}{y_1 y_2}\right) \cO_{\yng(2)} + q\frac{ y_3 y_4}{y_1 y_2}\cO_{\yng(2,1)}\\
            &\qquad \qquad+  \left(1 - \frac{y_4}{y_2} +\frac{y_4^2}{y_1y_2} - \frac{y_3}{y_1} - \frac{y_4}{y_1} + \frac{y_3 y_4}{y_1 y_2} - \frac{y_3 y_4^2}{y_1^2 y_2}+\frac{y_3 y_4}{y_1^2}\right)\cO_{\yng(2,2)}~,\\
    &\cO_{\yng(2,2)}\; \cO_{\yng(2,2)} = q^2 \frac{y_3 y_4 }{y_1 y_2}  + q\left(1-\frac{y_3}{y_1}\right)\left(1-\frac{y_3}{y_2}\right)\left(1-\frac{y_4}{y_2}\right)\frac{y_4}{y_1}\cO_{\yng(1)} \\
   &\qquad\qquad+ q \left(1-\frac{y_3}{y_1}\right)\left(1-\frac{y_3}{y_2}\right)\frac{y_4}{y_2}\cO_{\yng(1,1)} + \left(\frac{y_3}{y_1}-\frac{y_3^2}{y_1y_2}-\frac{y_3 y_4}{y_1 y_2} + \frac{y_3^2 y_4}{y_1 y_2^2}\right)\cO_{\yng(2)}\\
   &\qquad\qquad +  \left(\frac{y_3}{y_2} - \frac{y_3^2 y_4}{y_1 y_2^2}\right)\cO_{\yng(2,1)}+\left(1-\frac{y_3}{y_1}\right)\left(1-\frac{y_3}{y_2}\right)\left(1-\frac{y_4}{y_1}\right)\left(1-\frac{y_4}{y_2}\right)\cO_{\yng(2,2)}~.\\
  \end{split}
 \end{align}}
\caption{The equivariant QK product for Gr$(2,4)$.  \label{tab:Gr24equivQK}}
\end{table}
In the non-equivariant limit, this reduces to:
\begin{align}\label{nonequiv-QK-Gr24}
        \begin{split}
        &\mathcal{O}_{\yng(1)}\;\mathcal{O}_{\yng(1)} = \mathcal{O}_{\yng(1,1)}+\mathcal{O}_{\yng(2)}-\mathcal{O}_{\yng(2,1)}~,\qquad \qquad\mathcal{O}_{\yng(1,1)}\;\mathcal{O}_{\yng(2,2)} =  q\mathcal{O}_{\yng(2)}~,\\
    &\mathcal{O}_{\yng(1)}\;\mathcal{O}_{\yng(1,1)} = \mathcal{O}_{\yng(2,1)}~,\qquad \qquad \qquad \; \qquad \quad\mathcal{O}_{\yng(2)}\;\mathcal{O}_{\yng(2)} = 
   \mathcal{O}_{\yng(2,2)}~,\\
    &\mathcal{O}_{\yng(1)}\;\mathcal{O}_{\yng(2)} = 
   \mathcal{O}_{\yng(2,1)}~, \qquad \qquad \qquad \qquad \;\;\;\mathcal{O}_{\yng(2)}\;\mathcal{O}_{\yng(2,1)} =  q
   \mathcal{O}_{\yng(1)}~,\\
   &\mathcal{O}_{\yng(1)}\;\mathcal{O}_{\yng(2,1)} = 
    \mathcal{O}_{\yng(2,2)}+q  -q
   \mathcal{O}_{\yng(1)}~,\qquad\;\qquad\mathcal{O}_{\yng(2)}\;\mathcal{O}_{\yng(2,2)} =  q
   \mathcal{O}_{\yng(1,1)}~,\\
  & \mathcal{O}_{\yng(1)}\;\mathcal{O}_{\yng(2,2)} =  q \mathcal{O}_{\yng(1)}~,\qquad \qquad \qquad \qquad \;\;\;\mathcal{O}_{\yng(2,1)}\;\mathcal{O}_{\yng(2,1)} =  q
   \mathcal{O}_{\yng(1,1)}+q \mathcal{O}_{\yng(2)}-q
   \mathcal{O}_{\yng(2,1)}~.\\
      &\mathcal{O}_{\yng(1,1)}\;\mathcal{O}_{\yng(1,1)} = \mathcal{O}_{\yng(2,2)}~,\qquad \qquad \qquad \qquad \quad\;\mathcal{O}_{\yng(2,1)}\;\mathcal{O}_{\yng(2,2)}=  q
   \mathcal{O}_{\yng(2,1)}~,\\
   &\mathcal{O}_{\yng(1,1)}\;\mathcal{O}_{\yng(2)} =  q ~,\qquad\qquad \qquad\qquad\qquad\mathcal{O}_{\yng(2,2)}\;\mathcal{O}_{\yng(2,2)}=  q^2
    ~,\\
   &\mathcal{O}_{\yng(1,1)}\;\mathcal{O}_{\yng(2,1)} =  q\mathcal{O}_{\yng(1)}~,\\
   \end{split}
    \end{align}
in perfect agreement with~\cite{10.1215/00127094-2010-218}.%
\footnote{Our computations in the equivariant case also agree perfectly with the results of~\protect\cite{10.1215/00127094-2010-218}. We thank Leonardo Mihalcea for sharing some of their computations with us.}

\subsection{Correlations functions and enumerative invariants}
In the rest of this section, we further comment on the computation of the two- and three-point functions in this case. We can compute them by two disting methods, as explained in section~\ref{subsec: line observables}.

\medskip
\noindent
{\bf Sum over Bethe vacua and companion-matrix method.}
 In principle, we can evaluate the correlation function of any set of lines by performing the sum over Bethe vacua in~\eqref{sum bethe vacua formula}. Hence, focusing on the insertion of Grothendieck lines $\SL_\lambda \cong \CO_\lambda$, we have:
\be\label{sum bethe GrX}
       \Big\langle \CO_\mu \CO_\nu\cdots\Big\rangle_{\bbP^1\times S^1}  =  \sum_{\h x \in \CS_{\rm BE} } \CH(\h x, y)^{-1}\,  \mathfrak{G}_\mu(\h x, y)\mathfrak{G}_\nu(\h x, y)\cdots~,
\ee
with $\mathfrak{G}_\mu$ the Grothendieck polynomials, and the handle-gluing operator:
\be
\CH(x, y) = \det_{1\leq a, b \leq N_c}\left(\delta_{ab} \left( N_c+  \sum_{\alpha=1}^{n_f}{x_a y_\alpha^{-1}\ov 1- x_a y_\alpha^{-1}} \right)  -1\right)\,
\prod_{\alpha=1}^{n_f} (1- x_a y_\alpha) \prod_{\substack{a,b\\ a\neq b}} (1- x_a x_b^{-1})^{-1} ~.
\ee
The sum~\eqref{sum bethe GrX} can be performed using the companion matrix method~\cite{Closset:2023vos}. This amounts to writing each factor $Q(x)$ in the summand of~\eqref{sum bethe GrX} as a large square matrix $\mathfrak{M}_Q$ of size~$\left|\CS_{\rm BE}\right|$, in a convenient basis of the quotient ring~$\CR^{\rm 3d}$. The eigenvalues of $\mathfrak{M}_Q$ are equal to $Q(\h x)$, the operator $Q$ evaluated at the Bethe vacua. Hence the sum over Bethe vacua can be performed by taking the trace over a product of companion matrices, without having to solve for the eigenvalues themselves.

\medskip
\noindent
{\bf JK residue.}
 Specialising the JK residue formula~\eqref{KGW}-\eqref{Id explicit} to the CS levels~\eqref{QK levels}, we have:
\bea\label{JK for QK ordinary}
&\Big\langle\SL\Big\rangle_{\bbP^1\times S^1}   = \sum_{d= 0}^\infty  q^{d}\, \textbf{I}_{d} [{\SL}]~,\\
& \textbf{I}_{d} [{\SL}] =\sum_{\substack{\m_a \geq 0 \\ |\m|=d}}\;\; {(-1)^{N_c}\ov N_c!}\oint_{\rm JK} \prod_{a=1}^{N_c}\left[{d x_a\ov 2\pi i} {(\det{x})^{-1}\ov   \prod_{\alpha=1}^{n_f} (1- x_a y_\alpha^{-1})}\right]\, \Delta(x) F_\m(x,y) \SL(x,y)~,
\eea
where it is understood that the JK residue is a sum over all the residues at the codimension-$N_c$ singularities  $\{x_a=y_{\alpha_a}\}$, and we defined the flux-dependent factor:
\be
F_\m(x,y) ={(-1)^{|\m| (N_c-1)} \prod_{a=1}^{N_c} x_a^{N_c \m_a}\ov (\det{x})^{|\m|} \prod_{a=1}^{N_c}\prod_{\alpha=1}^{n_f} (1- x_a y_\alpha^{-1})^{\m_a} }~.
\ee
This is a completely explicit formula for the genus-zero degree-$d$ K-theoretic enumerative invariants of the Grassmannian. In the 2d limit (which gives us the quantum cohomology of $X$), the equivalence between the physical and the mathematical results is rigorously established~\cite{Kim:2016jye}, and similar considerations hold in 3d as well~\cite{Ueda:2019qhg, givental2021quantum}.

Obviously, the Bethe-vacua formula~\eqref{sum bethe GrX} is more powerful in principle, since it gives us the full answer directly as a rational function in $q$, while the JK residue formula gives us that same answer as a Taylor series in $q$, in which case we need to compute each term individually. The JK residue formula is nonetheless very practical to compute specific enumerative invariants~$\textbf{I}_{d} [{\SL}]$, at fixed degree, as opposed to the full correlation functions.

\subsubsection{Two-point functions and topological metric}
In the basis for QK$_T(X)$ spanned by the Schubert classes, 
$\{\CO_\lambda\}$, the topological metric is given by:
\be\label{top-metric-definition}
\eta_{\mu\nu} = \Big\langle \CO_\mu \,\CO_\nu \Big\rangle_{\bbP^1\times S^1}~, 
\ee
where we insert the double Grothendieck polynomials in the integrand of~\eqref{JK for QK ordinary} as indicated, plugging in $\SL(x,y)= \mathfrak{G}_\mu(x,y) \mathfrak{G}_\nu(x,y)$. Alternatively, we can use the Bethe-vacua formula~\eqref{sum bethe GrX}. Let us consider a few simple examples.

%\CyC{to do: give companion matrices for the computations below, using Grothendieck basis. (To be coded/ I will do it later.)}

\medskip
\noindent
{\bf Topological metric for ${\rm QK}_T(\mathbb{P}^1)$.}
For the abelian $U(1)_{-1}$ theory with $2$ matter multiplets with charge $1$, we find the following metric  in the non-equivariant limit:
\begin{equation}
    \eta_{\mu\nu} = \frac{1}{1-q}\left(
\begin{array}{cc}
 1&1 \\
 1& q \\
\end{array}
\right)~,
\end{equation}
in the basis $\{1, \CO_{\yng(1)}\}$. In the equivariant case, the components are the same as above except for:
\begin{equation}
    \eta_{\yng(1),\;\yng(1)} =  1-\frac{y_2}{y_1} + \frac{q}{1-q}~.
\end{equation}

\medskip
\noindent
{\bf Topological metric for ${\rm QK}_T(\mathbb{P}^2)$.} For the abelian theory  $U(1)_{-\frac{3}{2}}$ with $n_f=3$, we find the following $y$-dependent components  for the topological metric:
\begin{align}
    \begin{split}
        &\eta_{\yng(1),\;\yng(2)} =  1-\frac{y_3}{y_1} +\frac{q}{1-q}~,\\
        &\eta_{\yng(2),\;\yng(2)} = \left(1-\frac{y_3}{y_1}\right) \left(1-\frac{y_3}{y_2}\right)+\frac{q}{1-q}~.
    \end{split}
\end{align}
The other components (up to symmetry, $\eta_{\mu\nu}=\eta_{\nu\mu}$) are the same as in the non-equivariant limit, where we have:
\begin{equation}
\eta_{\mu\nu} = 
\frac{1}{1-q}\left(
\begin{array}{ccc}
 1 & 1 & 1 \\
 1 &1& q\\
 1 &q& q \\
\end{array}
\right)~,
\end{equation}
in the basis $\{1, \CO_{\yng(1)}, \CO_{\yng(2)}\}$.

% \begin{equation}
% \left(
% \begin{array}{ccc}
%  \frac{1}{1-q} & \frac{1}{1-q} & \frac{1}{1-q} \\
%  \frac{1}{1-q} &\frac{1}{1-q} & 1-\frac{y_3}{y_1} + \frac{q}{1-q} \\
%  \frac{1}{1-q} & 1-\frac{y_3}{y_1} +\frac{q}{1-q} & \left(1-\frac{y_3}{y_1}\right) \left(1-\frac{y_3}{y_2}\right)+\frac{q}{1-q} \\
% \end{array}
% \right)~.
% \end{equation}

\medskip
\noindent
{\bf Topological metric for ${\rm QK}_T(\mathbb{P}^4)$.}
For $U(1)_{-{5\ov 2}}$ with $n_f = 5$, we find that the components of the topological metric have the explicit form:
\begin{align}\label{eq-top-metric-P4}
    \begin{split}
        &\eta_{1,4} = 1-\frac{y_5}{y_1}+\frac{q}{1-q}~,\\
        &\eta_{2,3} = 1-\frac{y_4 y_5}{y_1 y_2} + \frac{q}{1-q}~,\\
        &\eta_{2,4} =\left(1-\frac{y_5}{y_1}\right) \left(1-\frac{y_5}{y_2}\right) +  \frac{q}{1-q}~,\\
        &\eta_{3,3} = 1 + \frac{y_4 y_5^2}{y_1 y_2 y_3} - \frac{y_4 y_5}{y_2 y_3} - \frac{y_4 y_5}{y_1 y_3} - \frac{y_4 y_5}{y_1 y_2} + \frac{y_4^2y_5}{y_1 y_2y_3}+\frac{q}{1-q}~,\\
        &\eta_{3,4} = \left(1-\frac{y_5}{y_1}\right) \left(1-\frac{y_5}{y_2}\right) \left(1-\frac{y_5}{y_3}\right)+ \frac{q}{1-q} ~,\\
        &\eta_{4,4} = \left(1-\frac{y_5}{y_1}\right) \left(1-\frac{y_5}{y_2}\right) \left(1-\frac{y_5}{y_3}\right) \left(1-\frac{y_5}{y_4}\right)+ \frac{q}{1-q}~,
    \end{split}
\end{align}
with the obvious index notation $\mu, \nu=0, \cdots, 4$.
The other components are the same as in the non-equivariant limit, in which case we find:
\begin{equation}\label{noneq-top-metric-P4}
\eta_{\mu\nu} = \frac{1}{1-q}\left(
\begin{array}{ccccc}
 1 &1 & 1 & 1 & 1 \\
1 & 1 & 1 & 1 & q \\
 1 & 1 & 1 & q & q \\
 1 & 1 &q & q & q \\
 1 & q & q & q & q \\
\end{array}
\right)~.
\end{equation}

\medskip
\noindent
{\bf Topological metric for ${\rm QK}_T({\rm Gr}(2,4))$.}
For the Gr$(2,4)$ case, the $y$-dependent components of the topological metric have the following explicit form:
\begin{align}\label{equiv-metric=ord-Gr24}
    \begin{split}
        &\eta_{\yng(1),\;\yng(2,2)} = \left(1-\frac{y_3 y_4}{y_1 y_2}\right) + \frac{q}{1-q}~,\\
    &\eta_{\yng(1,1),\;\yng(2)}  = \eta_{\yng(1,1),\;\yng(2,1)} = \eta_{\yng(2),\;\yng(2,1)} =  \left(1-\frac{ y_4}{y_1}\right) + \frac{q}{1-q}~,\\
    &\eta_{\yng(1,1),\;\yng(2,2)} (q,y) = \left(1-\frac{y_3}{y_1}\right) \left(1-\frac{y_4}{y_1}\right) + \frac{q}{1-q}~,\\
    &\eta_{\yng(2,1),\;\yng(2,1)} = \left(1-\frac{y_4}{y_1}\right)^2 +\frac{q}{1-q}~,\\
     & \eta_{\yng(2,1),\;\yng(2,2)}(q,y) = \left(1-\frac{y_3}{y_1}\right) \left(1-\frac{y_4}{y_1}\right) \left(1-\frac{y_4}{y_2}\right) + \frac{q}{1-q}~,\\
        &\eta_{\yng(2),\;\yng(2,2)}(q,y) = \left(1-\frac{y_4}{y_1}\right) \left(1-\frac{y_4}{y_2}\right) + \frac{q}{1-q}~,\\
   &\eta_{\yng(2,2),\;\yng(2,2)}(q,y) = \left(1-\frac{y_3}{y_1}\right)
   \left(1-\frac{y_3}{y_2}\right) \left(1-\frac{y_4}{y_1}\right) \left(1-\frac{y_4}{y_2}\right)+ \left(1-\frac{y_3 y_4}{y_1 y_2}\right) q+ \frac{q^2}{1-q}~.
    \end{split}
\end{align}
The other components take the same form as in the non-equivariant limit, namely:
\begin{equation}\label{metric-ord-Gr24}
        \eta_{\mu\nu} = \frac{1}{1-q}\left(
\begin{array}{cccccc}
 1 & 1 & 1 & 1 & 1 & 1 \\
 1 & 1 & 1 & 1 & 1 & q \\
 1 & 1 & 1 & q & q & q \\
 1 & 1 & q & 1 & q & q \\
 1 & 1 & q & q & q & q \\
 1 & q & q & q & q & q^2 \\
\end{array}\right)~,
\end{equation}
in the basis $\{1, \CO_{\yng(1)}, \CO_{\yng(1,1)}, \CO_{\yng(2)},\CO_{\yng(2,1)}, \CO_{\yng(2,2)} \}$.

\subsubsection{Three-point functions and structure constants}
The knowledge of the correlators implies the knowledge of the ring structure, so the JK residue formula gives us another way to compute the ring ${\rm QK}_T(X)$. Let us  decompose the ring structure constants as:
\be
{\CN_{\mu\nu}}^{\lambda} =  \sum_{d\geq 0} \CN^{(d)  \lambda}_{\mu\nu}\, q^d~, \qquad \CN^{(d)  \lambda}_{\mu\nu}\in \Z(y)~.
\ee
 These quantities can be computed using the JK residue formula~\eqref{JK for QK ordinary}, as:
 \be\label{deg-d-structure-constant}
\CN^{(d)  \lambda}_{\mu\nu} =  \textbf{I}_{d} \left[\CO_\mu\, \CO_\nu\, \CO^{\vee\, \lambda}\right]~,
 \ee
where we introduced the dual basis $\{\CO^{\vee\, \lambda}\}$, indexed by partitions $\lambda$, such that:
\be
\Big\langle \CO_\mu\, \CO^{\vee\, \nu} \Big\rangle_{\bbP^1\times S^1} = {\delta_\mu}^\nu~.
\ee
The dual structure sheaves $\mathcal{O}^{\vee\, \lambda}$~\cite{10.1215/00127094-2010-218} can be realised by dual Grothendieck lines whose 1d Witten indices give us the following {\it dual double Grothendieck polynomials}:
\begin{equation}\label{dual-double-groth-polys}
  \mathcal{O}^{\vee\, \lambda}(x,y) = \frac{\det x}{\prod_{a=1}^{N_c} y_{n_f-N_c+a-\lambda^\vee_a}} \, {\CO}_{\lambda^\vee}(x, y^D)~,
\end{equation}
where $\lambda^\vee$ is the partition dual to $\lambda$:
\be
[\lambda^\vee_1, \cdots, \lambda^\vee_{N_c}]= [N_c-n_f-\lambda_{N_c}, \cdots, N_c-n_f-\lambda_{1}]~,
\ee
and $y^D$ denotes the order-inverted $SU(n_f)$ parameters, $y_\alpha^D=y_{n_f+1-\alpha}$.

\medskip
\noindent
{\bf Dual double Grothendieck polynomials for $\bbP^{n_f-1}$.}
   For the Schubert cells of QK$_T(\bbP^{n_f-1})$, we have the following dual double Grothendieck polynomials: 
    \begin{align}\label{dual-Groth-lines-P4}
    \begin{split}
        &\mathcal{O}^{\vee \;\lambda}(x,y)\equiv\frac{x}{y_{\lambda+1}}\prod_{\alpha=\lambda+2}^{n_f-1}(1-x y_\alpha^{-1}) ~, \qquad \lambda = 1, \cdots, n_f-1~.
    \end{split}
    \end{align}

   %  \medskip
   %  \noindent
   %  {\bf For $n_f=5$.} For simplicity, let us take the $\bbP^4$ GLSM. In this case, we find the following 3-point functions:
   %  \begin{align}\label{str-const-P4-1}
   %  \begin{split}
   %     & {\mathcal{N}_{[l]\;[p]}}^{[0]}(q,y) = \begin{cases}
   %         1 ~,\qquad &(l, p ) = (0, 0)~,\\
   %         q\frac{y_5}{y_1}~, \qquad &(l , p) = (1,4)~,\\
   %        {q}\frac{ y_4 y_5}{y_1 y_2}~, \qquad &(l,p) = (2,3)~,\\
   %       {q}\frac{ \left(y_2-y_5\right) y_5}{y_1 y_2}~, \qquad &(l,p)  = (2,4)~, \\
   %       q\frac{\left(y_2-y_5\right) \left(y_3-y_5\right)
   % y_5}{y_1 y_2 y_3}~, \qquad &(l,p) = (3,4)~, \\
   %      {q} \frac{\left(y_2-y_5\right) \left(y_3-y_5\right)
   % \left(y_4-y_5\right) y_5}{y_1 y_2 y_3 y_4}~, \qquad &(l,p) = (4,4)~.
   %      \end{cases}\\
   %  & {\mathcal{N}_{[l]\;[p]}}^{[1]}(q,y) = \begin{cases}
   %           1~, \qquad &(l,p) = (0,1)~,\\
   %           \frac{y_1 - y_2}{y_1}~, \qquad &(l,p) = (1,1)~,\\
   %           q\frac{y_5}{y_2}~, \qquad &(l,p) = (2,4)~,\\
   %           q\frac{y_4 y_5}{y_2y_3}~, \qquad &(l,p)  = (3,3)~,\\
   %           q\frac{(y_3-y_5)y_5}{y_2 y_3}~, \qquad &(l,p) = (3,4)~, \\
   %          {q}\frac{ \left(y_3-y_5\right) \left(y_4-y_5\right)
   % y_5}{y_2 y_3 y_4}~, \qquad &(l,p) = (4,4)~.
   %       \end{cases}
   %       \end{split}
   %  \end{align}

\medskip
\noindent
{\bf Dual double Grothendieck polynomials for ${\rm Gr}(2,4)$.} For the case of Gr$(2,4)$ the dual double Grothendieck polynomials associated with the dual structure sheaves defined in \eqref{dual-double-groth-polys} read:
\begin{align}
    \begin{split}
        &\mathcal{O}^{\vee,\; \yng(1)}(x,y) = \frac{x_1 x_2}{y_1 y_3}-\frac{x_1 x_2^2 }{y_1 y_3 y_4}-\frac{x_1^2x_2^2
   }{y_1 y_2 y_3^2}+\frac{x_1^2x_2^2 }{y_1 y_3 y_4^2}-\frac{ x_1^2 x_2 }{y_1 y_3 y_4}+\frac{x_1^2 x_2^3 }{y_1 y_2 y_3^2 y_4} \\
   &\qquad\qquad\qquad+ \frac{x_1^3 x_2^2 }{y_1 y_2 y_3^2 y_4}-\frac{x_1^3 x_2^3 }{y_1 y_2 y_3^2 y_4^2}~,\\
   &\mathcal{O}^{\vee,\; \yng(1,1)}(x,y) =\frac{x_1 x_2}{y_2 y_3}-\frac{x_1 x_2^2 }{y_2 y_3 y_4}-\frac{x_1^2 x_2 }{y_2 y_3 y_4}+\frac{x_1^2 x_2^2 }{y_2 y_3 y_4^2}~,\\
   &\mathcal{O}^{\vee,\; \yng(2)}(x,y) = \frac{x_1 x_2}{y_1 y_4}-\frac{x_1^2x_2^2 }{y_1 y_2 y_3 y_4}-\frac{x_1^2 x_2^2}{y_1 y_2 y_4^2}-\frac{x_1^2 x_2^2}{y_1 y_3 y_4^2}+\frac{x_1^2x_2^3 }{y_1
   y_2 y_3 y_4^2} + \frac{ x_1^3x_2^2}{y_1 y_2 y_3 y_4^2}~,\\
   &\mathcal{O}^{\vee,\; \yng(2,1)}(x,y) = \frac{x_1 x_2}{y_2 y_4}-\frac{x_1^2 x_2^2}{y_2 y_3 y_4^2}~,\\
    &\mathcal{O}^{\vee,\; \yng(2,2)}(x,y) = \frac{x_1 x_2}{y_3 y_4}~.
    \end{split}
\end{align}

\medskip
\noindent
{\bf Structure constants for ${\rm QK}_T({\rm Gr}(2,4))$.} As an example of an application of the JK residue formula~\eqref{JK for QK ordinary}, we can compute the structure constants up to some degree $d$   using \eqref{deg-d-structure-constant}.  Given the explicit expression for the dual double Grothendieck polynomials, this is a straightforward computation (using {\it e.g.}  {\sc Mathematica}). For instance, computing up to order $q^4$, we find:
\begin{align}
    \begin{split}
        &\left<\mathcal{O}_{\yng(1)} \; \mathcal{O}_{\yng(1)} \; \mathcal{O}^{\vee \;\yng(1)}\right>_{\mathbb{P}^1\times S^1} = 1-\frac{y_3}{y_2}~, \\
        &\left<\mathcal{O}_{\yng(1,1)} \; \mathcal{O}_{\yng(2,2)} \; \mathcal{O}^{\vee \;\yng(2)}\right>_{\mathbb{P}^1\times S^1}  = \frac{y_3}{y_1} q~,\\
        &\left<\mathcal{O}_{\yng(2,1)} \; \mathcal{O}_{\yng(2,1)} \; \mathcal{O}^{\vee \;\yng(2,1)}\right>_{\mathbb{P}^1\times S^1}  = \left(1-\frac{y_2}{y_1}\right)\left(1-\frac{y_4}{y_1}\right)\left(1-\frac{y_4}{y_3}\right) - q\frac{y_4}{y_1}~,\\
        &\left<\mathcal{O}_{\yng(2,2)} \; \mathcal{O}_{\yng(2,2)} \; \mathcal{O}^{\vee \;\yng(1,1)}\right>_{\mathbb{P}^1\times S^1}  = q\left(1-\frac{y_3}{y_1}\right)\left(1-\frac{y_3}{y_2}\right)\frac{y_4}{y_2}~,
    \end{split}
\end{align}
which indeed agrees with the results already reported in table~\ref{tab:Gr24equivQK}.

\subsubsection{Classical limit ($q\rightarrow 0$): K-theoretic Littlewood–Richardson coefficients}
As a sanity check of our computations, it is interesting to consider the ``classical'' limit $q\rightarrow 0$ (that is, the large-volume limit), in which case the equivariant quantum K-theory QK$_T(X)$ reduces to the equivariant K-theory K$_T(X)$, with the ring structure~\eqref{classical Kring} given in terms of the K-theoretic LR coefficients:
\be\label{LR K gen}
{C_{\lambda\mu}}^\nu = \CN_{\lambda\mu}^{(0)\nu}~.
\ee
The JK-residue formula~\eqref{JK for QK ordinary} gives us an explicit expression for these LR coefficients:
\be
{C_{\lambda\mu}}^\nu={(-1)^{N_c}\ov N_c!}\oint_{\rm JK} \prod_{a=1}^{N_c}\left[{d x_a\ov 2\pi i} {(\det{x})^{-1} \, \prod_{b\neq a}(x_a-x_b)\ov   \prod_{\alpha=1}^{n_f} (1- x_a y_\alpha^{-1})}\right]\,\CO_\lambda(x,y) \CO_\mu(x,y) 
\CO^{\vee \;\nu}(x,y)~.
\ee
This is the general formula in the equivariant case. For $y_i=1$, this reduces to:
\be
{C_{\lambda\mu}}^\nu={(-1)^{N_c}\ov N_c!}\oint_{(x_a=1)} \prod_{a=1}^{N_c}\left[{d x_a\ov 2\pi i} { \prod_{b\neq a}(x_a-x_b)\ov   (1- x_a)^{n_f}}\right] (\det{x})^{1-N_c}\,\mathfrak{G}_\lambda(x) \mathfrak{G}_\mu(x) \mathfrak{G}_{\nu^\vee}(x)~,
\ee
in terms of the ordinary Grothendieck polynomials, with a single residue at $\{x_a=1\}$. We checked in many examples that this formula always returns an integer, as expected. It also appears to agree with the known K-theoretic LR coefficients -- see {\it e.g.}~\cite{buch2002littlewood, wheeler2019littlewood}. For instance, one can easily check that:
\bea
&{C_{[1,0],[1,0]}}^{[2,1]} &&=-1 \qquad && \text{for}\; \; N_c=2~,  && n_f=4~,\\
&{C_{[2,0,0],[2,1,0]}}^{[3,2,1]}  &&= -2 \qquad && \text{for}\; \; N_c=3~,  && n_f=9~,\\
&{C_{[3,2,1,0],[3,2,1,0]}}^{[5,4,2,2]}  &&= -9 \qquad && \text{for}\; \; N_c=4~,  && n_f=10~.
\eea
Here we picked examples with $|\nu|>|\lambda|+|\mu|$, which would vanish in the cohomological limit. It may be worthwhile to obtain a direct proof that this residue formula indeed gives us the K-theoretic LR coefficients. We leave this as a another challenge for the interested reader.

\subsection{Generalised QK rings}\label{subsec: gen-QK-rings}

Finally, let us comment on the generalisation of the above considerations when one chooses other Chern-Simons levels $(k,l)$ in the geometric window. The same computations can obviously be performed in such cases, and it is expected that the more general twisted chiral rings essentially corresponds to the level structure of Ruan and Zhang \cite{Ruan2018TheLS}:
\be\label{R3d to QK gen}
\CR^{\rm 3d}[k, l \; \text{in geometric window}] \qquad \overset{?}{\longleftrightarrow} \qquad {\rm QK}_T(X)\; \text{with level structure.}
\ee
To the best of our knowledge, the precise map between the physics of the CS levels $k,l$ and the mathematical notions of~\cite{Ruan2018TheLS} has not been worked out yet, and we hope to better address this point in future work -- see~\cite{Jockers:2018sfl, Jockers:2019lwe,Ueda:2019qhg} for some relevant past works. Here, we simply point out that we can easily compute the left-hand-side of~\eqref{R3d to QK gen} for any value of $k, l$. It would be interesting to compare these results to direct enumerative geometry computations of QK rings with non-trivial level structure.

\medskip
\noindent
{\bf Generalised QK rings for ${\rm Gr}(2,4)$.}  Recall that the ordinary QK ring is given by $\CR^{\rm 3d}[0,-1]$, which gives us~\eqref{nonequiv-QK-Gr24} in the non-equivariant limit.
 As an example of a generalised QK ring for ${\rm Gr}(2,4)$, consider $\CR^{\rm 3d}[1,-1]$, for the gauge theory $U(2)_{1,-1}$ with 4 fundamentals. By direct computation, one can work out its ring structure in the basis of Schubert classes, which reads:
\begin{align}
    \begin{split}
&\mathcal{O}_{\yng(1)}^2 = \mathcal{O}_{\yng(1,1)}+\mathcal{O}_{\yng(2)}-\mathcal{O}_{\yng(2,1)}~, \qquad\qquad \qquad \mathcal{O}_{\yng(1,1)}\; \mathcal{O}_{\yng(2,2)} = -q
   \mathcal{O}_{\yng(2)}+q \mathcal{O}_{\yng(2,1)}~,\\
&\mathcal{O}_{\yng(1)}
   \; \mathcal{O}_{\yng(1,1)} = \mathcal{O}_{\yng(2,1)}~,\qquad\qquad\qquad \qquad \qquad \mathcal{O}_{\yng(2)}^2 =-q \mathcal{O}_{\yng(1)} +q
\mathcal{O}_{\yng(2)}+\mathcal{O}_{\yng(2,2)}~,\\
   &\mathcal{O}_{\yng(1)}\;
   \mathcal{O}_{\yng(2)} =-q + q \mathcal{O}_{\yng(1)} + \mathcal{O}_{\yng(2,1)}~,\qquad \qquad\mathcal{O}_{\yng(2)}\;
   \mathcal{O}_{\yng(2,1)} = -q \mathcal{O}_{\yng(1)}+q \mathcal{O}_{\yng(2,1)}~,\\
   &\mathcal{O}_{\yng(1)} \;\mathcal{O}_{\yng(2,1)} =-q +q \mathcal{O}_{\yng(1)}+\mathcal{O}_{\yng(2,2)}~, \qquad \qquad\mathcal{O}_{\yng(2)}\; \mathcal{O}_{\yng(2,2)} = -q \mathcal{O}_{\yng(1,1)}+q \mathcal{O}_{\yng(2,2)}~,\\
   &\mathcal{O}_{\yng(1)}\;\mathcal{O}_{\yng(2,2)} = -q \mathcal{O}_{\yng(1)}+q
   \mathcal{O}_{\yng(1,1)}~, \qquad \qquad\quad\;\;\mathcal{O}_{\yng(2,1)}^2 = -q \mathcal{O}_{\yng(1,1)}-q \mathcal{O}_{\yng(2)}+2 q \mathcal{O}_{\yng(2,1)}~,\\
   &\mathcal{O}_{\yng(1,1)}^2 = \mathcal{O}_{\yng(2,2)}~,\qquad \qquad\qquad\qquad\quad\;\qquad \mathcal{O}_{\yng(2,1)} \;\mathcal{O}_{\yng(2,2)}  = -q \mathcal{O}_{\yng(2,1)}+q
   \mathcal{O}_{\yng(2,2)}~,\\
    &\mathcal{O}_{\yng(1,1)}\; \mathcal{O}_{\yng(2)} = -q+q \mathcal{O}_{\yng(1,1)}~, \qquad\qquad\qquad\;\;\;\mathcal{O}_{\yng(2,2)}^2 =q^2 - q^2 \mathcal{O}_{\yng(1)}~.\\
     &\mathcal{O}_{\yng(1,1)} \;\mathcal{O}_{\yng(2,1)} = -q \mathcal{O}_{\yng(1)}+q
   \mathcal{O}_{\yng(1,1)}~,\\
    \end{split}
\end{align}
All the other ${\rm Gr}(2,4)$ theories in the geometric window can be worked out similarly. We list a few other non-equivariant rings $\CR^{\rm 3d}[k,l]$ for ${\rm Gr}(2,4)$ in appendix~\ref{app: gen-QK-Gr24}.

%%%%%%%%%%%%%%%%%%%
\section{Quantum cohomology and Schubert defects in the 2d GLSM}\label{sec:2dlimit}

In this section, we briefly discuss the dimensional reduction of the 3d GLSM to the ordinary 2d GLSM onto the Grassmannian.  The 2d GLSM of interest is the 2d $\CN=(2,2)$ $U(N_c)$ gauge theory with $n_f$ fundamental chiral multiplets. Its twisted chiral ring gives us the quantum cohomology of $X= {\rm Gr}(N_c, n_f)$, and turning on the twisted masses $m_\alpha$ corresponds to the $T$-equivariant deformation:%
\footnote{Here we use the notation $m_\alpha$ instead of $m_i$, to diminish clutter in some formulas.}
 \be
\CR^{\rm 2d}\cong {\rm QH}^\bullet_T(X)~.
 \ee
We can similarly compute genus-zero Gromov-Witten (GW) invariants as correlation functions in the 2d GLSM~\cite{Witten:1993xi, Closset:2015rna}.

It is interesting to see how these 2d quantities emerge from our discussion above, by taking the 2d limit of the 3d GLSM. Let us consider the limit $\beta\rightarrow 0$ on $\Sigma\times S^1$, where $\beta$ is the radius of the $S^1$. The 3d and 2d variables are related as:
\be\label{3d 2d vars}
x_a= e^{-2 \pi \beta \sigma_a}~, \qquad \qquad
y_\alpha= e^{-2 \pi \beta m_\alpha}~, \qquad \qquad
q= (-2\pi \beta)^{n_f} q_{\rm 2d}~,
\ee
with $\sigma_a$ the 2d Coulomb-branch scalars, of mass dimension $1$. 
Note that $q_{\rm 2d}$  has mass dimension $n_f$ due to the non-trivial running of the 2d FI parameters.

\subsection{Defect point operators and Schubert polynomials}
The Poincar\'e duals of the Schubert varieties,  $\omega_\lambda\equiv [X_\lambda]$, are also called the Schubert classes:
 \be
 \omega_\lambda \in {\rm H}^{2|\lambda|}(X)~,
 \ee
and similarly in the equivariant setting. The equivariant Schubert classes in cohomology can be written as double Schubert polynomials $\mathfrak{S}_\lambda(\sigma, m)$, where $\sigma_a \in {\rm H}^2(X)$ correspond to the Chern roots of $S$ -- see {\it e.g.}~\cite{lam2011quantum, 2005math......6335C, 2011arXiv1110.5896A}. 
 The double Schubert polynomial $\mathfrak{S}_\lambda(\sigma, m)$ indexed by the partition $\lambda$ can be written as \cite{ikeda2011double}:
\begin{equation}\label{double-schubert-def}
    \mathfrak{S}_{\lambda}(\sigma, m) \equiv \frac{\det_{1\leq a,b\leq N_c}\left(\prod_{\alpha=1}^{\lambda_a +N_c - b}\left(\sigma_b - m_{\alpha}\right)\right)}{\prod_{1\leq b< a\leq N_c} \left(\sigma_a - \sigma_b\right)}~.
\end{equation}
In the non-equivariant limit, $m_\alpha \rightarrow 0$, they reduce to the Schur polynomials:
\begin{equation}\label{shur-poly-def}
    s_\lambda(\sigma) \equiv \frac{\det_{1\leq a,b\leq N_c} \left(\sigma_a^{\lambda_{N_c-b+1}-n_f+N_c} \right)}{\prod_{1\leq b< a\leq N_c} \left(\sigma_a - \sigma_b\right)}~.
\end{equation}
We should insert these polynomials in the 2d $A$-model computation, as we will review below.

\medskip
\noindent
Taking the 2d limit $\beta \rightarrow 0$ using the parameterisation~\eqref{3d 2d vars}, we easily check that the double Grothendieck polynomials~\eqref{doub-groth-poly} reduce to the double Schubert polynomials~\eqref{double-schubert-def}, with the scaling:
\be
  \mathfrak{G}_{\lambda}(x, y)  \rightarrow     (2\pi\beta)^{|\lambda|}\, \mathfrak{S}_{\lambda}(\sigma, m) 
\ee
up to higher-order terms in $\beta$. Therefore, we expect that the Schubert polynomials can be obtained from the point defects in 2d that we can obtain by wrapping the Grothendieck lines along the $S^1$.  

\subsubsection{Schubert point defects and 0d $\CN=2$ quivers}
Let us thus consider the 2d GLSM coupled to a defect operator, dubbed {\it Schubert defect}, defined by coupling the 2d theory to a 0d $\CN=2$ supersymmetric quiver -- that is, a supersymmetric matrix model (SMM) at the point $p\in \Sigma$ -- see~{\it e.g.}~\cite{Franco:2016tcm, Closset:2017yte} for discussions of such gauged SMMs. 

  The 0d quiver is defined exactly as in the 3d/1d case. We have the 0d gauge group
$G_{\rm 0d} = \prod_{l=1}^{n} U(r_l)$  with bifundamental chiral matter multiplets connecting each two consecutive nodes of the 0d defect, as well as $M_l$ 0d fundamental fermi multiplets at each $U(r_l)$ node, exactly as in figure~\ref{fig:line defect}.
 Similarly to \eqref{1d-defect}, we denote these defects by:
\begin{equation}\label{0d-defect}
 \omega \begin{bmatrix}
        \textbf{r}\\
        \textbf{M}
    \end{bmatrix}~, \qquad \begin{bmatrix}
        \textbf{r}\\
        \textbf{M}
    \end{bmatrix} \equiv \begin{bmatrix}
        r_1\; &  \cdots & r_n\\
        M_1\;& \cdots & M_n\\
    \end{bmatrix}~.
\end{equation}
The ``supersymmetric vacuum equations'' for this defect theory are similar to the ones of the 1d defect, namely we still have to solve \eqref{J-equaions} and \eqref{D-equations}. Therefore, the defect restricts the 2d field $\phi$ at $p\in \Sigma$ to the Schubert cell given as in~\eqref{phi-final}. 

The defect contributes to the 2d $A$-model according to its supersymmetric `partition function', which can be obtained by naive dimensional reduction of the 1d index. This gives us a polynomial in $\sigma_a$ and $m_\alpha$, which we denote by:
\be \label{0d-defect}
    \omega_{\lambda}(\sigma, m) \equiv Z^{\rm 0d}\begin{bmatrix}
        r_1\; & \;\cdots\;&  \; r_n\; \\
        M_1\;&\; \cdots \; &\;M_n
    \end{bmatrix}(\sigma,m)~.
\ee
For definiteness, let us consider the `generic Schubert defect' defined as in figure~\ref{fig: groth-defect}.  
Then the parameters $r_l$ and $M_l$ are given by \eqref{Ml-groth}.
 This supersymmetric matrix model can be reduced to a JK residue, similarly to the 1d index. One finds:
\begin{equation}\label{0dcontour}
	 \omega_{\lambda}(\sigma, m) = \prod_{l=1}^{n}\frac{1}{l!}\oint \frac{d^l s^{(l)}}{(2\pi i)^l}\;  \Delta^{(l)}(s) \;  \text{Z}^{\rm 0d}_{\text{matter}}(s, \sigma, m)~,
\end{equation}
where $s^{(l)}_{i_{l}}$ are the components of the adjoint complex scalar that live in the 0d $\mathcal{N}=2$ $U(r_l)$ vector multiplet.  The matter contribution is given by:
\begin{equation}\label{0d-matter-factor}
	\text{Z}^{\rm 0d}_{\text{matter}} (\sigma, m) = \prod_{l=1}^{n-1} \left(\prod_{i_l=1}^{r_l}\frac{\prod_{\alpha^{(l)} \in I_{l}}\left(s_{i_{l}}^{(l)} - m_{\alpha^{(l)}}\right)}{\prod_{j_{l+1}=1}^{l+1}\left(s_{i_{l}}^{(l)} - s_{j_{l+1}}^{(l+1)}\right)}\right) \prod_{i_{n}=1}^{r_n} \frac{\prod_{\alpha^{(n)}\in I_n}\left(s_{i_n}^{(n)} - m_{\alpha^{(n)}}\right)}{\prod_{a=1}^{N_c}\left(s_{i_n}^{(n)}-\sigma_a\right)}~,
\end{equation}
and we defined the Vandermonde determinant factor as in \eqref{vand-det-fac}:
\begin{equation}\label{vander-det}
	\Delta^{(l)}(s) \equiv \prod_{1\leq i_l \neq j_l \leq l} \left(s_{i_l}^{(l)} - s_{j_l}^{(l)}\right)~. 
\end{equation}
The factors appearing in the numerator of \eqref{0d-matter-factor} come from the Fermi multiplets at each node of the 0d quiver. These are indexed by the sets $I_l$ defined in \eqref{Il def Gline}. Meanwhile, the factors in the denominator originate from the bifundamental chiral multiplets of the 0d quiver. The contour integrals in \eqref{0dcontour} should be performed recursively,  starting with the $U(r_1)$ node. 

\medskip
\noindent
{\bf The case of 1-partitions.} In the case of the parition $\lambda = [\lambda_1, 0, \cdots, 0]$, the partition function \eqref{0dcontour} becomes:
\begin{equation}
	\omega_{[\lambda_1, 0, \cdots, 0]}(\sigma, m) = \oint \frac{ds}{2\pi {i}} \frac{\prod_{\alpha\in I_1}(s-m_\alpha)}{\prod_{a=1}^{N_c} (s-\sigma_a)} = \sum_{1\leq a \leq N_c} \frac{\prod_{\alpha\in I_1}\left(\sigma_a-m_\alpha\right)}{\prod_{b\neq a}^{N_c}\left(\sigma_a - \sigma_b\right)}~.
\end{equation}

\medskip
\noindent
{\bf The generic partition.} Jumping ahead to the general case, let us take the partition $\lambda = [\lambda_1, \cdots, \lambda_n, 0, \cdots, 0]$. Doing the contour integrals \eqref{0dcontour} recursively, as we did for \eqref{Z-after-cont-int}, we find the following explicit form for the partition function of the matrix model:
\begin{align}\label{0defectfinal}
\begin{split}
    	\omega_{\lambda}(\sigma, m) &= \sum_{\mathcal{J}} \prod_{l=1}^{n}\left[\Delta^{(J_l)}(\sigma)\prod_{i_l\in J_l}\frac{\prod_{\alpha^{(l)}\in I_l}(\sigma_{i_l}-m_{\alpha^{(l)}})}{\prod_{j_{l+1}\in J_{l+1}}\left(\sigma_{i_{l}}-\sigma_{{j}_{l+1}}\right)}\right]~,\\
     &= \sum_{\mathcal{J}} \prod_{l=1}^{n}\prod_{i_l\in J_l}\frac{\prod_{\alpha^{(l)}\in I_l}(\sigma_{i_l}-m_{\alpha^{(l)}})}{\prod_{j_{l+1}\in J_{l+1}\smallsetminus J_l}\left(\sigma_{i_{l}}-\sigma_{{j}_{l+1}}\right)}~,\\
\end{split}
\end{align}
with the indexing sets $\mathcal{J}$ defined in \eqref{J-index-set}. The Vandermonde factor appearing in the first line is defined exactly as in \eqref{Jl-vandermonde}.
 The expression~\eqref{0defectfinal} can be massaged into the determinant formula:
\begin{equation}\label{double-Schubert}
	\omega_{\lambda} (\sigma, m) = \frac{\det_{1\leq a,b\leq N_c} \left[\prod_{l=N_c-b+1}^{N_c}\prod_{\alpha^{(l)}\in I_{l}}\left(\sigma_a - m_{\alpha^{(l)}}\right)\right]}{\prod_{1\leq b< a\leq N_c} \left(\sigma_a - \sigma_b\right)} =\mathfrak{S}_{\lambda}(\sigma, m)~,
\end{equation}
which is none other that the double Schubert polynomial~\eqref{double-schubert-def}.

\medskip
\noindent
{\bf Double Schubert polynomials for ${\rm Gr}(2,4)$.} As an example, let us write down the Schubert polynomials for  Gr$(2,4)$:
\begin{align}\label{schubet-poly-Gr24}
    \begin{split}
        % &\mathfrak{S}_{[0,0]}(\sigma, m) =1 ~,\\
        &\mathfrak{S}_{\yng(1)}(\sigma, m) = \sigma _1+\sigma _2-m_1-m_2~,\\
        &\mathfrak{S}_{\yng(1,1)}(\sigma, m) = \sigma _1 \sigma _2 - m_1 \sigma _1-m_1 \sigma _2+m_1^2~,\\
        &\mathfrak{S}_{\yng(2)}(\sigma, m) = \sigma _1^2+\sigma _1 \sigma _2 +\sigma _2^2 -(m_1 +m_2 +m_3) \sigma _1-(m_1 +m_2+m_3) \sigma _2\\
        &\qquad\qquad\quad+m_1 m_2+m_1 m_3+m_2 m_3~,\\
        &\mathfrak{S}_{\yng(2,1)}(\sigma, m) = \sigma _1 \sigma _2^2+\sigma _1^2 \sigma _2+(m_1^2+m_1 m_2 + m_1m_3) \sigma _1+(m_1^2 + m_1m_2+m_1m_3) \sigma _2\\
        &\qquad \qquad\quad-m_1 \sigma _1^2-m_1 \sigma _2^2 -(2 m_1 +m_2+m_3) \sigma _1 \sigma _2-m_1^2 m_2 -m_1^2 m_3 ~,\\
        &\mathfrak{S}_{\yng(2,2)}(\sigma, m) =\sigma
   _1^2 \sigma _2^2-(m_1^2 m_2  + m_1m_2^2) \sigma _1-(m_1^2 m_2  +m_1m_2^2 )
   \sigma _2+(m_1 + m_2)^2 \sigma _1 \sigma _2\\
   &\qquad \qquad\quad +m_1 m_2 \sigma_1^2 + m_1 m_2\sigma_2^2 -(m_1 + m_2)\sigma_1^2 \sigma_2 - (m_1 +m_2)\sigma_1 \sigma_2^2+m_1^2 m_2^2~. 
    \end{split}
\end{align}
They are indeed obtained as the 2d limit of the Grothendieck polynomials~\eqref{grothendieck-polys-Gr24}.

\subsection{Localisation formula on $\mathbb{P}^1$ and GW invariants}
Any $A$-model correlation function of the Grassmannian GLSM on $\mathbb{P}^1$ can be obtained from a JK residue similar to~\eqref{JK for QK ordinary}, as derived in~\cite{Closset:2015rna,Benini:2015noa}:
\bea\label{JK for HK}
&\big\langle\omega \big\rangle_{\bbP^1}   = \sum_{d= 0}^\infty  q_{\rm 2d}^{d}\, \textbf{I}^{\rm 2d}_{d} [\omega]~,\\
&\textbf{I}^{\rm 2d}_{d} [\omega] =\sum_{\substack{\m_a \geq 0 \\ |\m|=d}}\;\; {(-1)^{d (N_c-1)}\ov N_c!}\oint_{\rm JK} \prod_{a=1}^{N_c}\left[{d \sigma_a\ov 2\pi i} {1\ov   \prod_{\alpha=1}^{n_f} (\sigma_a-m_\alpha)^{1+\m_a}}\right]\, \Delta(\sigma)\, \omega(x,y)~,
\eea
where $\Delta(\sigma) =\prod_{a\neq b}(\sigma_a-\sigma_b)$. This formula captures all the genus-$0$ GW invariants of Gr$(N_c, n_f)$. The 3d and 2d formulas are related by a naive scaling limit. Indeed, if we assume that the insertion in 3d gives a 2d insertion according to:
\be
\SL \rightarrow (2\pi \beta)^{d_\omega} \omega~,
\ee
where $d_\omega$ is the mass-dimension of the homogenous polynomial $\omega=\omega(\sigma)$, and if we further assume that we can commute the limit and the integration, then one finds:%
\footnote{Of course, integration and $\beta\rightarrow 0$ limit do not commute in general, which is why QK-theoretic invariants contain strictly more information than GW invariants. In the computation of the general 3d/1d observables in the small radius limit, one must generally consider the contribution of several `holonomy saddles'~\protect\cite{Ghim:2019rol}, which decomposes the 3d quantities into several 2d observables.}
\be
 \textbf{I}_{d} [{\SL}]\rightarrow
(2\pi \beta)^{-{\rm dim}(X)- d \, n_f + d_\omega} \; \textbf{I}^{\rm 2d}_{d} [\omega]~.
\ee
Let us write down the  quantum cohomology ring in the Schubert basis as:
\be
\omega_\lambda \omega_\mu = {c_{\lambda\mu}}^{\nu} \, \omega_\nu~,\qquad \qquad  {c_{\lambda\mu}}^{\nu}= \sum_{d\geq 0} c_{\lambda\mu}^{(d)\nu} q_{\rm 2d}^2~.
\ee
By dimensional analysis, we see that, in the non-equivariant limit $m_\alpha=0$, only one single degree $d$ can contribute to ${c_{\lambda\mu}}^{\nu}$ in 2d, with:
\be\label{munulam cond 2d}
|\lambda|+|\mu|= |\nu|+ d \, n_f~.
\ee 
In particular, using the fact that $|\lambda^\vee|= {\rm dim}(X)- |\lambda|$, we find that the structure ring constants in the Schubert basis can be deduced from the 3d results above whenever~\eqref{munulam cond 2d} holds true:
\be
 \CN^{(d)  \nu}_{\lambda\mu} \rightarrow c_{\lambda\mu}^{(d)\nu}~.
\ee
In particular, for $d=0$ and in the non-equivariant limit, the K-theoretic LR coefficients~\eqref{LR K gen} that satisfy $|\lambda|+|\mu|= |\nu|$ are equal to the ordinary LR coefficients for the product of Schur polynomials, as expected.

 %%%%%%%%%%%

\subsection*{Acknowledgements}
 
We thank Mathew Bullimore, Stefano Cremonesi,  Hans Jockers, Heeyeon Kim, Horia Magureanu, Qaasim Shafi, Leonardo Mihalcea, and Eric Sharpe for discussions and correspondence. CC is particularly indebted to Mathew Bullimore and Heeyeon Kim for seminal contributions and for collaboration on an earlier version of this project, and to Leonardo Mihalcea and Eric Sharpe for sharing their deep understanding of quantum K-theory. 
CC is a Royal Society University Research Fellow supported by the University Research Fellowship Renewal 2022 `Singularities, supersymmetry and SQFT invariants'. The work of OK is supported by the School of Mathematics at the University of Birmingham.

 %%%%%%%%%%
 \appendix
 
 %%%%%%%%%%
 
 %%%%%%%%%%%%%%%%%%%%%%%%%%%%%%%%%
 \section{The JK residue formula and the sum over 3d vacua}\label{app: JK residue}    

In this appendix, we further study the JK residue formula~\eqref{JK-residue-formula} for the genus-zero topologically twisted index of the 3d $\CN=2$ SQCD$[N_c, k, l, n_f,0]$ theory. We show that the JK residues that contribute non-trivially to the index can be organised in terms of the 3d vacua that contribute to the Witten index recently analysed in~\cite{Closset:2023jiq}. In particular, for $(k,l)$ in the geometric window as defined in section~\ref{subsec:geom window}, only the so-called Higgs-branch singularities~\eqref{x to y HB sing} contribute.

\subsection{Singularity structure of the twisted index integrand}

Let us first review the singularity structure of the twisted-index integrand~\eqref{JK form}.  Consider the set of all codimension-one singularities of this integrand on:
\be
\t {\mathfrak{M}}\cong \left(\mathbb{P}^1\right)^{N_c}~,
\ee
a natural compactification of the space spanned by the gauge variables $x_a$, $a=1, \cdots, N_c$. Here, $\t {\mathfrak{M}}$ is really a compactification of a covering space of the classical 3d Coulomb branch $\mathfrak{M}\cong (\C^\ast)^{N_c}/S_{N_c}$, and we are effectively dealing with an abelianised theory. Then, one can consider the 3d monopole operators $T_a^\pm$ for each factor $U(1)_a\subset U(N_c)$, whose charges are governed by the singularities of the integrand at $x_a=0$ and $x_a=\infty$.

The integrand then has two types of codimension-one singularities, from either the matter fields or the monopole operators, in any given topological sector indexed by the abelianised magnetic flux $\m$. To every hyperplane singularity $H$, we assign a gauge charge $Q=(Q^a)$ under $\prod_{a=1}^{N_c} U(1)_a$, as follows~\cite{Closset:2016arn}:

    \medskip 
    \noindent
    {\bf Matter field singularities.}
    Depending on the value of the magnetic fluxes $\mathfrak{m}\in \bbZ^{N_c}$, there can be singularites that arise from the chiral multiplet contribution to $Z_{\mathfrak{m}}(x,y)$ in \eqref{JK form}. They are characterised by the singular hyperplanes $H_{\rho_a, i}$ defined by:
    \begin{equation}\label{matter-hyperplane}
        H_{\rho_a, \alpha} \equiv \{x\in \t {\mathfrak{M}}:\quad x^{\rho}\;y^{-1}_{\alpha} = 1~\}~, \qquad \alpha = 1, \cdots, n_f~,
    \end{equation}
    with $\rho_a$ being the fundamental weights of $U(N_c)$. To   such 
 hyperplanes, we assign the gauge charges of the corresponding chiral multiplets:
    \begin{equation}\label{matter-sing}
        Q_{a, \alpha}  = \rho_a =  (0, \cdots, 0, \underbrace{1}_{a-\text{th}}, 0, \cdots, 0)~, \qquad \alpha = 1, \cdots, n_f~, \qquad a=1, \cdots, N_c~.
    \end{equation}
Note that the charge vectors are the same for the $n_f$ distinct fundamental chiral multiplets. 
    
    \medskip
    \noindent
    {\bf Monopole operators singularities.}
    These singularities as associated to the monopole operators that we can define semi-classically in  the asymptotic regions of the classical Coulomb branch $\mathfrak{M}$. Depending on the whether we consider the limit $x_a = \infty\; (\sigma_a = -\infty)$ or $x_a = 0 \;(\sigma_a = \infty)$, we have the monopole operators $T_a^+$ or $T_a^-$, respectively. We define the corresponding hyperplanes: 
    \begin{equation}\label{monopole-hyperplane}
        H_{a, \pm} \equiv \{x \in \t {\mathfrak{M}}: x_a = 0,~\infty\}~,\qquad a = 1, \cdots, N_c~.
    \end{equation}
 In these asymptotic regions, the JK form $\mathfrak{I}_{\mathfrak{m}}$ in \eqref{JK form} has the following behaviour:
    \begin{equation}\label{behaviour-at-mon-sing}
        \mathfrak{I}_{\mathfrak{m}} \sim x_a^{\pm \left(Q^{(\pm)}_{a} (\mathfrak{m})-r_{a, \pm}\right)} \frac{dx_a}{x_a}~,
    \end{equation}
    with $Q^{(\pm)}_{a}(\mathfrak{m}) \equiv  \sum_{b=1}^{N_c}{Q^{(\pm)\;b}_{a}} \mathfrak{m}_b$ being the 1-loop-exact gauge charges of the monopole operators under $\prod_{a=1}^{N_c} U(1)_a$. These are the gauge charges that we should assign to the hyperplanes \eqref{monopole-hyperplane}. They take the explicit form:
    \cite{Closset:2016arn}: 
    \begin{equation}\label{monopole-sing}
        {Q^{(\pm)b}_a} = \delta_a^b \left(\pm k - \frac{n_f}{2}\right) \pm l~, \qquad a, b = 1, \cdots, N_c~.
    \end{equation}
Note also that $r_{a,\pm}$ are the $R$-charges of the monopole operators.

\subsection{JK residue prescription and phases of SQCD$[N_c, k, l, n_f,0]$}
In order to define the JK residue, we need to consider the set of all codimension-$N_c$ singularites in each flux sector $\m$:
\be
\t {\mathfrak{M}}_{\text{sing}}^{\mathfrak{m}}\subset \t {\mathfrak{M}}~.
\ee
Such singularities arise from the intersection of $r_s \geq N_c$ hyperplanes -- for generic $y_i$ we have $r_s=N_c$ always, which is the case we will focus on. The JK residue prescription~\cite{ASENS_1999_4_32_5_715_0} instructs us to pick `the JK residue' at the singularity $x=x_\ast \in \t {\mathfrak{M}}_{\text{sing}}^{\mathfrak{m}}$:
\be
\underset{x = x_*}{\text{JK-Res}} \left[\textbf{Q}(x_*), \eta_\xi \right]\mathfrak{I}_{\mathfrak{m}}[\SL](x, y, q)~.
\ee
This is defined in terms of the gauge charges $Q$ assigned to the $N_c$ hyperplanes, as follows. 
Let us first define the positive cone generated by the charges $\textbf{Q}(x_*)=(Q_1, \cdots,Q_N)$ as:
    \begin{equation}
        \text{Cone}_+\left(\textbf{Q}(x_*)\right) \equiv \{c_1 Q_1 + \cdots + c_s Q_{N_c} : c_1, \cdots, c_{N_c} > 0 \}\subset \bbC^{N_c}~.
    \end{equation}
    We note that, for the definition of this cone to make sense, we need these charges to be \textit{projective} in the sense that they live in a half-space of $\bbC^{N_c}$. In the \textit{non-projective} case, the JK residue is ill-defined. 
 Then, we are instructed to pick an auxiliary parameter $\eta \in \mathbb{C}^{N_c}$, which could be arbitrary as long as it is not parallel to any of the JK charges. Here, we will choose to allign $\eta\equiv \eta_\xi$ with the real 3d FI parameter $\xi$, as follows:
    \begin{equation}\label{choice-eta}
        \eta_\xi \equiv  \xi (1, \cdots, 1) \in \bbC^{N_c}~.
    \end{equation}
We thus assume that $\xi\neq 0$ in the following. 
 Then, the singular points $x_*\in \t {\mathfrak{M}}^{\mathfrak{m}}_{\text{sing}}$ that contribute non-trivially to the JK residue are those such that $\eta\in\text{Cone}_+(\textbf{Q}(x_*))$ -- see {\it e.g.}~\cite{Closset:2016arn} for further details.

    Hence, to determine which singularities contribute to the correlation function \eqref{JK-residue-formula} for SQCD$[N_c, k, l, n_f, 0]$, we need to determine all the possible positive cones that contain $\eta=\eta_\xi$. That is, we need to find all possible abelianised gauge charges $Q^{(p_1)}_1, \cdots, Q^{(p_{N_c})}_{N_c}$ such that:
    \begin{equation}\label{sing-contr-equ}
        \eta  = \sum_{a=1}^{N_c} c_a \;Q^{(p_a)}_{a}~, \qquad c_a > 0~, ~\forall a = 1,~ \cdots, N_c~.
    \end{equation}
    Here, we use the labels $p_a \in \{+, -, 1, \cdots, n_f\}$ depending on whether the associated singularity comes from the monopole operators $T^\pm$ or the fundamental chiral multiplets $\Phi_\alpha$, $\alpha=1, \cdots, n_f$, respectively. Let us first work out the case $N_c=2$. We will then briefly discuss the general case. 

    \medskip
    \noindent
    {\bf For SQCD$[2, k, l, n_f, 0]$.}  Consider the $U(2)_{k, k+2l}$ gauge theory coupled with $n_f$ matter multiplets in the fundamental representation. In this case, we have the following charges defining the singular hyperplanes:
    \begin{equation}
        Q_1^{(\alpha)} = (1,\; 0)~, \qquad Q_2^{(\alpha)} = (0,\; 1)~ \qquad i = 1, \cdots, n_f~, 
    \end{equation}
    for the matter singularities \eqref{matter-sing}. And, 
    \begin{equation}
        Q^{(\pm)}_1 = \left(\pm k-\frac{n_f}{2}\pm l,\; \pm l\right)~, \qquad Q^{(\pm)}_2 = \left(\pm l,\; \pm k - \frac{n_f}{2}\pm l\right)~, 
    \end{equation}
    for the monopole singularities \eqref{monopole-sing}. 
    To find which singularities do contribute for different choices of CS levels $k$ and $l$, and for a given $n_f$, we need to study the  equations \eqref{sing-contr-equ}, namely:
    \begin{equation}\label{sing-contr-equ-U(2)}
        \xi (1,\; 1) = c_1 Q^{(p_1)}_1 + c_2 Q^{(p_2)}_2~, \qquad p_1, p_2 \in \{+, -, 1, \cdots, n_f\}~,
    \end{equation}    
    with the constraint that $c_1, c_2 > 0$.
The solutions to these equations are closely related to the 3d vacua that contribute to the 3d Witten index, hence we shall index the solutions as in~\cite{Closset:2023jiq}. We have the following possibilities,
    \begin{itemize}
        \item \textit{\textbf{Type I.}} In this case,  all the contributing singularities come from the matter multiplets. In this case, equations \eqref{sing-contr-equ-U(2)} become:
        \begin{equation}
            \xi (1,\;1) = c_1 (1,\; 0) + c_2 (0,\; 1)~,
        \end{equation}
        which has the unique solution $c_1 = c_2 = \xi$. Thus, this type of singularities does indeed contribute to the correlation function if $\xi>0$. 

        Let $\alpha_1, \alpha_2 \in \{1, \cdots, n_f\}$, $\alpha_1\neq \alpha_2$, denote two possible distinct matter singularities. In this case, the JK residue \eqref{JK-residue-formula} for $N_c=2$ becomes:
        \begin{equation}
            \left<\SL\right>_{\bbP^1\times S^1} = \Theta(\xi) \sum_{\mathfrak{m}\in \bbZ^2}\; \sum_{1\leq \alpha_1 < \alpha_2 \leq n_f}\; \underset{\substack{x_1 = y_{\alpha_1}\\x_2 = y_{\alpha_2}}}{\text{Res}} \mathfrak{I}_{\mathfrak{m}}[\SL](q,x,y)~,
        \end{equation}
        where we used the residual gauge symmetry to cancel the $2!$ factor. Here $\Theta$ is the Heaviside step function, as in~\cite{Closset:2023jiq}. We further note  that, although the residue formula is formally given in terms of a sum over all topological sectors $\mathfrak{m}\in \bbZ^2$,  only the topological sectors with $\m_1, \m_2\geq 0$ actually contribute singularities. 
        From the point of view of the semi-classical analysis of the vacua~\cite{Closset:2023jiq}, we get a 3d Gr$(2, n_f)$ Higgs-branch vacuum spanned by the matter multiplets.

         \item \textit{\textbf{Type II}.} In this case, we take one of the contributing singularities to be from the matter multiplets and the other from one of the monopoles $T^{(\pm)}_{1,2}$. The equations \eqref{sing-contr-equ-U(2)} become:
        \begin{equation}
            \xi(1,1) = c_1 (1,\; 0) + c^{(\pm)}_2 (\pm l,\; \pm k-\frac{n_f}{2}\pm l)~,
        \end{equation}
        which have the solution:
        \begin{equation}
            c_1 = \xi\frac{\left(\pm k-\frac{n_f}{2}\right)}{\pm k-\frac{n_f}{2} \pm l}>0~, \qquad c_2 = \frac{\xi}{\pm k-\frac{n_f}{2} \pm l}>0~.
        \end{equation}
    The corresponding 3d supersymmetric vacuum is a  hybrid topological-Higgs vacuum of the form: $\bbP^{n_f-1} \times U(1)_{k + l \pm \frac{n_f}{2}}$~\cite{Closset:2023jiq}. 
        
        \item \textit{\textbf{Type III.}} In this case,  all the contributing singularities come from the two monopole operators $T^{\pm}_{1,2}$ \eqref{monopole-sing}. We actually have three possible choices.
        The first two are where we take the singular hyperplanes $x^{(+)}_{1,2} = \infty$ or $x^{(-)}_{1,2} = 0$. The corresponding equations \eqref{sing-contr-equ-U(2)} take the following form:
        \begin{equation}
            \xi (1,\;1) = c^{(\pm)}_1 \left(\pm k-\frac{n_f}{2} \pm l ,\; \pm l\right) + c^{(\pm)}_2 \left(\pm l,\; \pm k-\frac{n_f}{2} \pm l\right)~.
        \end{equation}
        In either case, we have the unique solution:
        \begin{equation}
            c^{(\pm)}_1 = c^{(\pm)}_2 =  \frac{\xi}{\pm k-\frac{n_f}{2} \pm 2l}>0~, \qquad \text{iff}~\quad \xi \left(\pm k-\frac{n_f}{2}\pm 2l\right) >0~.
        \end{equation}
        From the point of view of moduli space of vacua, in this case, we find topological vacua of the form 3d TQFT with gauge group $U(2)_{\pm k,\; \pm k-\frac{n_f}{2}\pm 2l}$~\cite{Closset:2023jiq}.

       The third choice of intersecting hyperplanes is $x^{(+)}_1 = \infty$ and $x_2^{(-)} = 0$. This leads to the equations:
        \begin{equation}
            \xi(1,\;1) = c^{(+)}_1 Q_1^{(+)} + c^{(-)}_2 Q_2^{(-)}~.
        \end{equation}
        
        These two equations can be uniquely solved by:
        \begin{equation}
            c^{(+)}_1 = \xi\frac{k+\frac{n_f}{2}}{\left(k+\frac{n_f}{2}\right)\left(k-\frac{n_f}{2}\right)+2kl}>0~, \quad c^{(-)}_2 =  -\xi \frac{k-\frac{n_f}{2}}{\left(k+\frac{n_f}{2}\right)\left(k-\frac{n_f}{2}\right) + 2kl} >0~.
        \end{equation}
        We can simplify these constraints into the following:
        \begin{equation}
            |k| <\frac{n_f}{2}~, \qquad \xi\left(k^2 +2k l -\frac{n_f^2}{4}\right)>0~.
        \end{equation}
        In this case, the moduli space of vacua consists of topological ones of the form of a TQFT with a gauge group $U(\underbracket{1)_{k+l-\frac{n_f}{2}}\times U(1}_{l})_{k+l+\frac{n_f}{2}}$~\cite{Closset:2023jiq}.
    \end{itemize} 

    \noindent
    In the analysis above, we tacitly assumed that we are staying away from the \textit{marginal case}, which is the case when  $|k| = \frac{n_f}{2}$. In the marginal case, the monopole charges become parallel to each other and render the JK residue ill-defined -- that is, the singularities become {\it non-projective} --, as we can see from~\eqref{monopole-sing}.  These non-projective singularities are conjecturally associated with strongly-coupled 3d vacua which are not accounted for when solving the semi-classical 3d vacuum equations \cite{Closset:2023jiq}.  It may be interesting to apply the methods of~\cite{Bullimore:2019qnt} to better deal with the marginal case. We will not consider this issue further in this work.

    \medskip
    \noindent
    {\bf For SQCD$[N_c, k, l, n_f, 0]$.} It is straightforward to extend the analysis of the $N_c=2$ case above to the general case. One finds a one-to-one correspondence between the possible mixtures of the singularities and types of supersymmetric vacua that we get from the semi-classical analysis of~\cite{Closset:2023jiq}.

\section{{Other generalised {${\rm QK}({\rm Gr}(2,4))$} rings}}\label{app: gen-QK-Gr24}
 
In this appendix, we provide a few more examples for the non-equivariant rings $\mathcal{R}^{\text{3d}}[k,l]$ for the Gr$(2,4)$ 3d GLSM, for different values of the CS levels $k,l$ in the geometric window, in addition to the ones discussed in subsection \ref{subsec: gen-QK-rings}. Specifically, we write down the result for the cases $(k,l) = (0,0), (1,0)$ and $(2,-2)$ in the Schubert class basis. In each of these three cases, we give both the topological metric  \eqref{top-metric-definition} and the 3d ring structure.

\subsection{Case $(k,l) = (0,0)$}
In this case, we find that the 2-point function \eqref{top-metric-definition}, computed up to degree $d=4$, has the following form:
\begin{equation}
      \left<\mathcal{O}_\mu \;\mathcal{O}_\nu\right>^{(k,l) =(0,0)}_{\mathbb{P}^1\times S^1}=\left(
\begin{array}{cccccc}
 1 & 1 & 1 & 1 & 1 & 1 \\
 1 & 1 & 1 & 1 & 1 & 0 \\
 1 & 1 & 1 & 0 & 0 & 0 \\
 1 & 1 & 0 & 1 & 0 & 0 \\
 1 & 1 & 0 & 0 & 0 & 0 \\
 1 & 0 & 0 & 0 & 0 & 0 \\
\end{array}
\right)~.
\end{equation}

Meanwhile, the generalised QK ring $\mathcal{R}^{\text{3d}}[0,0]$ has the following structure:
\begin{align}
\begin{split}
&\mathcal{O}_{\yng(1)}^2 = \mathcal{O}_{\yng(1,1)}+\mathcal{O}_{\yng(2)}-\mathcal{O}_{\yng(2,1)}~. \\
 &\mathcal{O}_{\yng(1)}\; \mathcal{O}_{\yng(2)} = \mathcal{O}_{\yng(2,1)}~, \\
 &\mathcal{O}_{\yng(1)}\;\mathcal{O}_{\yng(1,1)} = \mathcal{O}_{\yng(2,1)}~, \\
  &\mathcal{O}_{\yng(1)}\;\mathcal{O}_{\yng(2,1)} = - q +2 q \mathcal{O}_{\yng(1)}-q \mathcal{O}_{\yng(1,1)}-q \mathcal{O}_{\yng(2)}+q \mathcal{O}_{\yng(2,1)} +\mathcal{O}_{\yng(2,2)}~, \\
  &\mathcal{O}_{\yng(1)}\;\mathcal{O}_{\yng(2,2)}  = -q \mathcal{O}_{\yng(1)}+q \mathcal{O}_{\yng(1,1)}+q\mathcal{O}_{\yng(2)}-q \mathcal{O}_{\yng(2,1)}~, \\
 &\mathcal{O}_{\yng(1,1)}^2 = \mathcal{O}_{\yng(2,2)}~, \\
  &\mathcal{O}_{\yng(1,1)}\; \mathcal{O}_{\yng(2)} = -q + q \mathcal{O}_{\yng(1)}~, \\
  &\mathcal{O}_{\yng(1,1)}\;\mathcal{O}_{\yng(2,1)} =-q \mathcal{O}_{\yng(1)}+q \mathcal{O}_{\yng(1,1)}+q \mathcal{O}_{\yng(2)}-q \mathcal{O}_{\yng(2,1)}~,\\
   &\mathcal{O}_{\yng(1,1)} \;\mathcal{O}_{\yng(2,2)} = - q \mathcal{O}_{\yng(2)}+q \mathcal{O}_{\yng(2,1)}~,\\
    &\mathcal{O}_{\yng(2)}^2 = \mathcal{O}_{\yng(2,2)}~, \\
    &\mathcal{O}_{\yng(2)}\; \mathcal{O}_{\yng(2,1)}  =  - q \mathcal{O}_{\yng(1)}+q \mathcal{O}_{\yng(1,1)}+q\mathcal{O}_{\yng(2)}-q \mathcal{O}_{\yng(2,1)}~,\\
     &\mathcal{O}_{\yng(2)}\; \mathcal{O}_{\yng(2,2)}  =  - q\mathcal{O}_{\yng(1,1)}+q\mathcal{O}_{\yng(2,2)}~,\\
      &\mathcal{O}_{\yng(2,1)}^2 = q^2- 2 q^2 \mathcal{O}_{\yng(1)}-\left(q-q^2\right) \mathcal{O}_{\yng(1,1)}-\left(q-q^2\right)\mathcal{O}_{\yng(2)}-\left(q^2-3 q\right)\mathcal{O}_{\yng(2,1)}-q
  \mathcal{O}_{\yng(2,2)}~, \\
  &\mathcal{O}_{\yng(2,1)}\; \mathcal{O}_{\yng(2,2)} = - q^2 +2 q^2 \mathcal{O}_{\yng(1)}-q^2 \mathcal{O}_{\yng(1,1)}-q^2\mathcal{O}_{\yng(2)}-\left(q-q^2\right) \mathcal{O}_{\yng(2,1)}+q \mathcal{O}_{\yng(2,2)}~, \\
&\mathcal{O}_{\yng(2,2)}^2 = q^2 -2 q^2 \mathcal{O}_{\yng(1)}+q^2 \mathcal{O}_{\yng(1,1)}+q^2 
\mathcal{O}_{\yng(2)}-q^2 \mathcal{O}_{\yng(2,1)}~.
 \end{split}
\end{align}
%%%%%%%%%%%%%%%%%%%%%%%%%%%%%%%%%%%%%%%%%%%%%%
%%%%%%%%%%%%%%%%%%%%%%%%%%%%%%%%%%%%%%%%%%%%%%
 %%%%%%%%%%%%%%%%%%%%%%%%%%%%%%%%%%%%%%%%%%%%%%
%%%%%%%%%%%%%%%%%%%%%%%%%%%%%%%%%%%%%%%%%%%%%%
%%%%%%%%%%%%%%%%%%%%%%%%%%%%%%%%%%%%%%%%%%%%%%
\subsection{Case $(k,l) = (1,0)$}
In this case, we find that the 2-point function takes the form:
%up degree $d=4$ has the following from:
\begin{equation}
    \left<\mathcal{O}_\mu\;\mathcal{O}_\nu\right>^{(k,l) = (1,0)}_{\bbP^1\times S^1} =\left(
\begin{array}{cccccc}
 1 & 1 & 1 & 1 & 1 & 1 \\
 1 & 1 & 1 & 1 & 1 & 0 \\
 1 & 1 & 1 & 0 & 0 & 0 \\
 1 & 1 & 0 & 1 & 0 & -q \\
 1 & 1 & 0 & 0 & 0 & -q \\
 1 & 0 & 0 & -q & -q & 0 \\
\end{array}
\right)~,
\end{equation}
 and we have the following $\mathcal{R}^{\text{3d}}[1,0]$ ring structure:
\begin{align}
\begin{split}
 &\mathcal{O}_{\yng(1)}^2 = \mathcal{O}_{\yng(1,1)}+\mathcal{O}_{\yng(2)}-\mathcal{O}_{\yng(2,1)}~. \\
  &\mathcal{O}_{\yng(1)}\; \mathcal{O}_{\yng(1,1)}  = \mathcal{O}_{\yng(2,1)}~,\\
   &\mathcal{O}_{\yng(1)}\; \mathcal{O}_{\yng(2)} = q-2 q \mathcal{O}_{\yng(1)}+q \mathcal{O}_{\yng(1,1)}+q \mathcal{O}_{\yng(2)}-(q-1) \mathcal{O}_{\yng(2,1)}~, \\
   &\mathcal{O}_{\yng(1)}\; \mathcal{O}_{\yng(2,1)} = q-2 q \mathcal{O}_{\yng(1)}+q \mathcal{O}_{\yng(1,1)}+q \mathcal{O}_{\yng(2)}-q \mathcal{O}_{\yng(2,1)}+\mathcal{O}_{\yng(2,2)}~,
   \\
    & \mathcal{O}_{\yng(1)} \;\mathcal{O}_{\yng(2,2)}= q \mathcal{O}_{\yng(1)}-2 q \mathcal{O}_{\yng(1,1)}-q \mathcal{O}_{\yng(2)}+2 q \mathcal{O}_{\yng(2,1)}~, \\
    &\mathcal{O}_{\yng(1,1)}^2 = \mathcal{O}_{\yng(2,2)}~, \\
    &\mathcal{O}_{\yng(1,1)} \;\mathcal{O}_{\yng(2)} = q-q \mathcal{O}_{\yng(1)}-q \mathcal{O}_{\yng(1,1)}+q \mathcal{O}_{\yng(2,1)}~, \\
     &\mathcal{O}_{\yng(1,1)} \;\mathcal{O}_{\yng(2,1)} =  \mathcal{O}_{\yng(1)}-2 q \mathcal{O}_{\yng(1,1)}-q \mathcal{O}_{\yng(2)}+2 q \mathcal{O}_{\yng(2,1)}~, \\
     &\mathcal{O}_{\yng(1,1)} \;\mathcal{O}_{\yng(2,2)} = q \mathcal{O}_{\yng(2)}-2 q \mathcal{O}_{\yng(2,1)}+q \mathcal{O}_{\yng(2,2)} ~,\\
     &\mathcal{O}_{\yng(2)}^2 =q^2 -\left(2 q^2-q\right) \mathcal{O}_{\yng(1)}-\left(q-q^2\right) \mathcal{O}_{\yng(1,1)}-\left(2 q-q^2\right)
   \mathcal{O}_{\yng(2)}-\left(q^2-2 q\right) \mathcal{O}_{\yng(2,1)}+\mathcal{O}_{\yng(2,2)}~, \\
    &\mathcal{O}_{\yng(2)}\; \mathcal{O}_{\yng(2,1)} =q^2 -\left(2 q^2-q\right) \mathcal{O}_{\yng(1)}-\left(q-q^2\right)
   \mathcal{O}_{\yng(1,1)}-\left(q-q^2\right) \mathcal{O}_{\yng(2)}- q^2 \mathcal{O}_{\yng(2,1)}+q \mathcal{O}_{\yng(2,2)} ~,\\
    &\mathcal{O}_{\yng(2)} \;\mathcal{O}_{\yng(2,2)} = q^2 \mathcal{O}_{\yng(1)}-\left(2 q^2-q\right) \mathcal{O}_{\yng(1,1)}-q^2 \mathcal{O}_{\yng(2)}-\left(q-2 q^2\right)
   \mathcal{O}_{\yng(2,1)}-q \mathcal{O}_{\yng(2,2)}~, \\
    &\mathcal{O}_{\yng(2,1)}^2 = q^2-2 q^2 \mathcal{O}_{\yng(1)}+\left(q^2+q\right) \mathcal{O}_{\yng(1,1)}+\left(q^2+q\right) \mathcal{O}_{\yng(2)}-\left(q^2+4 q\right)
   \mathcal{O}_{\yng(2,1)}+2 q \mathcal{O}_{\yng(2,2)}~, \\
 &\mathcal{O}_{\yng(2,1)}\; \mathcal{O}_{\yng(2,2)} = -q^2 +3 q^2 \mathcal{O}_{\yng(1)}-3 q^2 \mathcal{O}_{\yng(1,1)}-2 q^2 \mathcal{O}_{\yng(2)} + \left(3 q^2+q\right)
   \mathcal{O}_{\yng(2,1)}-2 q \mathcal{O}_{\yng(2,2)}~, \\
   &\mathcal{O}_{\yng(2,2)}^2 = q^2-3 q^2 \mathcal{O}_{\yng(1)}+3 q^2 \mathcal{O}_{\yng(1,1)}+3 q^2 \mathcal{O}_{\yng(2)}-5 q^2 \mathcal{O}_{\yng(2,1)}+q^2 \mathcal{O}_{\yng(2,2)}
   ~.\\
\end{split}
\end{align}
% %%%%%%%%%%%%%%%%%%%%%%%%%%%%%%%%%%%%%%%%%%%%%%
% %%%%%%%%%%%%%%%%%%%%%%%%%%%%%%%%%%%%%%%%%%%%%%
% %%%%%%%%%%%%%%%%%%%%%%%%%%%%%%%%%%%%%%%%%%%%%%
% \subsubsection{$(k,l) = (1,1)$:}
%%%%%%%%%%%%%%%%%%%%%%%%%%%%%%%%%%%%%%%%%%%%%%
%%%%%%%%%%%%%%%%%%%%%%%%%%%%%%%%%%%%%%%%%%%%%%
%%%%%%%%%%%%%%%%%%%%%%%%%%%%%%%%%%%%%%%%%%%%%%
\subsection{Case $(k,l) = (2,-2)$}
In this case, the 2-point function reads (here up to order $q^4$):
%\CyC{check/ give full answer!}
\begin{multline}
\left<\mathcal{O}_\mu\;\mathcal{O}_\nu\right>^{(k,l)=(2,-2)}_{\bbP^1\times S^1}\\ = \left(
\begin{array}{cccccc}
 q^4+q^2+1\; &\; q^4+q^2+1 \;& \;q^4+q^2+1\; & \;q^4+q^2+1\; &\; q^4+q^2+1\; & \;q^4+q^2+1 \\
 \;q^4+q^2+1\; &\; q^4+q^2+1\; & \;q^4+q^2+1 & \;q^4+q^2+1 & \;q^4+q^2+1 &\; q^4+q^2 \\
 \;q^4+q^2+1\; & \;q^4+q^2+1 \;& \;q^4+q^2+1\; & \;q^4+q^2\; &\; q^4+q^2 \;& \;q^4+q^2 \;\\
 \;q^4+q^2+1\; & \;q^4+q^2+1\; & \;q^4+q^2\; & \;q^4+q^2+1\; &\; q^4+q^2 \;&\; q^4+q^2\; \\
\; q^4+q^2+1\; &\; q^4+q^2+1\; &\; q^4+q^2 \;& \;q^4+q^2\; & \;q^4+q^2\; & \;q^4+q^2 \;\\
 \;q^4+q^2+1\; & \;q^4+q^2\; & \;q^4+q^2\; & \;q^4+q^2\; &\; q^4+q^2\; &\; q^4+q^2\; \\
\end{array}
\right)~.
\end{multline}
For the ring $\mathcal{R}^{\text{3d}}[2,-2]$, we find:
\begin{align}
\begin{split}
&\mathcal{O}_{\yng(1)}^2 = \mathcal{O}_{\yng(1,1)}+\mathcal{O}_{\yng(2)}-\mathcal{O}_{\yng(2,1)}~, \\
&\mathcal{O}_{\yng(1)}\; \mathcal{O}_{\yng(1,1)} =  \mathcal{O}_{\yng(2,1)} ~,\\
&\mathcal{O}_{\yng(1)}\; \mathcal{O}_{\yng(2)} = -2 q+q \mathcal{O}_{\yng(1)}+q \mathcal{O}_{\yng(1,1)}  + \mathcal{O}_{\yng(2,1)}~,\\
&\mathcal{O}_{\yng(1)} \mathcal{O}_{\yng(2,1)} = -q+q \mathcal{O}_{\yng(1,1)}+\mathcal{O}_{\yng(2,2)}~, \\
 &\mathcal{O}_{\yng(1)} \;\mathcal{O}_{\yng(2,2)} = q^2 -q \mathcal{O}_{\yng(1)}+q \mathcal{O}_{\yng(1,1)}-q \mathcal{O}_{\yng(2)}+q \mathcal{O}_{\yng(2,1)}~, \\
 &\mathcal{O}_{\yng(1,1)}^2  = \mathcal{O}_{\yng(2,2)}~,\\
 &\mathcal{O}_{\yng(1,1)}\; \mathcal{O}_{\yng(2)} = q^2-q-q \mathcal{O}_{\yng(1)}+2 q \mathcal{O}_{\yng(1,1)}-q \mathcal{O}_{\yng(2)}+q \mathcal{O}_{\yng(2,1)}~, \\
  &\mathcal{O}_{\yng(1,1)}\; \mathcal{O}_{\yng(2,1)} = q^2-q \mathcal{O}_{\yng(1)}+q \mathcal{O}_{\yng(1,1)}-q \mathcal{O}_{\yng(2)}+q \mathcal{O}_{\yng(2,1)}~, \\
   &\mathcal{O}_{\yng(1,1)} \;\mathcal{O}_{\yng(2,2)} = q^2-q \mathcal{O}_{\yng(2)}+q \mathcal{O}_{\yng(2,1)}~, \\
    &\mathcal{O}_{\yng(2)}^2 =-2 q \mathcal{O}_{\yng(1)}-q \mathcal{O}_{\yng(1,1)}+ q \mathcal{O}_{\yng(2)}+q \mathcal{O}_{\yng(2,1)} +(q+1)\mathcal{O}_{\yng(2,2)}~, \\
    &\mathcal{O}_{\yng(2)}\; \mathcal{O}_{\yng(2,1)} = q-q \mathcal{O}_{\yng(1)}-q \mathcal{O}_{\yng(1,1)}-q \mathcal{O}_{\yng(2)}+2 q \mathcal{O}_{\yng(2,1)}~,\\
    &\mathcal{O}_{\yng(2)} \;\mathcal{O}_{\yng(2,2)} = q^2-q \mathcal{O}_{\yng(1,1)}-q \mathcal{O}_{\yng(2,1)}+2 q \mathcal{O}_{\yng(2,2)} ~,\\
     &\mathcal{O}_{\yng(2,1)}^2 = -q \mathcal{O}_{\yng(1,1)}-q \mathcal{O}_{\yng(2)}+q \mathcal{O}_{\yng(2,1)}+q \mathcal{O}_{\yng(2,2)} ~,\\
     &\mathcal{O}_{\yng(2,1)} \;\mathcal{O}_{\yng(2,2)} =q^2 -q \mathcal{O}_{\yng(2,1)}+q \mathcal{O}_{\yng(2,2)}~, \\
  &\mathcal{O}_{\yng(2,2)}^2 = q^2~. \\
  \end{split}
\end{align}
All the other theories in the geometric window can be worked out similarly. 

%%%%%%%%%%%%%%%%%%%%%%%%%%%%%%%%%%%%%%%%%%%%%%
%%%%%%%%%%%%%%%%%%%%%%%%%%%%%%%%%%%%%%%%%%%%%%
%%%%%%%%%%%%%%%%%%%%%%%%%%%%%%%%%%%%%%%%%%%%%%
 
%%%%%%%%%%%%%%%%%%%%%%%%%%%%%%
    
%%%%%%%%%%%%%%%%%%%%%%%%%%%%%%%
 \section{Another example: The 3d GLSM for ${\rm Gr}(3,5)$}\label{app: Gr(3,5)}
\begin{figure}\label{fig: Hasse Gr35}
    \centering
    
\scalebox{0.8}{ \begin{tikzpicture}[baseline=15mm]

%%%%%%% For partition 000
\node[node3ds] (0001) []{$3$};
\node[squareMini] (0002) [below= 0.85 of 0001]{$5$};
\draw[->-=0.6,very thick] (0001) --  (0002) node[midway,right]{$\; \; \phi =  \begin{pmatrix} 1&0&0&\star&\star\\0&1&0&\star&\star\\0&0&1&\star&\star\end{pmatrix}$};

%%%%%%%%%%%%%For partition 100
\node[node3ds] (1001) [below = 0.851 of 0002]{$3$};
\node[squareMini] (1002) [below= 0.85 of 1001]{$5$};
\draw[->-=0.6,very thick] (1001) --  (1002) node[midway,right]{$\; \; \phi =  \begin{pmatrix} 1&0&\star&0&\star\\0&1&\star&0&\star\\0&0&0&1&\star\end{pmatrix}$};

\node[node1ds] (100r1) [left= 0.95 of 1001]{$1$};
\draw[->-=0.6, thick] (100r1) --  (1001) node[midway,above]{$\varphi_1^{2}$};
\draw[->-=0.6,red, dashed, thick] (100r1) -- (1002) node[midway,below]{$ (123)\;\;\;\;\;\;\;\;\;$};
 \node[fit=(100r1),  dashed, blue, draw, inner sep=9pt, minimum width=2cm,minimum height=2.6cm, shift={(0.3cm,-0.65cm)}] (1001d) {};

\node[ dashed, blue, draw, inner sep=9pt, minimum width=2cm,minimum height=2.6cm, shift={(0cm,-0.65cm)}] (0001d) [above = 1.35 of 1001d] {};

\node[] (yng100) [left = 0.05 of 1001d]{$\yng(1)$};
%%%%%%%%%%%%%%%%%%%

\node[](110200mid) [below = 0.851 of 1002] {};

%%%%%%%%%%%%%For partition 110
\node[node3ds] (1101) [right = 4 of 110200mid]{$3$};
\node[squareMini] (1102) [below= 0.85 of 1101]{$5$};
\draw[->-=0.6,very thick] (1101) --  (1102) node[midway,right]{$\; \; \phi =  \begin{pmatrix}1&\star&0&0&\star\\0&0&1&0&\star\\0&0&0&1&\star\end{pmatrix}$};

\node[node1ds] (110r2) [left= 0.95 of 1101]{$2$};
\draw[->-=0.6, thick] (110r2) --  (1101) node[midway,above]{$\varphi_1^{2}$};
\draw[->-=0.6,red, dashed, thick] (110r2) -- (1102) node[midway,above]{$\;(12)$};
 \node[fit=(110r2), dashed, blue, draw, inner sep=9pt, minimum width=2cm,minimum height=2.6cm, shift={(0.3cm,-0.65cm)}] (1101d) {};

\node[] (yng110) [left = 0.05 of 1101d]{$\yng(1,1)$};

%%%%%%%%%%%%%%For partition 200
\node[node3ds] (2001) [left= 5 of 110200mid]{$3$};
\node[squareMini] (2002) [below= 0.85 of 2001]{$5$};
\draw[->-=0.6,very thick] (2001) --  (2002) node[midway,right]{$\; \; \phi =  \begin{pmatrix} 1&0&\star&\star&0\\0&1&\star&\star&0\\0&0&0&0&1\end{pmatrix}$};

\node[node1ds] (200r1) [left= 0.95 of 2001]{$1$};
\draw[->-=0.6, thick] (200r1) --  (2001) node[midway,above]{$\varphi_1^{2}$};
\draw[->-=0.6,red, dashed, thick] (200r1) -- (2002) node[midway,below]{$(1234)\;\;\;\;\;\;\;\;$};
 \node[fit=(200r1),  dashed, blue, draw, inner sep=9pt, minimum width=2cm,minimum height=2.6cm, shift={(0.3cm,-0.65cm)}] (2001d) {};

\node[] (yng200) [left = 0.05 of 2001d]{$\yng(2)$};

%%%%%%%%%%%%%%%
\node[] (111210mid) [below = 0.851 of 1102] {};

%%%%%%%%%%%%%For partition 210
\node[node3ds] (2101) [left= 10 of 111210mid]{$3$};
\node[squareMini] (2102) [below= 0.85 of 2101]{$5$};
\draw[->-=0.6,very thick] (2101) --  (2102) node[midway,right]{$\; \; \phi =  \begin{pmatrix}1&\star&0&\star&0\\0&0&1&\star&0\\0&0&0&0&1 \end{pmatrix}$};

\node[node1ds] (210r2) [left= 0.95 of 2101]{$2$};
\node[node1ds] (210r1) [left= 0.95 of 210r2]{$1$};
\draw[->-=0.6, thick] (210r2) --  (2101) node[midway,above]{$\varphi_2^{3}$};
\draw[->-=0.6, thick] (210r1) -- (210r2) node[midway,above]{$\varphi_1^{2}$};
\draw[->-=0.6,red, dashed, thick] (210r1) -- (2102) node[midway,below]{$(34)\;\;\;\;$};
\draw[->-=0.6,red, dashed, thick] (210r2) -- (2102) node[midway,above]{$\;(12)$};
 \node[fit=(210r1)(210r2), dashed, blue, draw, inner sep=9pt, minimum width=3.5cm,minimum height=2.6cm, shift={(0.4cm,-0.65cm)}] (2101d) {};

\node[] (yng210) [left = 0.05 of 2101d]{$\yng(2,1)$};

%%%%%%%%%%%%%For partition 111
\node[node3ds] (1111) [right= 0 of 111210mid]{$3$};
\node[squareMini] (1112) [below= 0.85 of 1111]{$5$};
\draw[->-=0.6,very thick] (1111) --  (1112) node[midway,right]{$\; \; \phi =  \begin{pmatrix} 0&1&0&0&\star\\0&0&1&0&\star\\0&0&0&1&\star\end{pmatrix}$};

\node[node1ds] (111r3) [left= 0.95 of 1111]{$3$};
\draw[->-=0.6, thick] (111r3) --  (1111) node[midway,above]{$\varphi_1^{2}$};
\draw[->-=0.6,red, dashed, thick] (111r3) -- (1112) node[midway,above]{$(1)$};
 \node[fit=(111r3), dashed, blue, draw, inner sep=9pt, minimum width=2cm,minimum height=2.6cm, shift={(0.3cm,-0.65cm)}] (1111d) {};

\node[] (yng111) [left = 0.05 of 1111d]{$\yng(1,1,1)$};

%%%%%%%%%%%%%For partition 220
\node[node3ds] (2201) [below= 3.7 of 2002]{$3$};
\node[squareMini] (2202) [below= 0.85 of 2201]{$5$};
\draw[->-=0.6,very thick] (2201) --  (2202) node[midway,right]{$\; \; \phi =  \begin{pmatrix}1&\star&\star&0&0\\0&0&0&1&0\\0&0&0&0&1\end{pmatrix}$};

\node[node1ds] (220r2) [left= 0.95 of 2201]{$2$};
\draw[->-=0.6, thick] (220r2) --  (2201) node[midway,above]{$\varphi_1^{2}$};
\draw[->-=0.6,red, dashed, thick] (220r2) -- (2202) node[midway,above]{$\;\;(123)$};
 \node[fit=(220r2), dashed, blue, draw, inner sep=9pt,minimum width=2cm,minimum height=2.6cm, shift={(0.3cm,-0.65cm)}] (2201d) {};

\node[] (yng220) [left = 0.05 of 2201d]{$\yng(2,2)$};

%%%%%%%%%%%%%For partition 211
\node[node3ds] (2111) [below = 3.7 of 1102]{$3$};
\node[squareMini] (2112) [below= 0.85 of 2111]{$5$};
\draw[->-=0.6,very thick] (2111) --  (2112) node[midway,right]{$\; \; \phi =  \begin{pmatrix}0&1&0&\star&0\\0&0&1&\star&0\\0&0&0&0&1\end{pmatrix}$};

\node[node1ds] (211r3) [left= 0.95 of 2111]{$3$};
\node[node1ds] (211r1) [left= 0.95 of 211r3]{$1$};
\draw[->-=0.6, thick] (211r3) --  (2111) node[midway,above]{$\varphi_2^{3}$};
\draw[->-=0.6, thick] (211r1) -- (211r3) node[midway,above]{$\varphi_1^{2}$};
\draw[->-=0.6,red, dashed, thick] (211r1) -- (2112) node[midway,below]{$(234)\;\;\;$};
\draw[->-=0.6,red, dashed, thick] (211r3) -- (2112) node[midway,above]{$(1)$};
 \node[fit=(211r1)(211r3), dashed, blue, draw, inner sep=9pt, minimum width=3.5cm,minimum height=2.6cm, shift={(0.4cm,-0.65cm)}] (2111d) {};

\node[] (yng211) [left = 0.05 of 2111d]{$\yng(2,1,1)$};

%%%%%%%%%%%%%For partition 221
\node[node3ds] (2211) [below = 4 of 2102,shift={(5cm,0cm)}]{$3$};
\node[squareMini] (2212) [below= 0.85 of 2211]{$5$};
\draw[->-=0.6,very thick] (2211) --  (2212) node[midway,right]{$\; \; \phi =  \begin{pmatrix}0&1&\star&0&0\\0&0&0&1&0\\0&0&0&0&1\end{pmatrix}$};

\node[node1ds] (221r3) [left= 0.95 of 2211]{$3$};
\node[node1ds] (221r2) [left= 0.95 of 221r3]{$2$};
\draw[->-=0.6, thick] (221r3) --  (2211) node[midway,above]{$\varphi_2^{3}$};
\draw[->-=0.6, thick] (221r2) -- (221r3) node[midway,above]{$\varphi_1^{2}$};
\draw[->-=0.6,red, dashed, thick] (221r2) -- (2212) node[midway,below]{$(23)\;\;\;$};
\draw[->-=0.6,red, dashed, thick] (221r3) -- (2212) node[midway,above]{$(1)$};
 \node[fit=(221r2)(221r3), dashed, blue, draw, inner sep=9pt, minimum width=3.5cm,minimum height=2.6cm, shift={(0.4cm,-0.65cm)}] (2211d) {};

\node[] (yng221) [left = 0.05 of 2211d]{$\yng(2,2,1)$};

%%%%%%%%%%%%%For partition 222
\node[node3ds] (2221) [below = 0.851 of 2212]{$3$};
\node[squareMini] (2222) [below= 0.85 of 2221]{$5$};
\draw[->-=0.6,very thick] (2221) --  (2222) node[midway,right]{$\; \; \phi =  \begin{pmatrix}0&0&1&0&0\\0&0&0&1&0\\0&0&0&0&1\end{pmatrix}$};

\node[node1ds] (222r3) [left= 0.95 of 2221]{$3$};
\draw[->-=0.6, thick] (222r3) --  (2221) node[midway,above]{$\varphi_1^{2}$};
\draw[->-=0.6,red, dashed, thick] (222r3) -- (2222) node[midway,above]{$\;(12)$};
 \node[fit=(222r3),dashed, blue, draw, inner sep=9pt,minimum width=2cm,minimum height=2.6cm, shift={(0.3cm,-0.65cm)}] (2221d) {};
 
 \node[] (yng222) [left = 0.05 of 2221d]{$\yng(2,2,2)$};
%%%%%%%%%%connecting them
\draw[-, black, dashed] (2211d) -- (2221d);
\draw[-, black, dashed] (2111d) -- (2211d);
\draw[-, black, dashed] (2201d.south) -- (2211d);
\draw[-, black, dashed] (2101d) -- (2201d);
\draw[-, black, dashed] (2101d) -- (2111d);
\draw[-, black, dashed] (1111d.south) -- (2111d.north);
\draw[-, black, dashed] (1101d) -- (1111d);
\draw[-, black, dashed] (1101d) -- (2101d);
\draw[-, black, dashed] (2001d.south) -- (2101d);
\draw[-, black, dashed] (1001d.south) -- (2001d.north);
\draw[-, black, dashed] (1001d.south) -- (1101d);
\draw[-, black, dashed] (1001d) -- (0001d);
 \end{tikzpicture}
 }
    \caption{The Hasse diagram associated with the Schubert subvarieties of Gr$(3,5)$, with the 1d quivers simplified using the duality moves \eqref{duality move}.}
    \label{fig:Gr35 hass diag bis}
\end{figure}
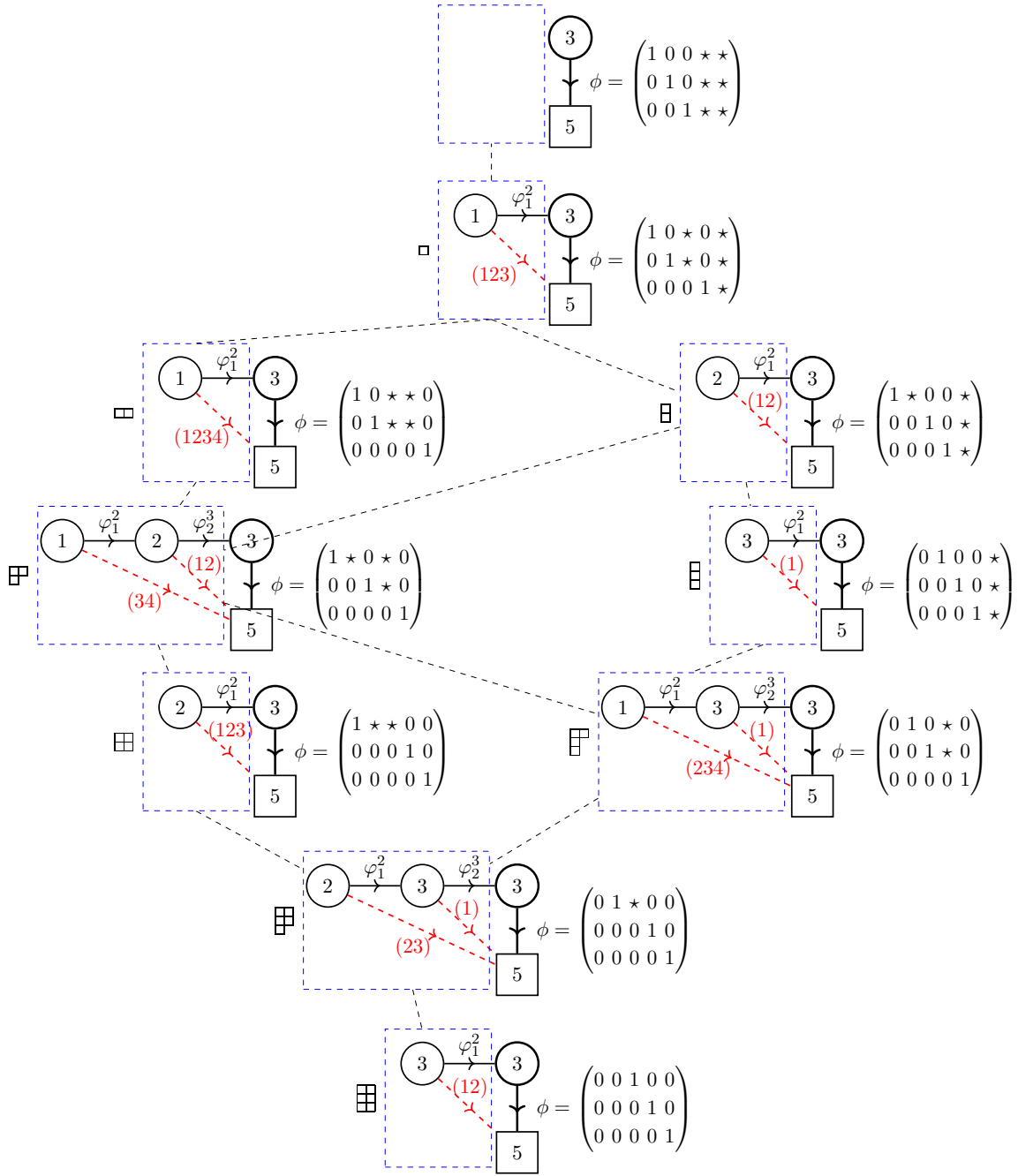

 In this appendix, we briefly display   another example, namely the 3d GLSM onto Gr$(3,5)$ that gives us its ordinary QK ring. This is the $U(3)_{\half, -{5\ov 2}}$ gauge theory with $n_f=5$. We restrict ourselves to the non-equivariant limit for simplicity of exposition. The Grothendieck lines are shown explicitly in figures~\ref{fig:Gr35 hass diag} and~\ref{fig:Gr35 hass diag bis}.

%\CyC{TBC}

\subsection{QK$($Gr$(3,5))$}

\begin{align}
    \begin{split}
        &\mathcal{O}_{\yng(1)}^2 = \mathcal{O}_{\yng(1,1)}+\mathcal{O}_{\yng(2)}-\mathcal{O}_{\yng(2,1)}~, \qquad\qquad 
 \mathcal{O}_{\yng(1)}\; \mathcal{O}_{\yng(1,1)} = \mathcal{O}_{\yng(2,1)}+\mathcal{O}_{\yng(1,1,1)}-\mathcal{O}_{\yng(2,1,1)}~, \\
 & \mathcal{O}_{\yng(1)}\; \mathcal{O}_{\yng(2)} = \mathcal{O}_{\yng(2,1)} ~, \qquad\qquad \quad\;\; \qquad 
  \mathcal{O}_{\yng(1)}\; \mathcal{O}_{\yng(2,1)} = \mathcal{O}_{\yng(2,2)} + \mathcal{O}_{\yng(2,1,1)}-\mathcal{O}_{\yng(2,2,1)} ~,\\
   &\mathcal{O}_{\yng(1)}\; \mathcal{O}_{\yng(1,1,1)} = \mathcal{O}_{\yng(2,1,1)}~, \qquad \qquad \qquad\qquad\mathcal{O}_{\yng(1)}\; \mathcal{O}_{\yng(2,2)} =\mathcal{O}_{\yng(2,2,1)}~, \\
  &\mathcal{O}_{\yng(1)}\; \mathcal{O}_{\yng(2,1,1)} = q-q \mathcal{O}_{\yng(1)}+\mathcal{O}_{\yng(2,2,1)}~, \qquad\quad
  \mathcal{O}_{\yng(1)} \; \mathcal{O}_{\yng(2,2,1)} = q \mathcal{O}_{\yng(1)}-q \mathcal{O}_{\yng(1,1)} + \mathcal{O}_{\yng(2,2,2)}~, \\
   &\mathcal{O}_{\yng(1)}\; \mathcal{O}_{\yng(2,2,2)} =  q \mathcal{O}_{\yng(1,1)}~, \qquad \qquad \qquad\qquad
 \mathcal{O}_{\yng(1,1)}^2 = \mathcal{O}_{\yng(2,2)}+\mathcal{O}_{\yng(2,1,1)}-\mathcal{O}_{\yng(2,2,1)}~, \\
 &\mathcal{O}_{\yng(1,1)} \;\mathcal{O}_{\yng(2)} = \mathcal{O}_{\yng(2,1,1)}~, \qquad \qquad \qquad \qquad
  \mathcal{O}_{\yng(1,1)} \;\mathcal{O}_{\yng(2,1)} = q-q \mathcal{O}_{\yng(1)}+\mathcal{O}_{\yng(2,2,1)}~, \\
%    \end{split}
%\end{align}
%
%
%
%\begin{align}
%\begin{split}    
 &\mathcal{O}_{\yng(1,1)} \;\mathcal{O}_{\yng(1,1,1)} = \mathcal{O}_{\yng(2,2,1)}~, \qquad\qquad \qquad \qquad \;\;\;
  \mathcal{O}_{\yng(1,1)}\; \mathcal{O}_{\yng(2,2)} = q \mathcal{O}_{\yng(1)}~, \\
  &\mathcal{O}_{\yng(1,1)}\; \mathcal{O}_{\yng(2,1,1)} = \mathcal{O}_{\yng(2,2,2)}+q \mathcal{O}_{\yng(1)}-q \mathcal{O}_{\yng(1,1)}~,  \qquad\;\;
  \mathcal{O}_{\yng(1,1)}\;\mathcal{O}_{\yng(2,2,1)} = q \mathcal{O}_{\yng(1,1)}+q \mathcal{O}_{\yng(2)}-q \mathcal{O}_{\yng(2,1)}~, \\
 & \mathcal{O}_{\yng(1,1)}\; \mathcal{O}_{\yng(2,2,2)}  = q \mathcal{O}_{\yng(2,1)}~,\qquad \qquad \qquad\quad\;\;\;\;
  \mathcal{O}_{\yng(2)}^2  = \mathcal{O}_{\yng(2,2)}~,\\
  &\mathcal{O}_{\yng(2)} \;\mathcal{O}_{\yng(2,1)} = \mathcal{O}_{\yng(2,2,1)}~, \qquad \qquad \qquad \quad\;\;\;\;\;
  \mathcal{O}_{\yng(2)} \;\mathcal{O}_{\yng(1,1,1)} = q ~,\\
   &\mathcal{O}_{\yng(2)} \;\mathcal{O}_{\yng(2,2)}=\mathcal{O}_{\yng(2,2,2)}~, \qquad \qquad \qquad\quad\;\;\;\;\;
    \mathcal{O}_{\yng(2)} \;\mathcal{O}_{\yng(2,1,1)} =  q \mathcal{O}_{\yng(1)} \\
    & \mathcal{O}_{\yng(2)} \;\mathcal{O}_{\yng(2,2,1)} = q \mathcal{O}_{\yng(1,1)}~, \qquad \qquad\qquad\qquad\;\;
    \mathcal{O}_{\yng(2)}\; \mathcal{O}_{\yng(2,2,2)} = q \mathcal{O}_{\yng(1,1,1)}~, \\
     &\mathcal{O}_{\yng(2,1)}^2 = q \mathcal{O}_{\yng(1)}-q \mathcal{O}_{\yng(1,1)} + \mathcal{O}_{\yng(2,2,2)} ~,\qquad\qquad\;
      \mathcal{O}_{\yng(2,1)} \;\mathcal{O}_{\yng(1,1,1)}   = q \mathcal{O}_{\yng(1)}~,\\
       &\mathcal{O}_{\yng(2,1)}\; \mathcal{O}_{\yng(2,2)} = q \mathcal{O}_{\yng(1,1)}~, \qquad \qquad \qquad\qquad\;\;
        \mathcal{O}_{\yng(2,1)}\; \mathcal{O}_{\yng(2,1,1)}=q \mathcal{O}_{\yng(1,1)}+q \mathcal{O}_{\yng(2)}-q \mathcal{O}_{\yng(2,1)}~, \\
        &\mathcal{O}_{\yng(2,1)} \;\mathcal{O}_{\yng(2,2,1)} = q \mathcal{O}_{\yng(1,1,1)}+q \mathcal{O}_{\yng(2,1)}-q \mathcal{O}_{\yng(2,1,1)}~, \qquad
         \mathcal{O}_{\yng(2,1)}\; \mathcal{O}_{\yng(2,2,2)} =q \mathcal{O}_{\yng(2,1,1)}\\
& \mathcal{O}_{\yng(1,1,1)}^2 = \mathcal{O}_{\yng(2,2,2)} ~,\qquad\qquad\qquad\qquad\;\;\qquad\;\;\;
\mathcal{O}_{\yng(1,1,1)} \;\mathcal{O}_{\yng(2,1,1)} =  q \mathcal{O}_{\yng(1,1)}~, \\
 &\mathcal{O}_{\yng(1,1,1)}\; \mathcal{O}_{\yng(2,2)} = q \mathcal{O}_{\yng(2)} ~,\qquad\qquad \qquad\;\;\qquad\;
  \mathcal{O}_{\yng(1,1,1)}\; \mathcal{O}_{\yng(2,2,1)} = q \mathcal{O}_{\yng(2,1)}~, \\
  &\mathcal{O}_{\yng(1,1,1)}\; \mathcal{O}_{\yng(2,2,2)} = q \mathcal{O}_{\yng(2,2)}~, \qquad \qquad \qquad \qquad\;\;
 \mathcal{O}_{\yng(2,1,1)}^2 = q \mathcal{O}_{\yng(2,1)}+q \mathcal{O}_{\yng(1,1,1)}-q \mathcal{O}_{\yng(2,1,1)}~,\\
&\mathcal{O}_{\yng(2,1,1)}\; \mathcal{O}_{\yng(2,2)} =  q \mathcal{O}_{\yng(2,1)}~,\qquad \qquad \qquad \qquad
 \mathcal{O}_{\yng(2,1,1)} \;\mathcal{O}_{\yng(2,2,1)}  = q \mathcal{O}_{\yng(2,1,1)}+q \mathcal{O}_{\yng(2,2)}-q \mathcal{O}_{\yng(2,2,1)}~, \\
  &\mathcal{O}_{\yng(2,1,1)}\; \mathcal{O}_{\yng(2,2,2)} = q \mathcal{O}_{\yng(2,2,1)}~, \qquad \qquad \qquad 
  \qquad\mathcal{O}_{\yng(2,2)}^2  =  q \mathcal{O}_{\yng(1,1,1)}~,\\
  &\mathcal{O}_{\yng(2,2)}\; \mathcal{O}_{\yng(2,2,1)} = q \mathcal{O}_{\yng(2,1,1)}~, \qquad \qquad \qquad\qquad
  \mathcal{O}_{\yng(2,2)} \;\mathcal{O}_{\yng(2,2,2)} = q^2~, \\
   &\mathcal{O}_{\yng(2,2,1)}^2 = q^2-q^2\mathcal{O}_{\yng(1)}+q \mathcal{O}_{\yng(2,2,1)}~, \qquad \qquad
 \mathcal{O}_{\yng(2,2,1)}\; \mathcal{O}_{\yng(2,2,2)}=   q^2\mathcal{O}_{\yng(1)}~, \\
 &\mathcal{O}_{\yng(2,2,2)}^2 =q^2 \mathcal{O}_{\yng(2)} \\
\end{split}
\end{align}

%%%%%%%%%
\begin{figure}\label{fig: Hasse Gr35}
    \centering
    
\scalebox{0.8}{ \begin{tikzpicture}[baseline=15mm]

%%%%%%% For partition 000
\node[node3ds] (0001) []{$3$};
\node[squareMini] (0002) [below= 0.85 of 0001]{$5$};
\draw[->-=0.6,very thick] (0001) --  (0002) node[midway,right]{$\; \; \phi =  \begin{pmatrix} 1&0&0&\star&\star\\0&1&0&\star&\star\\0&0&1&\star&\star\end{pmatrix}$};

%%%%%%%%%%%%%For partition 100
\node[node3ds] (1001) [below = 0.851 of 0002]{$3$};
\node[squareMini] (1002) [below= 0.85 of 1001]{$5$};
\draw[->-=0.6,very thick] (1001) --  (1002) node[midway,right]{$\; \; \phi =  \begin{pmatrix} 1&0&\star&0&\star\\0&1&\star&0&\star\\0&0&0&1&\star\end{pmatrix}$};

\node[node1ds] (100r1) [left= 0.95 of 1001]{$1$};
\draw[->-=0.6, thick] (100r1) --  (1001) node[midway,above]{$\varphi_1^{2}$};
\draw[->-=0.6,red, dashed, thick] (100r1) -- (1002) node[midway,below]{$ (123)\;\;\;\;\;\;\;\;\;$};
 \node[fit=(100r1),  dashed, blue, draw, inner sep=9pt, minimum width=2cm,minimum height=2.6cm, shift={(0.3cm,-0.65cm)}] (1001d) {};

\node[ dashed, blue, draw, inner sep=9pt, minimum width=2cm,minimum height=2.6cm, shift={(0cm,-0.65cm)}] (0001d) [above = 1.35 of 1001d] {};

\node[] (yng100) [left = 0.05 of 1001d]{$\yng(1)$};
%%%%%%%%%%%%%%%%%%%

\node[](110200mid) [below = 0.851 of 1002] {};

%%%%%%%%%%%%%For partition 110
\node[node3ds] (1101) [right = 4 of 110200mid]{$3$};
\node[squareMini] (1102) [below= 0.85 of 1101]{$5$};
\draw[->-=0.6,very thick] (1101) --  (1102) node[midway,right]{$\; \; \phi =  \begin{pmatrix}1&\star&0&0&\star\\0&0&1&0&\star\\0&0&0&1&\star\end{pmatrix}$};

\node[node1ds] (110r2) [left= 0.95 of 1101]{$2$};
\node[node1ds] (110r1) [left= 0.95 of 110r2]{$1$};
\draw[->-=0.6, thick] (110r2) --  (1101) node[midway,above]{$\varphi_2^{3}$};
\draw[->-=0.6, thick] (110r1) -- (110r2) node[midway,above]{$\varphi_1^{2}$};
\draw[->-=0.6,red, dashed, thick] (110r1) -- (1102) node[midway,below]{$(3)\;$};
\draw[->-=0.6,red, dashed, thick] (110r2) -- (1102) node[midway,above]{$\;(12)$};
 \node[fit=(110r1)(110r2), dashed, blue, draw, inner sep=9pt, minimum width=3.5cm,minimum height=2.6cm, shift={(0.4cm,-0.65cm)}] (1101d) {};

\node[] (yng110) [left = 0.05 of 1101d]{$\yng(1,1)$};

%%%%%%%%%%%%%%For partition 200
\node[node3ds] (2001) [left= 5 of 110200mid]{$3$};
\node[squareMini] (2002) [below= 0.85 of 2001]{$5$};
\draw[->-=0.6,very thick] (2001) --  (2002) node[midway,right]{$\; \; \phi =  \begin{pmatrix} 1&0&\star&\star&0\\0&1&\star&\star&0\\0&0&0&0&1\end{pmatrix}$};

\node[node1ds] (200r1) [left= 0.95 of 2001]{$1$};
\draw[->-=0.6, thick] (200r1) --  (2001) node[midway,above]{$\varphi_1^{2}$};
\draw[->-=0.6,red, dashed, thick] (200r1) -- (2002) node[midway,below]{$(1234)\;\;\;\;\;\;\;\;$};
 \node[fit=(200r1),  dashed, blue, draw, inner sep=9pt, minimum width=2cm,minimum height=2.6cm, shift={(0.3cm,-0.65cm)}] (2001d) {};

\node[] (yng200) [left = 0.05 of 2001d]{$\yng(2)$};

%%%%%%%%%%%%%%%
\node[] (111210mid) [below = 0.851 of 1102] {};

%%%%%%%%%%%%%For partition 210
\node[node3ds] (2101) [left= 10 of 111210mid]{$3$};
\node[squareMini] (2102) [below= 0.85 of 2101]{$5$};
\draw[->-=0.6,very thick] (2101) --  (2102) node[midway,right]{$\; \; \phi =  \begin{pmatrix}1&\star&0&\star&0\\0&0&1&\star&0\\0&0&0&0&1 \end{pmatrix}$};

\node[node1ds] (210r2) [left= 0.95 of 2101]{$2$};
\node[node1ds] (210r1) [left= 0.95 of 210r2]{$1$};
\draw[->-=0.6, thick] (210r2) --  (2101) node[midway,above]{$\varphi_2^{3}$};
\draw[->-=0.6, thick] (210r1) -- (210r2) node[midway,above]{$\varphi_1^{2}$};
\draw[->-=0.6,red, dashed, thick] (210r1) -- (2102) node[midway,below]{$(34)\;\;\;\;$};
\draw[->-=0.6,red, dashed, thick] (210r2) -- (2102) node[midway,above]{$\;(12)$};
 \node[fit=(210r1)(210r2), dashed, blue, draw, inner sep=9pt, minimum width=3.5cm,minimum height=2.6cm, shift={(0.4cm,-0.65cm)}] (2101d) {};

\node[] (yng210) [left = 0.05 of 2101d]{$\yng(2,1)$};

%%%%%%%%%%%%%For partition 111
\node[node3ds] (1111) [right= 0 of 111210mid]{$3$};
\node[squareMini] (1112) [below= 0.85 of 1111]{$5$};
\draw[->-=0.6,very thick] (1111) --  (1112) node[midway,right]{$\; \; \phi =  \begin{pmatrix} 0&1&0&0&\star\\0&0&1&0&\star\\0&0&0&1&\star\end{pmatrix}$};

\node[node1ds] (111r3) [left= 0.95 of 1111]{$3$};
\node[node1ds] (111r2) [left= 0.95 of 111r3]{$2$};
\node[node1ds] (111r1) [left= 0.95 of 111r2]{$1$};
\draw[->-=0.6, thick] (111r3) --  (1111) node[midway,above]{$\varphi_3^{4}$};
\draw[->-=0.6, thick] (111r2) --  (111r3) node[midway,above]{$\varphi_2^{3}$};
\draw[->-=0.6, thick] (111r1) -- (111r2) node[midway,above]{$\varphi_1^{2}$};
\draw[->-=0.6,red, dashed, thick] (111r1) -- (1112) node[midway,below]{$(3)\;$};
\draw[->-=0.6,red, dashed, thick] (111r2) -- (1112) node[midway,above]{$(2)\;\;\;\;\;\;$};
\draw[->-=0.6,red, dashed, thick] (111r3) -- (1112) node[midway,above]{$(1)$};
 \node[fit=(111r1)(111r3), dashed, blue, draw, inner sep=9pt, minimum width=5.3cm,minimum height=2.6cm, shift={(0.4cm,-0.65cm)}] (1111d) {};

\node[] (yng111) [left = 0.05 of 1111d]{$\yng(1,1,1)$};

%%%%%%%%%%%%%For partition 220
\node[node3ds] (2201) [below= 3.7 of 2002]{$3$};
\node[squareMini] (2202) [below= 0.85 of 2201]{$5$};
\draw[->-=0.6,very thick] (2201) --  (2202) node[midway,right]{$\; \; \phi =  \begin{pmatrix}1&\star&\star&0&0\\0&0&0&1&0\\0&0&0&0&1\end{pmatrix}$};

\node[node1ds] (220r2) [left= 0.95 of 2201]{$2$};
\node[node1ds] (220r1) [left= 0.95 of 220r2]{$1$};
\draw[->-=0.6, thick] (220r2) --  (2201) node[midway,above]{$\varphi_2^{3}$};
\draw[->-=0.6, thick] (220r1) -- (220r2) node[midway,above]{$\varphi_1^{2}$};
\draw[->-=0.6,red, dashed, thick] (220r1) -- (2202) node[midway,below]{$(4)\;$};
\draw[->-=0.6,red, dashed, thick] (220r2) -- (2202) node[midway,above]{$\;\;(12 3)$};
 \node[fit=(220r1)(220r2), dashed, blue, draw, inner sep=9pt, minimum width=3.5cm,minimum height=2.6cm, shift={(0.4cm,-0.65cm)}] (2201d) {};

\node[] (yng220) [left = 0.05 of 2201d]{$\yng(2,2)$};

%%%%%%%%%%%%%For partition 211
\node[node3ds] (2111) [below = 3.7 of 1102]{$3$};
\node[squareMini] (2112) [below= 0.85 of 2111]{$5$};
\draw[->-=0.6,very thick] (2111) --  (2112) node[midway,right]{$\; \; \phi =  \begin{pmatrix}0&1&0&\star&0\\0&0&1&\star&0\\0&0&0&0&1\end{pmatrix}$};

\node[node1ds] (211r3) [left= 0.95 of 2111]{$3$};
\node[node1ds] (211r2) [left= 0.95 of 211r3]{$2$};
\node[node1ds] (211r1) [left= 0.95 of 211r2]{$1$};
\draw[->-=0.6, thick] (211r3) --  (2111) node[midway,above]{$\varphi_3^{4}$};
\draw[->-=0.6, thick] (211r2) --  (211r3) node[midway,above]{$\varphi_2^{3}$};
\draw[->-=0.6, thick] (211r1) -- (211r2) node[midway,above]{$\varphi_1^{2}$};
\draw[->-=0.6,red, dashed, thick] (211r1) -- (2112) node[midway,below]{$(34)\;$};
\draw[->-=0.6,red, dashed, thick] (211r2) -- (2112) node[midway,above]{$(2)\;\;\;\;\;\;$};
\draw[->-=0.6,red, dashed, thick] (211r3) -- (2112) node[midway,above]{$(1)$};
 \node[fit=(211r1)(211r3), dashed, blue, draw, inner sep=9pt, minimum width=5.3cm,minimum height=2.6cm, shift={(0.4cm,-0.65cm)}] (2111d) {};

\node[] (yng211) [left = 0.05 of 2111d]{$\yng(2,1,1)$};

%%%%%%%%%%%%%For partition 221
\node[node3ds] (2211) [below = 4 of 2102,shift={(5cm,0cm)}]{$3$};
\node[squareMini] (2212) [below= 0.85 of 2211]{$5$};
\draw[->-=0.6,very thick] (2211) --  (2212) node[midway,right]{$\; \; \phi =  \begin{pmatrix}0&1&\star&0&0\\0&0&0&1&0\\0&0&0&0&1\end{pmatrix}$};

\node[node1ds] (221r3) [left= 0.95 of 2211]{$3$};
\node[node1ds] (221r2) [left= 0.95 of 221r3]{$2$};
\node[node1ds] (221r1) [left= 0.95 of 221r2]{$1$};
\draw[->-=0.6, thick] (221r3) --  (2211) node[midway,above]{$\varphi_3^{4}$};
\draw[->-=0.6, thick] (221r2) --  (221r3) node[midway,above]{$\varphi_2^{3}$};
\draw[->-=0.6, thick] (221r1) -- (221r2) node[midway,above]{$\varphi_1^{2}$};
\draw[->-=0.6,red, dashed, thick] (221r1) -- (2212) node[midway,below]{$(4)\;$};
\draw[->-=0.6,red, dashed, thick] (221r2) -- (2212) node[midway,above]{$(23)\;\;\;\;\;\;$};
\draw[->-=0.6,red, dashed, thick] (221r3) -- (2212) node[midway,above]{$(1)$};
 \node[fit=(221r1)(221r3), dashed, blue, draw, inner sep=9pt, minimum width=5.3cm,minimum height=2.6cm, shift={(0.4cm,-0.65cm)}] (2211d) {};

\node[] (yng221) [left = 0.05 of 2211d]{$\yng(2,2,1)$};

%%%%%%%%%%%%%For partition 222
\node[node3ds] (2221) [below = 0.851 of 2212]{$3$};
\node[squareMini] (2222) [below= 0.85 of 2221]{$5$};
\draw[->-=0.6,very thick] (2221) --  (2222) node[midway,right]{$\; \; \phi =  \begin{pmatrix}0&0&1&0&0\\0&0&0&1&0\\0&0&0&0&1\end{pmatrix}$};

\node[node1ds] (222r3) [left= 0.95 of 2221]{$3$};
\node[node1ds] (222r2) [left= 0.95 of 222r3]{$2$};
\node[node1ds] (222r1) [left= 0.95 of 222r2]{$1$};
\draw[->-=0.6, thick] (222r3) --  (2221) node[midway,above]{$\varphi_3^{4}$};
\draw[->-=0.6, thick] (222r2) --  (222r3) node[midway,above]{$\varphi_2^{3}$};
\draw[->-=0.6, thick] (222r1) -- (222r2) node[midway,above]{$\varphi_1^{2}$};
\draw[->-=0.6,red, dashed, thick] (222r1) -- (2222) node[midway,below]{$(4)\;$};
\draw[->-=0.6,red, dashed, thick] (222r2) -- (2222) node[midway,above]{$(3)\;\;\;\;$};
\draw[->-=0.6,red, dashed, thick] (222r3) -- (2222) node[midway,above]{$\;(12)$};
 \node[fit=(222r1)(222r3),dashed, blue, draw, inner sep=9pt, minimum width=5.3cm,minimum height=2.6cm, shift={(0.4cm,-0.65cm)}] (2221d) {};
 
 \node[] (yng222) [left = 0.05 of 2221d]{$\yng(2,2,2)$};
%%%%%%%%%%connecting them
\draw[-, black, dashed] (2211d) -- (2221d);
\draw[-, black, dashed] (2111d) -- (2211d);
\draw[-, black, dashed] (2201d) -- (2211d);
\draw[-, black, dashed] (2101d) -- (2201d);
\draw[-, black, dashed] (2101d) -- (2111d);
\draw[-, black, dashed] (1111d.south) -- (2111d.north);
\draw[-, black, dashed] (1101d) -- (1111d);
\draw[-, black, dashed] (1101d) -- (2101d);
\draw[-, black, dashed] (2001d.south) -- (2101d);
\draw[-, black, dashed] (1001d.south) -- (2001d.north);
\draw[-, black, dashed] (1001d.south) -- (1101d);
\draw[-, black, dashed] (1001d) -- (0001d);
 \end{tikzpicture}
 }
    \caption{The Hasse diagram associated with the Schubert subvarieties of Gr$(3,5)$. The defining partitions are displayed at Young tableaux, and the `generic' Grothendieck line defects are shown explicitly.}
    \label{fig:Gr35 hass diag}
\end{figure}
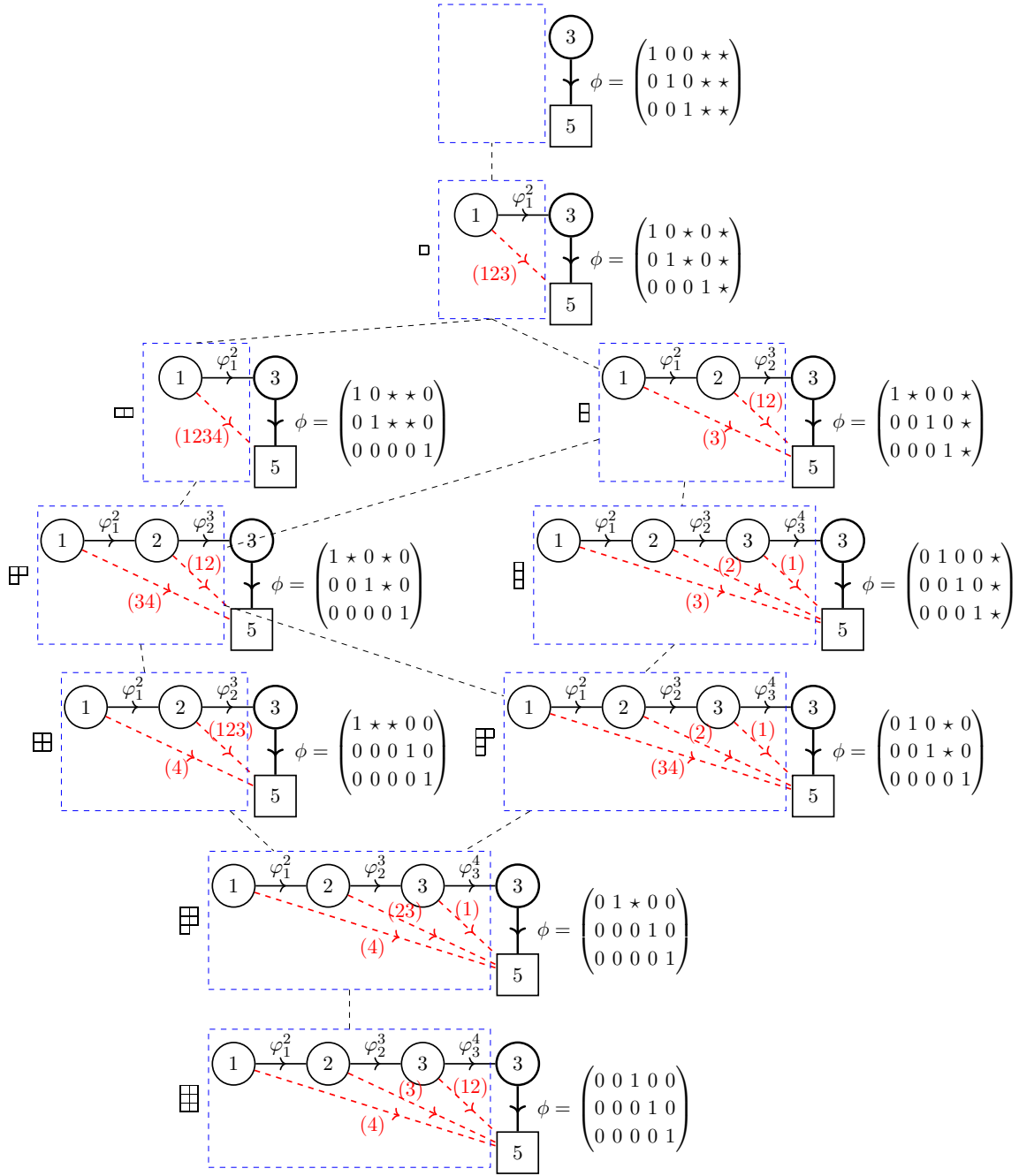
%%%%%%%%%%%%%%%%%%%%%%%%%%%%%%%%%%%%%%%%%%%%%%%%%%%%%%%
%%%%%%%%%%%%%%%%%%%%%%%%%%%%%%%%%%%%%%%%%%%%%%%%%%%%%%%
%%%%%%%%%%%%%%%%%%%%%%%%%%%%%%%%%%%%%%%%%%%%%%%%%%%%%%%
%%%%%%%%%%%%%%%%%%%%%%%%%%%%%%%%%%%%%%%%%%%%%%%%%%%%%%%
%%%%%%%%%%%%%%%%%%%%%%%%%%%%%%%%%%%%%%%%%%%%%%%%%%%%%%%
%%%%%%%%%%%%%%%%%%%%%%%%%%%%%%%%%%%%%%%%%%%%%%%%%%%%%%%

%%%%%%%%%

\bibliographystyle{JHEP}
\bibliography{3dbib}

\end{document}